 \documentclass[acmsmall]{acmart}

\AtBeginDocument{%
  \providecommand\BibTeX{{%
    \normalfont B\kern-0.5em{\scshape i\kern-0.25em b}\kern-0.8em\TeX}}}

\setcopyright{acmlicensed}
\acmJournal{TOIS}
\acmYear{2020} \acmVolume{1} \acmNumber{1} \acmArticle{1} \acmMonth{1} \acmPrice{15.00}\acmDOI{10.1145/3426723}

\usepackage[utf8]{inputenc}

\usepackage{threeparttable}
\usepackage{multirow}
\usepackage{tikz}
\usepackage{verbatim}
\usepackage{pgfplots}
\usetikzlibrary{plotmarks}
\usepackage{pgfplots}
\pgfplotsset{compat=1.12}
\usepackage{subcaption}
\usepackage{graphicx}
\usepackage{longtable}

\usepackage{amsthm}
\theoremstyle{definition}
\newtheorem{definition}{Definition}[section]

\begin{document}

\title{Deep Learning for Sequential Recommendation: Algorithms, Influential Factors, and Evaluations}

\author{Hui Fang}
\affiliation{
  \institution{RIIS \& SIME, Shanghai University of Finance and Economics}
  \country{China}}
\email{fang.hui@mail.shufe.edu.cn}

\author{Danning Zhang}
\authornote{Corresponding author}
\affiliation{
  \institution{SIME, Shanghai University of Finance and Economics}
  \country{China}}
\email{zhangdanning5@gmail.com}

\author{Yiheng Shu}
\affiliation{%
  \institution{Software College, Northeastern University}
  \country{China}}
\email{shuyiheng29@gmail.com}

\author{Guibing Guo}
\affiliation{%
  \institution{Software College, Northeastern University}
  \country{China}
}
\email{guogb@swc.neu.edu.cn}



\begin{abstract}
  In the field of sequential recommendation, deep learning (DL)-based methods have received a lot of attention in the past few years and surpassed traditional models such as Markov chain-based and factorization-based ones.
  However, there is little systematic study on DL-based methods, especially regarding how to design an effective DL model for sequential recommendation.
  In this view, this survey focuses on DL-based sequential recommender systems by taking the aforementioned issues into consideration.
  Specifically, we illustrate the concept of sequential recommendation, propose a categorization of existing algorithms in terms of three types of behavioral sequences, summarize the key factors affecting the performance of DL-based models, and conduct corresponding evaluations to showcase and demonstrate the effects of these factors. We conclude this survey by systematically outlining future directions and challenges in this field.
\end{abstract}

\ccsdesc[500]{Information systems~Recommender systems}

\keywords{sequential recommendation, session-based recommendation, deep learning, influential factors, survey, evaluations}

\maketitle


\section{Introduction}
With the prevalence of information technology (IT), recommender system has long been acknowledged as an effective and powerful tool for addressing information overload problem. It makes users easily filter and locate information in terms of their preferences, and allows online platforms to widely publicize the information they produce.
Most traditional recommender systems are content-based and collaborative filtering based ones. They strive to model users' preferences towards items on the basis of either explicit or implicit interactions between users and items.
Specifically, they incline to utilize a user's historical interactions to learn his/her static preference with the assumption that all user-item interactions in the historical sequences are equally important. However, this might not hold in real-world scenarios, where the user's next behavior not only depends on the static long-term preference, but also relies on the current intent to a large extent, which might be probably inferred and influenced by a small set of the most recent interactions. On the other side, the conventional approaches always ignore to consider the sequential dependencies among the user's interactions, leading to inaccurate modeling of the user's preferences.
Therefore, sequential recommendation has become increasingly popular in academic research and practical applications.

Sequential recommendation (identical to sequence-aware recommendation in \cite{quadrana2018sequence}) is also related to session-based, or session-aware recommendation. Considering that the latter two terms can be viewed as the sub-types of sequential recommendation \cite{quadrana2018sequence}, we thus use the much broader term \textit{sequential recommendation} to describe the task that explores the sequential data.

For sequential recommendation, besides capturing users' long-term preferences across different sessions as the conventional recommendation does, it is also extremely important to simultaneously model users' short-term interest in a session (or a short sequence) for accurate recommendation.
Regarding the time dependency among different interactions in a session as well as the correlation of behavior patterns among different sessions, traditional sequential recommender systems are particularly interested in employing appropriate
and effective machine learning (ML) approaches to model sequential data, such
as Markov Chain \cite{Davidson2010The} and session-based KNN \cite{he2016fusing,hu2020modeling}, which are criticized by their incomplete modeling problem, as they fail to thoroughly model users' long-term patterns by combining different sessions.

\begin{figure}[htbp]
   \centering
   \includegraphics[width=0.7\linewidth]{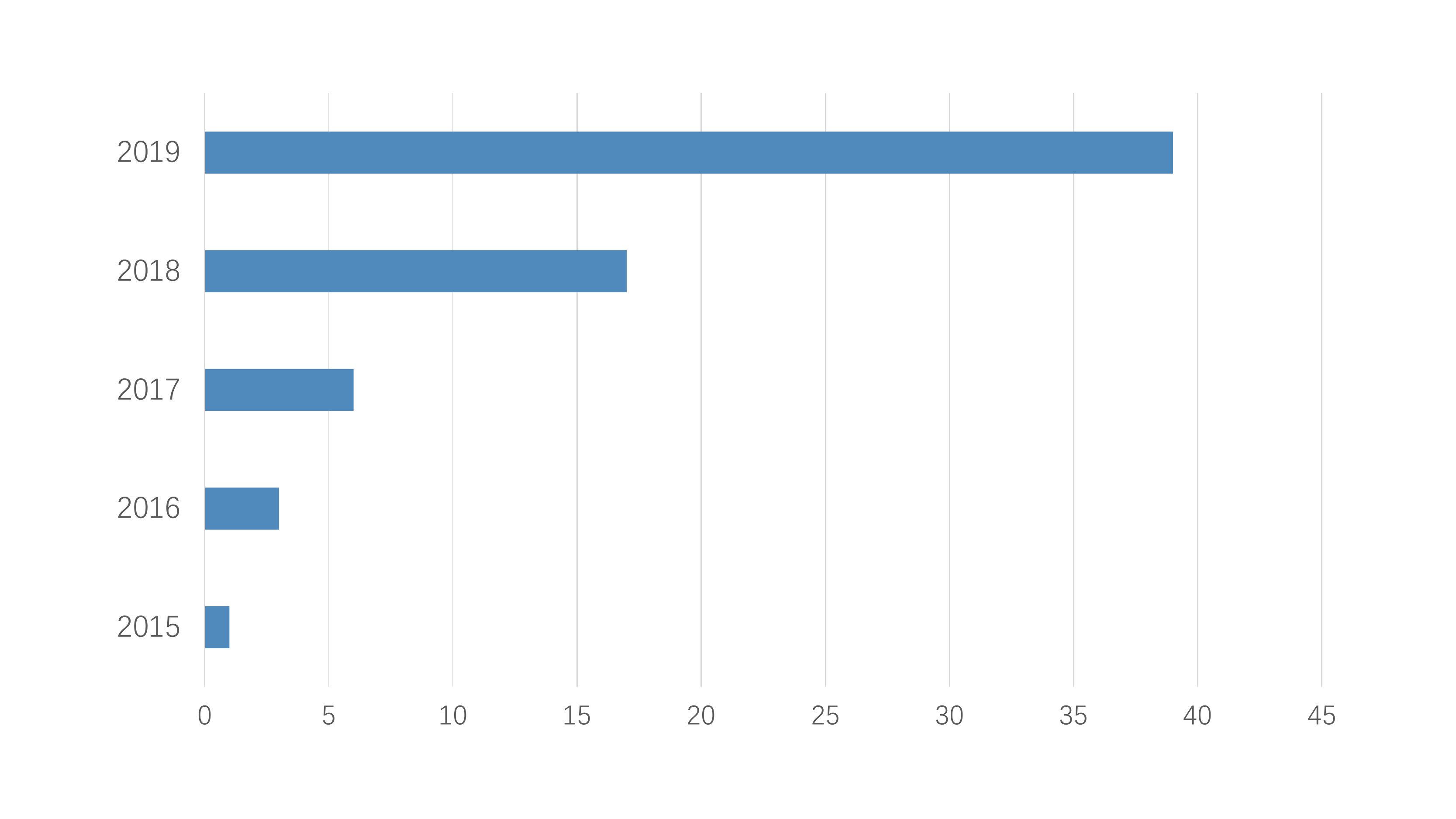}
   \vspace{-4mm}
    \caption{The number of arXiv articles on DL-based sequential recommendation in 2015-2019.}
    \label{fig:the number of relevant arXiv articles}
\end{figure}
\begin{figure}[htbp]
    \centering
   \includegraphics[width=0.85\linewidth]{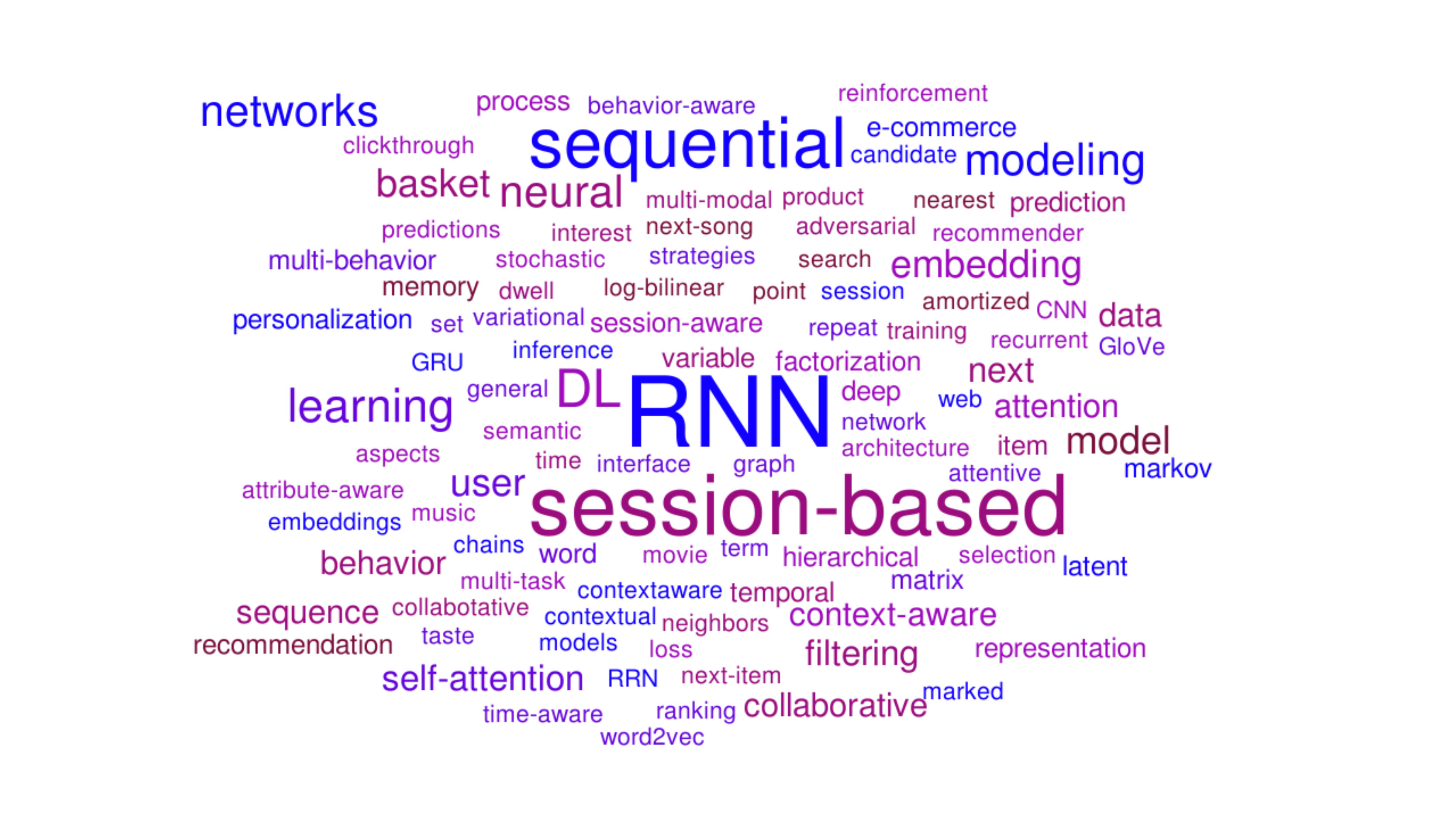}\vspace{-4mm}
    \caption{Word cloud of the keywords for sequential recommendation related articles.}
    \label{fig:word cloud}
\end{figure}
In recent years, deep learning (DL) techniques, such as recurrent neural network (RNN), obtain tremendous achievements in natural language processing (NLP), demonstrating their effectiveness in processing sequential data. Thus, they have attracted increasing interest in the sequential recommendation, and many DL-based models have achieved state-of-the-art performance \cite{zhang2019deep}.
The number of relevant arXiv articles posted in 2015-2019 is shown in Figure \ref{fig:the number of relevant arXiv articles}\footnote{We searched on arXiv.org for articles using the related keywords as well as their combinations, such as \textit{sequential recommendation}, \textit{deep learning} and \textit{session-based recommendation} in June 2020. We did not report the data of 2020 in Figure \ref{fig:the number of relevant arXiv articles} due to its incomplete. Figure \ref{fig:word cloud} considers all articles.}, where we can see that the interest in DL-based sequential recommendation has increased phenomenally. Besides, common application domains of sequential recommendation include e-commerce (e.g., RecSys Challenge 2015\footnote{\url{www.kaggle.com/chadgostopp/recsys-challenge-2015}.}), POI (Point-of-Interest), music (e.g., Last.fm\footnote{\url{labrosa.ee.columbia.edu/millionsong/lastfm}.}), and movie/video (e.g., MovieLens\footnote{\url{movielens.org}.}).
Figure \ref{fig:word cloud} depicts the word cloud of keywords in the sequential recommendation related articles, which to some extent reflects the hot topics in the field of sequential recommendation.

More and more new techniques and improved structures have been applied to facilitate the DL-based sequential recommendation. However, we find that the DL-based models still have some common drawbacks. For example, many existing works do not take items and users into consideration at the same important level, i.e., they largely emphasize item representation, but lack of a careful design on user representation. Besides, they merely consider all interactions into one type, instead of distinguishing different types. More importantly, although more advanced techniques have been increasingly adopted, it is still unclear the real progress for this area, as only complex DL structures could not possibly yield better performance \cite{Dacrema2019recsys}. In this case, it becomes extremely necessary to unveil the influential factors leading to a useful DL-based sequential recommender systems. Therefore, we tend to thoroughly discuss the improved techniques and the aforementioned issues in this survey.
The contributions of this survey are concluded as follows:
\begin{itemize}
    \item We provide a comprehensive overview of the sequential recommendation in terms of the deployed DL techniques. To the best of our knowledge, it is the first survey article on DL-based sequential recommendation.

    \item We propose an original classification framework for the sequential recommendation, corresponding to three different recommendation scenarios, which can contribute to the existing taxonomies as regards sequential recommendation.

    \item We summarize the influential factors for typical DL-based sequential recommendation and demonstrate their impacts on the sequential recommendation with regard to recommendation accuracy via a well-designed empirical study, which can serve as a guidance for sequential recommendation research and practices. Besides, our experimental settings and experimental repositories (including the source codes and datasets) can be viewed as a valuable testbed for future research.

    \item We summarize quite a few open issues in existing DL-based sequential recommendation and outline future directions.
\end{itemize}

\subsection{Related Survey}
There have been some surveys on either DL-based recommendation or sequential recommendation.
For DL-based recommendation, Singhal et al. \cite{singhal2017use} summarized DL-based recommender systems and categorized them into three types: collaborative filtering, content-based, and hybrid ones.
Batmaz et al. \cite{Batmaz2018} classified and summarized the DL-based recommendation from the perspectives of DL techniques and recommendation issues, and also gave a brief introduction of the session-based recommendations.
Zhang et al. \cite{zhang2019deep} further discussed the state-of-the-art DL-based recommender systems, including several RNN-based sequential recommendation algorithms. For sequential recommendation, Quadrana et al. \cite{quadrana2018sequence} proposed a categorization of the recommendation tasks and goals, and summarized existing solutions. Wang et al. \cite{wang2019sequential} summarized the key challenges, progress and future directions for sequential recommender systems. In a more comprehensive manner \cite{wang2019asurvey}, they further illustrated the value and significance of the session-based recommender systems (SBRS), and proposed a hierarchical framework to categorize issues and methods, including some DL-based ones.

However, to the best of our knowledge, our survey is the first to specifically and systematically summarize and explore DL-based sequential recommendation, and discuss the common influential factors using a thorough demonstration of experimental evaluations on several real datasets. The experiment results and conclusions can further guide the future research on how to design an effective DL model for the sequential recommendation.

\subsection{Structure of This Survey}
The rest of the survey is organized as follows. In Section \ref{sec:overview}, we provide a comprehensive overview of DL-based sequential recommender systems, including a careful refinement of sequential recommendation tasks.
In Section \ref{sec:algorithms_for_sequential_recommendation}, we present the details of the representative algorithms for each recommendation task. In Section \ref{sec:influential factors of models}, we summarize the influential factors for existing DL-based sequential recommendation followed by a thorough evaluation on real datasets in Section \ref{sec:evaluation of deep learning based sequential recommender system}. Finally, we conclude this survey by presenting open issues and future research directions of DL-based sequential recommendation in Section \ref{sec:future directions}.

\section{Overview of Sequential Recommendation}
\label{sec:overview}

In this section, we provide a comprehensive overview of the sequential recommendation. First, we clarify the related concepts, and then formally describe the sequential recommendation tasks.
Finally, we elaborate and compare the traditional ML and DL techniques for the sequential recommendation.

\subsection{Concept Definitions}
\label{subsec:problem description}
To facilitate the understanding, we first formally define \emph{behavior object} and \emph{behavior type} to distinguish different user behaviors in sequential data.
\begin{definition}
\textbf{behavior object} refers to the items or services that a user chooses to interact with, which is usually presented as an ID of an item or a set of items. It may be also associated with other information including text descriptions, images and interaction time. For simplicity, we often use \emph{item}(s) to describe behavior object(s) in the following sections.
\end{definition}
\begin{definition}
\textbf{behavior type} refers to the way that a user interacts with items or services, including \textit{search}, \textit{click}, \textit{add-to-cart}, \textit{buy}, \textit{share}, etc.
\end{definition}
\begin{figure}[htbp]
  \centering
  \includegraphics[width=0.7\linewidth]{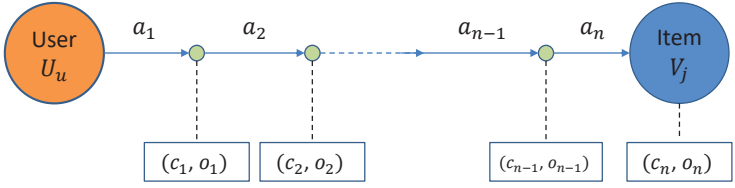}
  \caption{A schematic diagram of the sequential recommendation. $c_i$: behavior type, $o_i$: behavior object. A behavior $a_i$ is represented by a 2-tuple, i.e., $a_i=(c_i,o_i)$. A behavior sequence (i.e., behavior trajectory) is a list of 2-tuples in the order of time.}
  \label{fig:sequential recommendation}
\end{figure}
Given these concepts, a \textbf{behavior} can be considered as a combination of a behavior type and a behavior object, i.e., a user interacting with a behavior object by a behavior type. A \textbf{behavior trajectory} can be thus defined as a \emph{behavior sequence} (or behavior session) consisting of multiple user behaviors.
A typical behavior sequence is shown in Figure \ref{fig:sequential recommendation}. Specifically,
a behavior ($a_i$) is represented by a 2-tuple $(c_i,o_i)$, i.e., a behavior type $c_i$ and behavior object $o_i$. A user who generates the sequence can either be anonymous or identified by his/her ID. The behaviors in the sequence are sorted in time order. When a single behavior involves with multiple objects (e.g., items recorded in a shopping basket), objects within the basket may not be ordered by time, and then multiple baskets together form a behavior sequence. It should be noted that \emph{sequence} and \emph{session} are interchangeably used in this paper.

Thus, a \textbf{sequential
recommender system} is referred to a system which takes a user's behavior trajectories as input, and
then adopts recommendation algorithms to recommend appropriate items or services to the user.
The input behavior sequence $\{a_1,a_2,a_3,...,a_t\}$ is polymorphic, which can thus be divided into three types\footnote{We discuss the sequence in a finer granularity in the purpose of better understanding the sequential recommendation task.
We name the three types mainly according to the behavior types and objects involved in a sequence. We argue that it can promote better designs of network structures to process the corresponding sequences for sequential recommendation.}: \emph{experience-based}, \emph{transaction-based} and \emph{interaction-based} behavior sequence, and the details are elaborated as follows:

\begin{figure}[htbp]
  \includegraphics[width=0.7\linewidth]{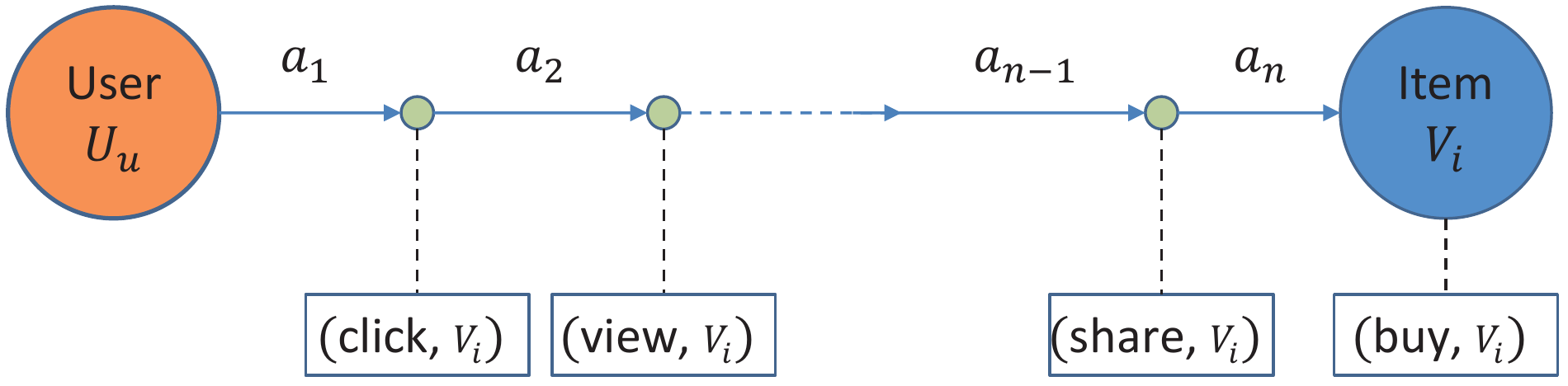}
  \caption{Experience-based behavior sequence.}
  \label{fig:Experience-based Behavior Sequence}
\end{figure}
\textbf{Experience-based behavior sequence.}
In an experience-based behavior sequence (see Figure \ref{fig:Experience-based Behavior Sequence}), a user may interact with a \emph{same object} (e.g., item $v_i$) multiple times by \emph{different behavior types}. For example, a user's interaction history with an item might be as follows: first \textit{searches} related keywords, then \textit{clicks} the item of interest on the result pages followed by \textit{viewing} the details of the item. Finally, the user may \textit{share} the item with his/her friends and \textit{add it to cart} if he/she likes it.
Different behavior types as well as their orders might indicate users' different intentions. For instance, \textit{click} and \textit{view} can only show a user's interest of a low degree, while \textit{share} behavior appears before (or after) \textit{purchase} might imply a user's strong desire (or satisfaction) to obtain (or have) the item.
\emph{For this type of behavior sequence, a model is expected to capture a user's underlying intentions indicated by different behavior types. The goal here is to predict the next behavior type that the user will exert given an item.}

\begin{figure}[htbp]
  \includegraphics[width=0.7\linewidth]{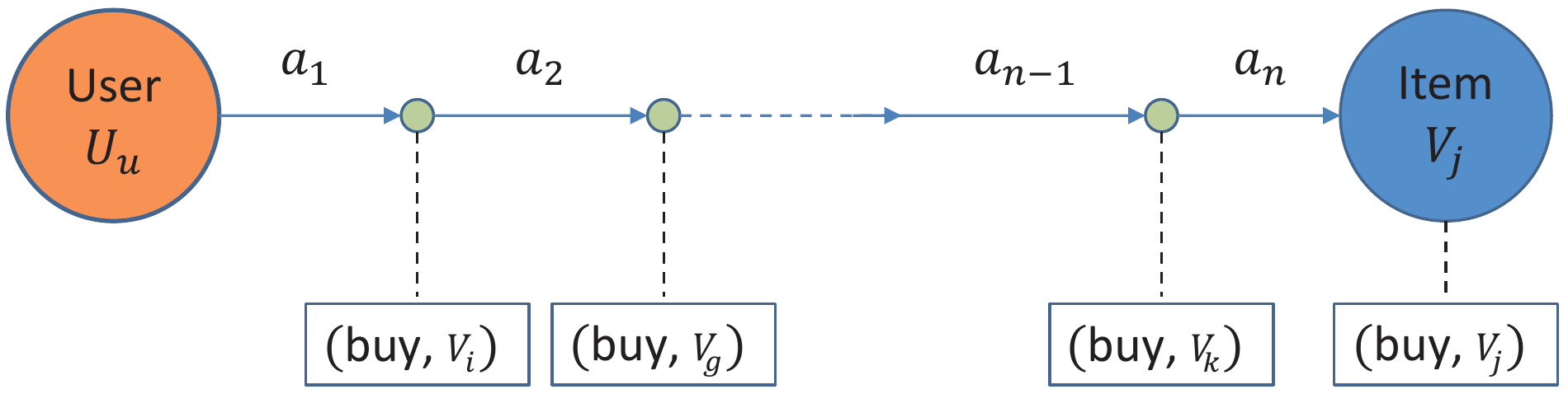}
  \caption{Transaction-based behavior sequence.}
  \label{fig:Transaction-based Behavior Sequence}
\end{figure}
\textbf{Transaction-based behavior sequence.}
A transaction-based behavior sequence (see Figure \ref{fig:Transaction-based Behavior Sequence}) records a series of \emph{different behavior objects} that a user interacts with, but with a \emph{same behavior type} (i.e., \emph{buy}). In practice, \textit{buy} is the most concerned one for online sellers. Therefore, \emph{with the transaction-based behavior sequence as input, the goal of a sequential recommender system is to recommend the next object (item) that a user will buy in view of the historical transactions of the user}.

\begin{figure}[htbp]
  \includegraphics[width=0.7\linewidth]{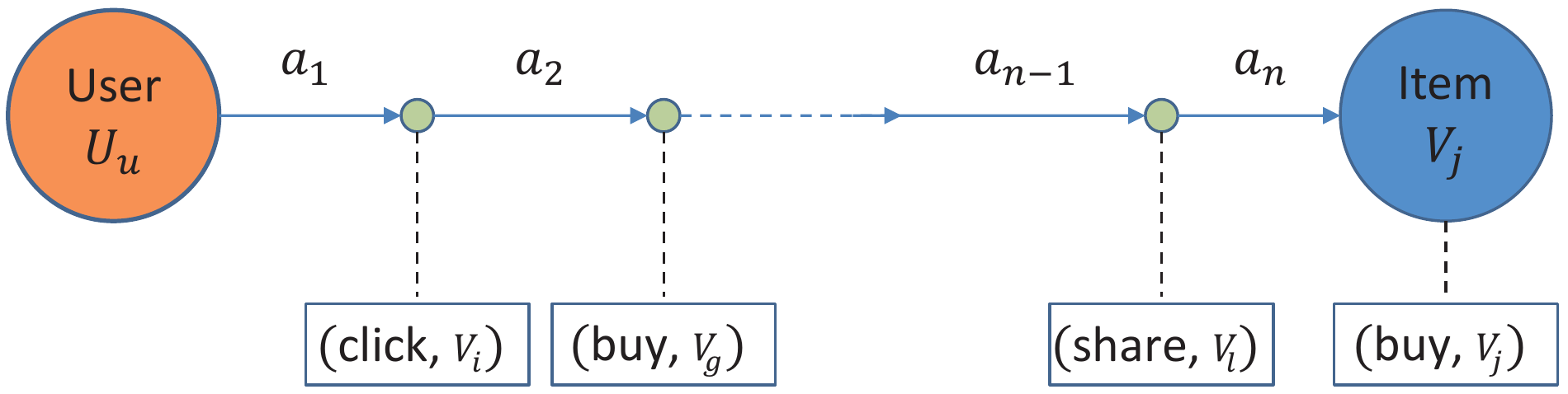}
  \caption{Interaction-based behavior sequence.}
  \label{fig:User Preference Learning}
\end{figure}
\textbf{Interaction-based behavior sequence.}
An interaction-based behavior sequence could be viewed as a mixture of experience-based and transaction-based behavior sequences (see Figure \ref{fig:User Preference Learning}), i.e., a generalization of previous two types and much closer to the real scenarios. That is to say, it consists of \emph{different behavior objects} and \emph{different behavior types} simultaneously.
\emph{In interaction-based behavioral sequence modeling, a recommender system is expected to understand user preferences more realistically, including different user intents expressed by different behavior types and preferences implied by different behavior objects. Its major goal is to predict the next behavior object that a user will interact with.}

\subsection{Sequential Recommendation Tasks}
\label{subsec:overview_task}
\begin{figure}[htbp]
    \centering
    \includegraphics[width=0.7\linewidth]{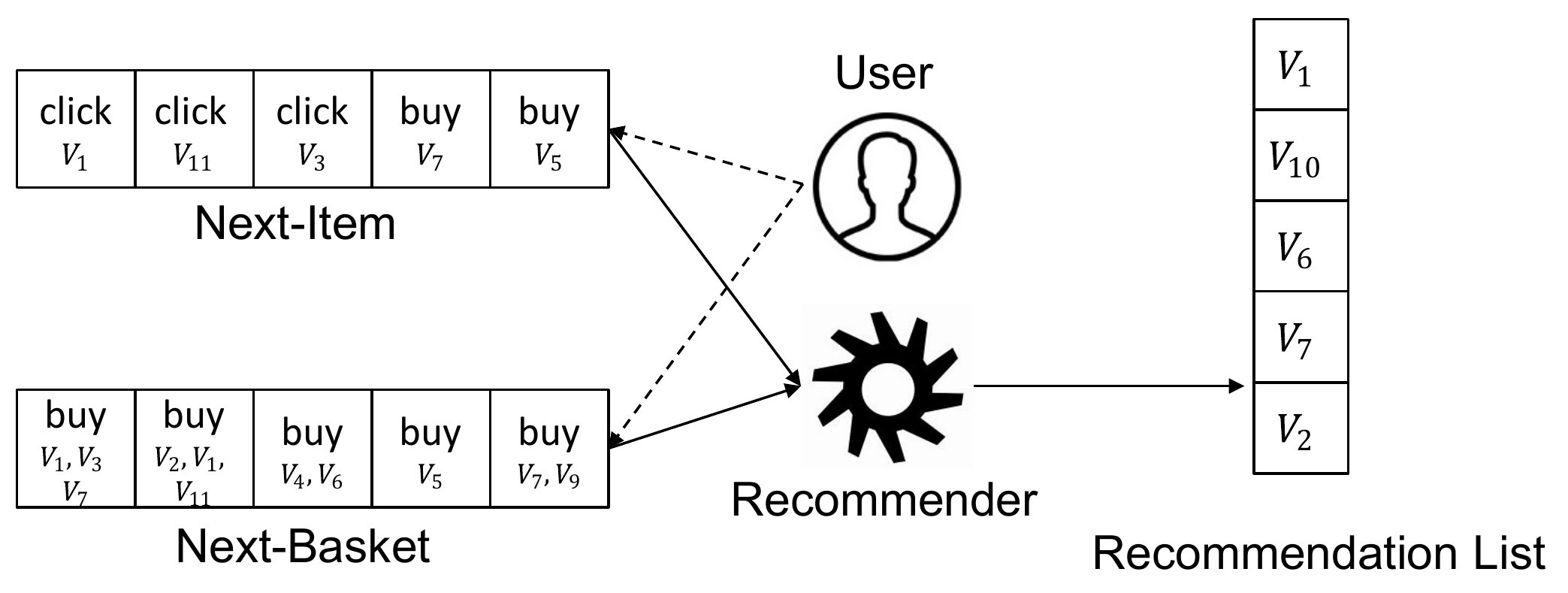}\vspace{-2mm}
    \caption{Next-item and next-basket recommendation.}
    \label{fig:session and basket}
\end{figure}

Before formally defining the sequential recommendation tasks, we firstly summarize the two representative tasks in the literature (as depicted in Figure \ref{fig:session and basket}): \emph{next-item recommendation} and \emph{next-basket recommendation}.
In \textbf{next-item recommendation}, a behavior contains only one object (i.e., item), which could be a product, a song, a movie, or a location. In contrast, in \textbf{next-basket recommendation}, a behavior contains more than one object.

However, although the input of the aforementioned recommendation tasks is varied, their goals are mostly identical. Specifically, both of them strive to predict the next item(s) for a user, whilst the most popular form of the output is the top-N ranked item list. The rank could be determined by probabilities, absolute values or relative rankings, while in most cases \emph{softmax function} is adopted to generate the output. Tan et al. \cite{tan2016improved} further proposed an embedding version of the softmax output for fast prediction to accommodate the large volume of items in recommendation.

In this paper, we consider the task of sequential recommendation as to generate a personalized ranked item list on the basis of the three types of user behavior sequences (input), which can be formally defined as:
\begin{equation}
(p_1,p_2,p_3,...,p_I)= f(a_1,a_2,a_3,...,a_t,u)
\label{equ:task1}
\end{equation}
where the input is the behavior sequence $\{a_1,a_2,a_3,...,a_t\}$, $u$ refers to the corresponding user of the sequence, and $p_i$ denotes the probability that item $i$ will be liked by user $u$ at time $t+1$. $I$ represents the number of candidate items. In other words, the sequential recommendation task is to learn a complex function $f$ for accurately predicting the probability that user $u$ will choose each item $i$ at time $t+1$ based on the input behavior sequence and the user profile.

According to the definition, and given the three types of behavior sequences, we thus divide the sequential recommendation tasks into three categories: \emph{experience-based sequential recommendation}, \emph{transaction-based sequential recommendation}, and \emph{interaction-based sequential recommendation}. We will comprehensively discuss these tasks as well as the specific DL-based recommendation models in Section \ref{sec:algorithms_for_sequential_recommendation}.

\subsection{Related Models}
In this subsection, we first review the traditional ML methods applied to the sequential recommendation and also briefly discuss their advantages and disadvantages. Second, we summarize related DL techniques for the sequential recommendation and elaborate how they overcome the issues involved in traditional methods.

\subsubsection{Traditional Methods}
Conventional popular methods for the sequential recommendation include frequent pattern mining, K-nearest neighbors, Markov chains, matrix factorization, and reinforcement learning \cite{quadrana2018sequence}. They generally adopt matrix factorization for addressing users' long-term preferences across different sequences, whilst use first-order Markov Chains for capturing users' short-term interest within a sequence \cite{he2016fusing}.
We next introduce the traditional methods as well as the representative algorithms for the sequential recommendation.

\textbf{Frequent pattern mining.} As we know, association rule \cite{ludewig2018evaluation} strives to use frequent pattern mining to mine frequent patterns with sufficient support and confidence. In the sequential recommendation, patterns refer to the sets of items which are frequently co-occurred within a sequence, and then are deployed to make recommendations.
Although these approaches
are easy to implement, and relatively explicable for users, they suffer from the limited scalability problem as matching patterns for recommendation is extremely strict and time-consuming.

Besides, determining suitable thresholds for support and confidence is also challenging, where a low minimum support or confidence value will lead to too many identified patterns, while a large value will merely mine co-occurred items with very high frequency, resulting in that only few items can be recommended or few users could get effective recommendation.

\textbf{K-nearest neighbors (KNN).} It includes item-based KNN and session-based KNN for the sequential recommendation. Item-based KNN \cite{Davidson2010The, Linden2003Amazon} only considers the last behavior in a given session and recommends items that are most similar to its behavior object (item), where the similarities are usually calculated via the cosine similarity or other advanced measurements \cite{Sottocornola2018session-based}.

 In contrast, session-based KNN
 \cite{Linden2003Amazon,lerche2016value,jannach2017recurrent} compares the entire existing session with all the past sessions to recommend items via calculating similarities using Jaccard index or cosine similarity on binary vectors over the item space.
KNN methods can generate highly explainable recommendation.
Besides, as the similarities can be pre-calculated, KNN-based recommender systems could generate recommendations promptly. However, this kind of algorithms generally fails to consider the sequential dependency among items.

\textbf{Markov chains (MC).}
In the sequential recommendation, Markov models assume that future user behaviors only depend on the last or last few behaviors. For example, \cite{he2017translation} merely considered the last behavior with first-order MC, while \cite{he2016fusing,he2016vista} adopted high-order MCs, which take the dependencies with more previous behaviors into account.
Considering only the last behavior or several behaviors makes the MC-based models unable to leverage the dependencies among behaviors in a relatively long sequence and thus fails to capture intricate dynamics of more complex scenarios. Besides, they might also suffer from data sparsity problems.

\textbf{Factorization-based methods.}
Matrix factorization (MF) tries to decompose the user-item interaction matrix into two low-rank matrices.
For example, BPR-MF \cite{rendle2009bpr} optimizes a pairwise ranking objective function via stochastic gradient descent (SGD).
Twardowski \cite{twardowski2016modelling} proposed a MF-based sequential recommender system (a simplified version of Factorization Machines \cite{rendle2011fast}), where only the interaction between a session and a candidate item is considered for generating recommendations.
FPMC \cite{rendle2010factorizing} is a representative baseline for next-basket recommendation, which integrates the MF with first-order MCs.
FISM \cite{kabbur2013fism} conducts matrix factorization on an item-item matrix, and thus no explicit user representation is learned.
On the basis of FISM, FOSSIL \cite{he2016fusing} tackles the sequential recommendation task by combining similarity-based methods and high-order Markov Chains. It performs better on sparse datasets in comparison with the traditional MC methods and FPMC.
The main drawbacks of MF-based methods lie in: 1) most of them only consider the low-order interactions (i.e., first-order and second-order) among latent factors, but ignore the possible high-order interactions;
and 2) excepts for a handful of algorithms considering temporal information (e.g., TimeSVD++ \cite{koren2009collaborative}), they generally ignore the time dependency among behaviors both within a session and across different sessions.

\textbf{Reinforcement learning (RL).} The essence of RL methods is to update recommendations according to the interactions between users and the recommender systems. When a system recommends an item to a user, a positive reward is assigned if the user expresses his/her interest on the item (via behaviors such as click or view). It is usually formulated as a Markov decision process (MDP) with the goal of maximizing the cumulative rewards in a set of interactions \cite{zhao2019leveraging,shih2018automatic}. With RL frameworks, sequential recommender systems can dynamically adapt to users (changing) preferences. However, similar to DL-based approaches, this kind of works is also lack of interpretability. Besides, more importantly, there is few appropriate platforms or resources for developing and testing RL-based methods in academia .

\subsubsection{Deep Learning Techniques}
\label{sec:Deep Learning Techniques}
In this subsection, we summarize the DL models (e.g., RNN and CNN) that have been adopted in the sequential recommendation in the literature.

\textbf{Recurrent neural networks (RNNs).} The effectiveness of RNNs in sequence modeling have been widely demonstrated in the field of natural language processing (NLP). In the sequential recommendation, RNN-based models are in the majority of DL-based models \cite{cho2014properties}. In comparison with the traditional models, RNN-based sequential recommendation models can well capture the dependencies among items within a session or across different sessions.
The main limitation of RNNs for the sequential recommendation is that it is relatively difficult to model dependencies in a longer sequence (although could be somehow mitigated by other techniques), and training is burdened with the high cost especially with the increase of sequence length.

\textbf{Convolutional neural networks (CNNs).} CNN is commonly applied to process time series data (e.g., signals) and image data, where a typical structure consists of convolution layers, pooling layers, and feed-forward full-connected layers. It is suitable to capture the dependent relationship across local information (e.g., the correlation between pixels in a certain part of an image or the dependencies between several adjacent words in a sentence). In the sequential recommendation, CNN-based models can well capture local features within a session, and also could take the time information into consideration in the input layer \cite{tuan20173d,tang2018personalized}.

\textbf{Multi-layer perceptrons (MLPs).} MLPs refer to feed-forward neural networks with multiple hidden layers, which can thus well learn the nonlinear relationship between the input and output via nonlinear activation functions (e.g., tanh and ReLU). Therefore, MLP-based sequential recommendation models are expected to well capture the complex and nonlinear relationships among users behaviors \cite{wu2017session}.

\textbf{Attention mechanisms.} Attention mechanism in deep learning is intuited from visual attentions of human-beings (incline to be attracted by more important parts of a target object). It is originated from the work of Bahdanau et al. \cite{bahdanau2014neural}, which proposes an attention mechanism in neural machine translation task to focus on modeling the importance of different parts of the input sentence on the output word.
Grounded on the work, \emph{vanilla attention} is proposed by applying the work as a decoder of the RNN, and has been widely used in the sequential recommendation \cite{li2017neural}.
On the other hand, \emph{self-attention mechanism} (originated in transformer \cite{vaswani2017attention} for neural machine translation by Google 2017) has also been deployed in the sequential recommendation. In contrast with vanilla attention, it does not include RNN structures, but performs much better than RNN-based models in recommender systems \cite{zhang2019next}.

\textbf{Graph neural networks (GNNs).} GNN \cite{zhou2018graph} can collectively aggregate information from the graph structure. Due to its effectiveness and superior performance in many applications, it has also obtained increasing interest in recommender systems. For example, Wu et al. \cite{wu2019session} first used GNN for session-based recommendation by capturing more complex relationships between items in a sequence, and each session is represented as the composition of the long-term preference and short-term interests within a session using an attention network.

\begin{figure}[htbp]
    \centering
    \includegraphics[width=0.9\linewidth]{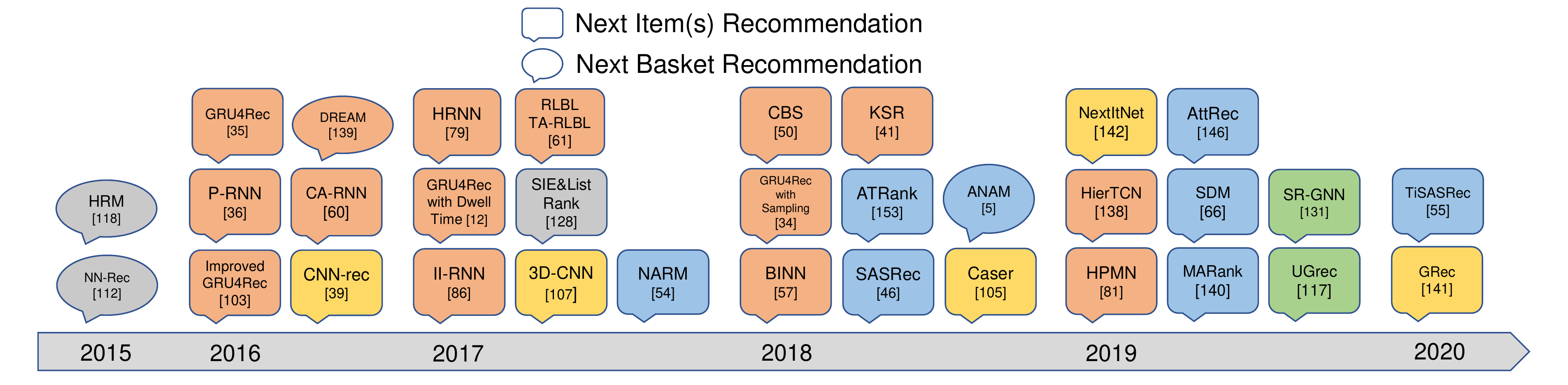}\vspace{-2mm}
    \caption{Some recent and representative DL-based sequential recommendation models. Different colors indicate different DL techniques (grey: MLP; orange: RNN; yellow: CNN; blue: attention mechanism; green: GNN).}
    \label{fig:sequential_recommender_system_timeline}
\end{figure}

\subsubsection{Concluding Remarks.}
Compared with conventional methods, DL-based methods are a much more active research area in the recent years. The MC- and MF-based models assume that a user's next behavior is related to only a few recent behavior(s), while DL methods utilize a much longer sequence for prediction \cite{wang2019asurvey}, as they are able to effectively learn the \textit{theme} of the whole sequence. Thus, they generally obtain better performance (in terms of accuracy measurement) than traditional models. Meanwhile, DL methods are more robust to sparse data and can adapt to varied length of the input sequence.
The representative DL-based sequential recommendation algorithms are presented in Figure \ref{fig:sequential_recommender_system_timeline}, which will be introduced in details in next sections.

The major problems of DL-based sequential recommendation methods include: 1) they are lack of explanability for the generated recommendation results. Besides, it is also difficult to calibrate why the recommendation models are effective, and thus to yield a robust DL-based model for varied scenarios; 2) the optimization is generally extremely challenging and more training data is required for complex networks.

\section{Sequential Recommendation Algorithms}
\label{sec:algorithms_for_sequential_recommendation}

In this section, in order to figure out whether sequential recommendation tasks have been sufficiently explored, we classify sequential recommendation algorithms in terms of the three tasks (Section \ref{subsec:overview_task}): \emph{experience-based sequential recommendation}, \emph{transaction-based sequential recommendation}, and \emph{interaction-based sequential recommendation}.

\subsection{Experience-based Sequential Recommendation}

As we have introduced, in an experience-based behavior sequence, a user interacts with a same item with different behavior types. The goal of experience-based sequential recommendation is to predict the next behavior type that the user will implement on the item, and thus it is also referred to as \emph{multi-behavior recommendation}. Accordingly, we first explore the studies on multi-behavior recommendation and then present DL-based models that leverage multi-behavior information in the sequential recommendation.

\subsubsection{Conventional models for multi-behavior recommendation}
Ajit el al. \cite{singh2008relational} first proposed a collective matrix factorization model (CMF) to simultaneously factorize multiple user-item interaction matrices (in terms of different behavior types) by sharing the item-side latent matrix (item embedding) across matrices. %
Other studies \cite{krohn2012multi, zhao2015improving} extended CMF to handle different user behaviors (e.g., social relationships). Besides, there are also some models addressing multi-behavior recommendation with Bayesian learning. For example, Loni et al. \cite{loni2016bayesian} proposed multi-channel BPR to adapt the sampling rule for different behavior types. Qiu et al. \cite{qiu2018bprh} further proposed an adaptive sampling method for BPR by considering the co-occurrence of multiple behavior types. Guo et al. \cite{guo2017resolving} aimed to resolve the data sparsity problem by sampling unobserved items as positive items based on item-item similarity, which is calculated by multiple behavior types. Ding et al. \cite{ding2018improving} developed a margin-based learning framework to model the pairwise ranking relations among purchase, view, and non-view behaviors.

\subsubsection{DL-based multi-behavior recommendation}
DL techniques have also been applied in multi-behavior recommendation. For example, \emph{NMTR} \cite{gao2019learning} is proposed to tackle some representative problems of conventional models for multi-behavior recommendation, e.g., lack of behavior semantics, unreasonable embedding learning and incapability in modeling complicated interactions.
To capture the sequential relationships between behavior types, NMTR \cite{gao2019learning} cascades predictions of different behavior types by considering the sequential dependency relationship among different behaviors in practice\footnote{For example, the \emph{search}, \emph{click}, and \emph{purchase} operations for the same item are usually sequentially ordered in e-commerce.}, which thus translates the heterogeneous behavior problem into the experience-based sequential recommendation problem as we have defined.
It should be noted that this cascaded prediction, which could be regarded as pre-training embedding layers of other behavior types before learning a recommendation model for the target behavior, only considers the connections between target behavior and previous behaviors but ignores the ones between target behavior and subsequent behaviors. Thus, it does not fully explore the relationship on various behavior types.
In this view, multi-task learning (MTL) can address this problem by providing a paradigm to predict multiple tasks simultaneously which also exploits similarities and differences across tasks. The performance of the MTL model proposed in \cite{gao2019learning} is generally better than those using the sequential training.
Besides, Xia et al. \cite{xia2018modeling} proposed a multi-task model with LSTM to explicitly model users' purchase decision process by predicting the stage and decision of a user at a specific time with the assistance of a pre-defined set of heuristic rules, and thus obtaining more accurate recommendation results.

\subsection{Transaction-based Sequential Recommendation}
\label{sec:transaction-based sequential recommendation}
In transaction-based sequential recommendation, there is only a single behavior type (transaction-related, e.g., purchase), and recommendation models generally consider the sequential dependency relationships between different objects (items) as well as user preferences. As there are a substantial amount of DL-based models for this task, we further summarize the existing models in terms of the employed specific DL techniques.

\subsubsection{RNN-based Models}
RNN structures have been well exploited in transaction-based sequential recommendation task, and we summarize RNN-based approaches from the following perspectives.

\begin{figure} [htbp]
    \centering
    \includegraphics[width=0.55\linewidth]{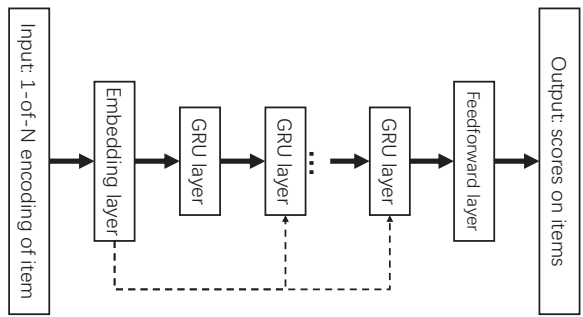}\vspace{-2mm}
    \caption{Architecture of GRU4Rec.}
    \label{fig:GRU4Rec_architecture}
\end{figure}

\textbf{(1) GRU4Rec-related models.}
 Hidasi et al. \cite{hidasi2015session} proposed a GRU-based RNN model for sequential recommendation (i.e., \emph{GRU4Rec}), which is the first model that applies RNN to sequential recommendation, and does not consider a user's identity (i.e., anonymous user). On its basis, a set of improved models \cite{tan2016improved,bogina2017incorporating,hidasi2018recurrent} have been proposed, which also use RNN architectures for modeling behavior sequence.
The architecture of GRU4Rec is shown in Figure \ref{fig:GRU4Rec_architecture}. As introduced in \cite{hidasi2015session}, the input of GRU4Rec is a session (behavior sequence), which could be a single item, or a set of items appeared in the session. It uses one-hot encoding to represent the current item, or a weighted sum of encoding to represent the set of items. The core of the model is the GRU layer(s), where the output of each layer is the input for the next layer, but each layer can also be connected to a deeper non-adjacent GRU layer in the network. Feedforward layers are added between the last GRU layer and the output layer. The output is the probability of each candidate item that will appear in the next behavior.

GRU4Rec employs \emph{session-parallel mini-batches} and \emph{popularity-based negative sampling} for training.
The reason for using session-parallel mini-batches is to form sessions with equal length while the length of actual sessions can be greatly varied. On the other hand, if simply breaking a session into different parts to force them into equal length, we could not well model the behavior sequence and fail to capture how a session evolves over time \cite{hidasi2015session}.

The extended studies strive to improve the model performance from the perspectives of model training and designing more advanced model structures for better learning item information. For example, for \textbf{facilitating training}, \cite{tan2016improved} applied \emph{data augmentation} to enhance training of GRU4Rec. \cite{bogina2017incorporating} considered the \emph{dwell time} to modify the generation of mini-batch, which has been verified to greatly improve performance.
On the other hand,
popularity-based sampling suffers from the problem that model learning slows down after all the candidate items have been ranked above popular ones, which could be a relatively serious problem for long-tail items recommendation. Thus, \cite{hidasi2018recurrent} proposed the \emph{additional sampling} (a combination of uniform sampling and popularity sampling) for negative sampling in GRU4Rec, which can enormously improve performance.

For \textbf{better modeling item information},
\cite{hidasi2016parallel} considered additional item information other than IDs (e.g., text descriptions and images) for improving prediction performance. Specifically, they introduced a number of parallel RNN (p-RNN) architectures to model sessions based on click behaviors and other features of the clicked items (e.g., pictures and text descriptions).
Moreover, they particularly proposed alternative but more suitable training strategies for p-RNNs: \emph{simultaneous}, \emph{alternating}, \emph{residual}, and \emph{interleaving} training. In simultaneous training (baseline), every parameter of each subnet is trained simultaneously. In alternating training, subnets are trained in an alternating fashion per epoch. In residual training, subnets are trained one by one by the residual error of the ensemble of the previously trained subnets, while interleaving training is alternating training per mini-batch.
Furthermore, \cite{jannach2017recurrent} combined the session-based KNNs with GRU4Rec using the methods of \emph{switching}, \emph{cascading}, and \emph{weighted hybrid}.

 \textbf{(2) With user representation}. There are also some studies that aim to better model users' preference. For example, \cite{zhang2014sequential} proposed a RNN-based framework for click-through rate (CTR) prediction in sponsor search, which considers the impact of the click dwell time with the assumption that the longer a user stays on an ad page, the more attractive the ad is for the user.
 In total, three categories of features are considered: ad features (ad ID, position and query text), user features (user ID, user's query) and sequential features (time interval, dwell time and click sequence).
\cite{soh2017deep} took one-hot encoding of items in users' behavior sequences as input of a GRU-based RNN to learn users' historical embeddings.
\emph{RRN} \cite{wu2017recurrent} is the first recurrent recommender network that attempts to capture the dynamics of both user and item representation.
\cite{bharadhwaj2018explanations} further improved the RRN's interpretability by devising a time-varying neighborhood style explanation scheme, which jointly optimizes prediction accuracy and interpretability of the sequential recommendation.

Considering that simply embedding a user's historical information into a single vector may lose the per-item or feature-level correlation information between a user's historical sequences and long-term preference, Chen et al. \cite{chen2018sequential} thus proposed a memory-augmented neural network for the sequential recommendation. The model explicitly stores and updates every user's historical information by leveraging an external memory matrix. Huang et al. \cite{huang2018improving} further improved \cite{chen2018sequential} by adopting a separate GRU component for capturing sequential dependency and incorporated knowledge base (KB) information for better learning attribute (feature)-level user preference.
Towards better modeling lifelong sequential patterns for each user, Ren et al. \cite{ren2019lifelong} proposed a Hierarchical Periodic Memory Network (\emph{HPMN}) to capture multi-scale sequential patterns, where periodic memory updating mechanism is designed to avoid unexpected knowledge drifting and hierarchical memory slots are used to deal with different update periods.

\emph{HRNN} \cite{quadrana2017personalizing}\footnote{\url{github.com/mquad/hgru4rec}.} uses GRU to model users and sessions respectively. The session-level GRU considers a user's activities within a session and thus generates recommendations, while the user-level GRU models the evolution of a user's preference across sessions.
Given that the length of sessions of different users are varied, it deploys user parallel mini-batch training, which is extended from session parallel mini-batch of GRU4Rec.
Donkers et al. \cite{donkers2017sequential} further proposed a user-based GRU framework (including linear user-based GRU, rectified linear user-based GRU, and attentional user-based GRU) to integrate user information for giving better user representations.
\emph{HierTCN} \cite{you2019hierarchical} also involves a GRU-based high-level model to aggregate users' evolving long-term preferences across different sessions, while \emph{SDM} \cite{lv2019sdm} particularly designs a gated fusion module to effectively integrate users' short-term and long-term preferences.

 \textbf{(3) Context-aware sequential recommendation}. Most of the previous models have ignored the huge amount of context information in real-word scenarios. In this view,
\cite{Liu2016ContextAwareSR} summarized two types of contexts: \emph{input contexts} and \emph{transition contexts}. Input contexts refer to the ones by which users conduct their behaviors, e.g., location, time and weather, whilst transition contexts mean the transitions between two adjacent input elements in historical sequences (e.g., time intervals between the adjacent behaviors). It further designed the context-aware recurrent neural networks (\emph{CA-RNN}) to simultaneously model the sequential and contextual information.
Besides,
\cite{song2018augmenting} proposed \emph{ARNN} to consider the user-side contexts, e.g., age, gender and location. Specifically, \emph{ARNN} extracts high-order user-contextual preferences using a product-based neural network, which is capable of being incorporated with any existing RNN-based sequential recommendation models.

 \textbf{(4) Other models.} Other than the aforementioned three categories, there are other RNN-based models (e.g., \emph{DREAM} \cite{yu2016dynamic}) for transaction-based sequential recommendation in the literature. For example,
\cite{devooght2017long} used RNN for the collaborative filtering task and considered two different objective functions in the RNN model: categorical cross-entropy (\emph{CCE}) and \emph{Hinge}, where CCE has been widely used in language modeling, and Hinge is extended from the objective function of SVMs.
\cite{ruocco2017inter-session} deployed a multi-layer GRU network to capture sequential dependencies and user interest from both the inter-session and intra-session levels.
In view of that existing studies assume that there is only an implicit purpose for users in a session, Wang et al. \cite{DBLP:conf/ijcai/Wang0WSOC19} proposed a mixture-channel purpose routing networks (MCPRNs) to capture the possible multi-purposes of users in a session (a channel implies a latent purpose). MCPRNS consists of a purpose router (PRN) and a multi-channel recurrent framework with purpose-specific recurrent units.

\subsubsection{CNN-based Models}
RNN models are limited to model relatively short sequences due to their network structures and relatively expensive computing costs, which can be partially alleviated by CNN models \cite{shi2018neural}. For example, \emph{3D-CNN} \cite{tuan20173d} designs an embedding matrix to concatenate the embedding of item ID, name, and category. \emph{Caser} \cite{tang2018personalized} views the embedding matrix of $L$ previous items as an `image', and thus uses a horizontal convolutional layer and a vertical convolutional layer to capture point-level and union-level sequential patterns respectively. Using convolution, the perception of relevant skip behaviors becomes possible. It also captures long-term user preferences through user embedding. The network structure of \emph{CNN-Rec} \cite{hsu2016neural} is highly similar to Caser in terms of user embedding and horizontal convolution, but it does not deploy vertical convolution. \emph{NextItNet} \cite{yuan2019simple} is a generative CNN model with the residual block structure for the sequential recommendation. It is capable of capturing both long and short-term item dependencies.
\emph{GRec} \cite{yuan2020future} further extends \emph{NextItNet} by utilizing a gap-filling based encoder-decoder framework with masked-convolution operations to jointly consider the past and future contexts (data) without the data leakage issue.
In \emph{HierTCN} \cite{you2019hierarchical}, a low-level model implemented with Temporal Convolutional Networks (TCN) unitizes the long-term user preference learned from GRU module and the short-term user preference within a session to generate the final recommendation.

\subsubsection{Attention-based Models}
The attention mechanisms have been largely applied to the sequential recommendation, and are capable of identifying more `relevant' items to a user given the user's historical experience. We conclude these models according to the deployed attention mechanism types: \emph{vanilla attention} and \emph{self-attention} (see Section \ref{sec:Deep Learning Techniques}).

 \textbf{(1) Vanilla attention mechanisms.} \emph{NARM}\footnote{\url{github.com/lijingsdu/sessionRec_NARM}.} \cite{li2017neural} is an encoder-decoder framework for transaction-based sequential recommendation. In the local encoder, RNN is combined with vanilla attention to capture the major purposes (or interest) of a user in the current sequence. With the attention mechanism, NARM is able to eliminate noises from unintended behaviors, such as accidental (unintended) clicks.
\cite{wang2018attention} applied the vanilla attention mechanism to weight each item in a sequence to reduce the negative impact of unintended interactions.
Liu et al. \cite{liu2018stamp} proposed a short-term attention/memory priority model, which uses vanilla attention to calculate attention scores of items in a sequence as well as the attention correlations between previous items and the most recent item in the sequence.
Ren et al. \cite{ren2018repeatnet} considered the repeated consumption issue, and thus proposed \textit{RepeatNet} which evaluates the recommendations from both the repeat mode and the explore mode, which refer to the old item from a user's history and the new item, respectively.
\cite{sachdeva2018attentive} incorporated vanilla attention with a Bi-GRU network to model user's short-term interest for music recommendation.
\cite{bai2018an} proposed a unified attribute-aware neural attentive model (\emph{ANAM}), which applies vanilla attention mechanism on feature level.
To better capture users' short-term preferences, Yu et al. \cite{yu2019multi} designed a multi-order attention network which is instantiated with two k-layer residual networks to model individual-level and union-level item dependencies respectively.

 \textbf{(2) Self-attention mechanisms.}
Self-attention mechanisms have also obtained increasing interest in the sequential recommendation. For example,
Zhang et al. \cite{zhang2019next} utilized the self-attention mechanism to infer the item-item relationship from the user's historical interactions.
With self-attention, it is capable of estimating weights of each item in the user's interaction trajectories to learn more accurate representations of the user's short-term intention, while it uses a metric learning framework to learn the user's long-term interest.
In \emph{SDM} \cite{lv2019sdm}, a multi-head self-attention module is incorporated to capture a user multiple interests in a session (i.e., short-term preference), while long-term preference of the user is also encoded through attention and dense fully connected networks based on various types of side information, e.g., item ID, first level category, leaf category, brand and shop in the user's historical transactions.
Similarly, \emph{SASRec} \cite{kang2018self-attentive} adopts a self-attention layer to balance short-term intent and long-term preference, and seeks to identify items relevant to the next behavior from the user's historical behavior sequences.

\emph{BERT4Rec} \cite{BERT4Rec2019} is the improved version of SASRec, which introduces the transformer architecture for the sequential recommendation and trains the bidirectional model to model sequential data using Cloze task.
\emph{TiSASRec} \cite{li2020time} further improves SASRec by taking time intervals between items in a sequence into consideration. Specifically, it models a relation matrix between items for each user according to the time interval information between each two items in the historical sequences.
Besides, to overcome the drawbacks of RNN-based sequential recommender systems, such as not supporting parallelism and only modeling one-way transitions between consecutive items, \emph{SANSR} \cite{DBLP:journals/access/SunTDZ19} incorporates the transformer framework \cite{vaswani2017attention} to speed up the training process and learn the relations between items in the session regardless the distance and the direction.
%

As we know, most of the previous studies on the sequential recommendation focus on recommendation accuracy, but ignore the \emph{diversity} of recommendation results, which is also a quite important measurement for effective recommendation. With respect to this issue, Chen et al. \cite{DBLP:journals/corr/abs-1908-10171} proposed an intent-aware sequential recommendation algorithm, which uses the self-attention mechanism to model a user's multi-intents in a given session.

\subsubsection{Other Models}
There are also some other DL-structures (e.g., MLP \cite{wang2015learning}, GNN \cite{wang2020global} and autoencoder) that have been adopted in the sequential recommendation. For example, \emph{NN-rec} \cite{wan2015next} is the first work considering neural network for next-basket recommendation, which is inspired by NLPM \cite{bengio2003a}.
Wu et al. \cite{wu2019session} used GNN for session-based recommendation to capture complex transitions among items. In this model, each session is modeled as a directed graph, and proceeded by a gated graph neural network to obtain session representations (local session embedding and global session embedding).
\emph{GACOforRec} \cite{GACOforRec} utilizes graph convolutional neural networks to learn the item order within a session as well as the spatiality within the network to handle a users' short-term intents, while it designs ConvLSTM to capture the user's long-term preference.
Besides, considering that different behaviors may have different impacts, it proposes a new pair of attention mechanisms which consider the different propagation distances in the graph convolutional network to obtain the different weights.
\emph{UGrec} \cite{wang2019unified} models user and item interactions as a graph network, defines sequential recommendation paths from users' purchased histories, and further aggregates different paths using an attention mechanism. Finally, a particularly designed translation learning objective function in graph embedding is designed for model learning and inference.

Sachdeva et al. \cite{sachdeva2019sequential} explored the variational autoencoder for modeling a user's preference through his/her historical sequence, which combines latent variables with temporal dependencies for preference modeling. Ma at al. \cite{ma2019hierarchical} specifically designed a hierarchical gating network (\emph{HGN}) with BPR to capture both the long-term and short-term user preferences.

\subsection{Interaction-based Sequential Recommendation} 
Compared to the aforementioned two tasks, the interaction-based one is much more complicated as each behavior sequence consists of both different behavior types and different behavior objects. Thus, the recommendation models are expected to capture both the sequential dependencies among different behaviors, different items as well as behaviors and items, respectively. Next, we summarize the related models according to the deployed DL techniques.

\subsubsection{RNN-based Models} RNN-based models still take the majority role in this task \cite{loyola2017modeling}. For example, \cite{twardowski2016modelling} proposed a RNN-based model without explicitly learning user representation.
Given the task of predicting the next item expected to appear in terms of a target behavior type, Le et al. \cite{le2018modeling} firstly divided a session into a target sequence and a supporting sequence according to the target behavior type.
Its basis idea is that the target behavior type (e.g., purchase) contains the most efficient information for the prediction task, and the remaining behaviors (e.g., click) can thus be utilized as the supporting sequences that can facilitate the next-item prediction task for the target behavior type.
Besides, in order to better model the dependencies among different behavior types, some studies would also assume that there is a cascading relationship (as in Section 3.1.2) among different types of behaviors (i.e., different behavior types are sequentially ordered). For example, Li et al. \cite{li2018learning} proposed a model that consists of two main components: \emph{neural item embedding} \cite{barkan2016item2vec} and \emph{discriminative behavior learning}.
For behavior learning, it utilizes all types of behavior (e.g., click, purchase and collect) to capture a user's present consumption motivation. Meanwhile, it selects purchase related behaviors (e.g, purchase, collect and add-to-cart) from user's historical experience to model the user's underlying long-term preference.
Considering that RNN cannot well handle users' short-term intent in a sequence whereas log-bilinear model (LBL) cannot capture users' long-term preference, Liu et al. \cite{liu2017multi} combined RNN with LBL to construct two models (\emph{RLBL} and \emph{TA-RLBL}) for modeling multi-behavioral sequences.
TA-RLBL is an extension of RLBL \cite{liu2017multi} which considers the continuous time difference information between input behavior objects and thus further improve the performance of RLBL.
\cite{smirnova2017contextual} took context information (e.g., behavior type) into consideration by modifying the structures of RNN.

\subsubsection{Other Models}
There are also some other DL techniques applied in the interaction-based sequential recommendation, including attention mechanisms, MLPs and Graph-based models.
For example,
\cite{zhou2018atrank} proposed \emph{ATRank}\footnote{\url{github.com/jinze1994/ATRank}.} which adopts both \emph{self-attention} and \emph{vanilla attention} mechanisms. Considering the heterogeneity of behaviors, ATRank models the influence among behaviors via self-attention, while it
uses vanilla attention to model the impact of different behaviors on the recommendation task.
\emph{CSAN} \cite{huang2018csan} is the improved version of ATRank by also considering side information and polysemy of behavior types.
Wu et al. \cite{wu2017session} proposed a deep listNet ranking framework (MLP-based) to jointly consider user's clicks and views.
Ma et al. \cite{ma2018graph} proposed a graph-based broad-aware network (\emph{G-BBAN}) for news recommendation, which considers multiple user behaviors, behavioral sequence representations, and user representation.
\begin{table}[!htbp]
\centering
\caption{Categories of representative algorithms regarding sequential recommendation tasks and DL models.}
\label{tab:categorization}
\begin{threeparttable}
\renewcommand{\multirowsetup}{\centering}
\begin{tabular}{ccc}
\toprule
\textbf{Task}& \textbf{DL Model}& \textbf{Papers}\\
\cline{1-3}
\multirow{2}{*}{Experienced-based} & MLP & \cite{gao2019learning}\\
\cline{2-3}
& RNN &\cite{xia2018modeling}\\
\cline{1-3}
\multirow{6}{*}{Transaction-based} & \multirow{3}{*}{RNN} & \cite{song2018augmenting,quadrana2017personalizing,donkers2017sequential,yu2016dynamic,tan2016improved,hidasi2015session,bogina2017incorporating,Liu2016ContextAwareSR,chen2018sequential}\\
&& \cite{huang2018improving,DBLP:journals/corr/abs-1908-10171,DBLP:journals/access/SunTDZ19,ruocco2017inter-session,devooght2017long,jannach2017recurrent,hidasi2016parallel,hidasi2018recurrent,zhang2014sequential} \\
& & \cite{soh2017deep,wu2017recurrent,bharadhwaj2018explanations,wu2019dual,ren2019lifelong,you2019hierarchical,lv2019sdm}\\
\cline{2-3}
& CNN & \cite{tuan20173d,yuan2019simple,hsu2016neural,tang2018personalized,you2019hierarchical,yuan2020future}\\
\cline{2-3}
& MLP & \cite{wang2015learning,wan2015next}\\
\cline{2-3}
& \multirow{2}{*}{Attention mechanism} & \cite{bai2018an,li2017neural,wang2018attention,liu2018stamp,kang2018self-attentive,sachdeva2018attentive,ren2018repeatnet,ying2018sequential}\\
& & \cite{lv2019sdm,zhang2019next,BERT4Rec2019,DBLP:journals/corr/abs-1908-10171,yu2019multi,wu2019dual,li2020time}\\
\cline{2-3}
& GNN & \cite{wu2019session,GACOforRec,wang2019unified,wang2020global}\\
\cline{2-3}
& Other networks & \cite{sachdeva2019sequential,ma2019hierarchical}\\
\cline{1-3}
\multirow{4}{*}{Interaction-based} & RNN & \cite{loyola2017modeling,le2018modeling,smirnova2017contextual,liu2017multi,li2018learning,twardowski2016modelling}\\
\cline{2-3}
& MLP & \cite{wu2017session}\\
\cline{2-3}
& Attention mechanism & \cite{loyola2017modeling,huang2018csan,zhou2018atrank}\\
\cline{2-3}
& GNN & \cite{ma2018graph}\\

\bottomrule

\end{tabular}
\end{threeparttable}
\end{table}

\subsection{Concluding Remarks}
In this section, we have introduced representative algorithms for the three sequential recommendation tasks. We list the representative algorithms in terms of tasks and DL techniques in Table \ref{tab:categorization}.
In summary, RNNs and attention mechanisms have been greatly explored in both transaction and interaction-based sequential recommendation tasks, where the effectiveness of other DL models (e.g., GNN and generative models) needs much further investigation. Besides, there are also some issues for the existing models especially for the complicated interaction-based sequential recommendation: (1) the behavior type and the item in a behavior 2-tuple $(c_i,o_i)$ are mostly equally treated. For example, ATRank \cite{zhou2018atrank} and CSAN \cite{huang2018csan} adopt the same attention score for the item and the corresponding behavior type; (2) different behavior types are not distinguished successfully. For example, \cite{twardowski2016modelling} used the same network to model different types of behaviors, assuming that different behavior types have similar patterns; (3) the correlation between behaviors in a sequence is easily ignored. For example, \cite{wu2017session} used pooling operation to model multi-type behavior in a sequence.
In view of these issues, more advanced approaches are needed for much more effective sequential recommendation, especially for the task of interaction-based sequential recommendation. In the next two sections, we will further summarize and evaluate the factors that might impact the performance of a DL-based model in regard to recommendation accuracy, which are expected to better guide future research.

\section{Influential Factors on DL-based Models}
\label{sec:influential factors of models}

\begin{figure*}[htbp]
    \centering
    \includegraphics[width=1\textwidth]{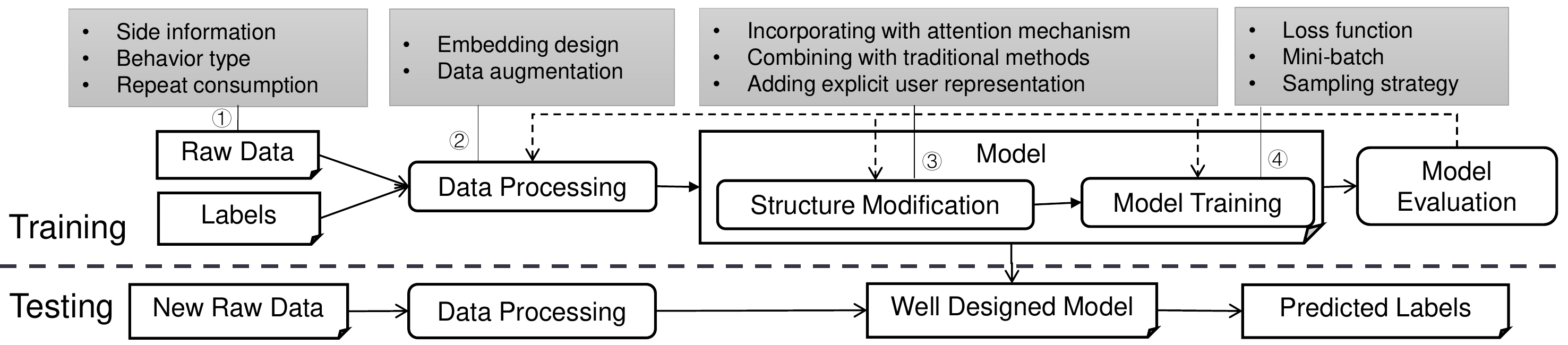}
    \caption{Influential factors of DL-based models.}
    \label{fig:training_testing}
\end{figure*}

\begin{table}[htb]
    \centering
    \small
    \caption{The influential factors on DL-based sequential recommender systems. }
\label{tab:influential_factors}\vspace{-2mm}
    \begin{tabular}{cccc}
        \toprule
        \textbf{Module} & \textbf{Factor} & \textbf{Method} & \textbf{Papers} \\
        \midrule
        \multirow{6}{*}{Input} & \multirow{2}{*}{side information} & utilize image/text  &\cite{hidasi2016parallel,huang2018csan}\\
         \cmidrule(r){3-4}
        && utilize dwell time &\cite{bogina2017incorporating,zhang2014sequential,dallmann2017improving}\\
        \cmidrule(r){2-4}
        & \multirow{3}{*}{behavior type}& simple behavior embedding &\cite{zhou2018atrank}\\
         \cmidrule(r){3-4}
        & & divide session into groups  &\multirow{2}{*}{\cite{le2018modeling,li2018learning}}\\
        & & for different purposes &\\
        \cmidrule(r){2-4}
        & repeat consumption& consider repeat behavior&\cite{wan2018representating,ren2018repeatnet,hu2020modeling}\\
        \midrule

        \multirow{4}{*}{Data processing}&\multirow{3}{*}{embedding design}& item embedding& \cite{greenstein2017session}\\
        \cmidrule(r){3-4}
        & &w-item2vec&\cite{li2018learning}\\
        \cmidrule(r){3-4}
        & & session embedding&\cite{wu2017session}\\
        \cmidrule(r){2-4}
        & data augmentation & &\cite{tan2016improved,tuan20173d}\\
        \midrule
        \multirow{8}{*}{Model structure} &\multirow{3}{*}{incorporating  }& only attention mechanism &\cite{zhang2019next,zhou2018atrank,huang2018csan,bai2018an}\\
        \cmidrule(r){3-4}
        &  & incorporating vanilla  &\multirow{3}{*}{\cite{li2017neural,wang2018attention,liu2018stamp,kang2018self-attentive}}\\
        & attention mechanism & attention mechanism  &\\
        & & with other DL methods& \\
        \cmidrule(r){2-4}
        & combining with & KNN &\cite{jannach2017recurrent}\\
        \cmidrule(r){3-4}
        & conventional methods& metric learning &\cite{zhang2019next}\\
        \cmidrule(r){2-4}
        &adding explicit & user embedded models&\cite{tang2018personalized,wang2015learning}\\
        \cmidrule(r){3-4}
        &  user representation& user recurrent models&\cite{chen2018sequential,quadrana2017personalizing,donkers2017sequential,li2018learning,wu2017recurrent,bharadhwaj2018explanations,ren2019lifelong}\\
        \midrule
        \multirow{10}{*}{Model training} &\multirow{4}{*}{negative sampling}&
        uniform &\cite{hidasi2015session,hidasi2018recurrent}\\
        \cmidrule(r){3-4}
        & & popularity-based &\cite{hidasi2015session,hidasi2018recurrent}\\
        \cmidrule(r){3-4}
        & & additional&\cite{hidasi2018recurrent}\\
        \cmidrule(r){3-4}
        & & sample size&\cite{hidasi2018recurrent}\\
        \cmidrule(r){2-4}
        & \multirow{3}{*}{mini-batch creation}& session parallel &\cite{hidasi2015session}\\
        \cmidrule(r){3-4}
        & & item boosting &\cite{bogina2017incorporating}\\
       \cmidrule(r){3-4}
        & & user parallel &\cite{quadrana2017personalizing}\\
          \cmidrule(r){2-4}
         &\multirow{3}{*}{loss function} & TOP1 &\cite{hidasi2015session}\\
           \cmidrule(r){3-4}
           & & TOP1-max \& BPR-max&\cite{hidasi2018recurrent}\\
        \cmidrule(r){3-4}
           & & CCE \& Hinge &\cite{devooght2017long}\\
        \bottomrule

    \end{tabular}
\end{table}

Figure \ref{fig:training_testing} shows the \emph{training} and \emph{testing} process of a sequential recommender system. In the training, the input includes raw data and label information, which are then fed into the data processing module, mainly including \emph{feature extraction} and \emph{data augmentation}. Feature extraction refers to converting raw data into structured data, while data augmentation is normally used to deal with data sparsity and cold-start problems, especially in DL-based models. Third, a model is trained and evaluated based on the processed data, and the model structure or training method (e.g., learning rate, loss function) can be updated in an iterated way based on the evaluation results till satisfactory performance is reached. In the testing, the data processing module only includes feature extraction, and then the obtained trained model is used to make recommendations.

On the basis of a thorough literature study, we identify some representative factors (listed in \emph{grey} boxes in Figure \ref{fig:training_testing} and Table \ref{tab:influential_factors}) that might impact the performance of DL-based models in terms of recommendation accuracy. The details of these factors are discussed subsequently.

\subsection{Input Module}
\emph{Side information} and \emph{behavior types} are critical factors to DL-based models in the input module.

\subsubsection{Side Information}
Side information has been well recognized to be effective in facilitating recommendation performance \cite{sun2019research}. It refers to information about items (other than IDs), e.g., \emph{category}, \emph{images}, \emph{text descriptions}, and \text{reviews}, or information related to transactions (behaviors) like \emph{dwell time}. Text and image information about items have been widely explored in DL-based collaborative filtering systems \cite{bansal2016ask,covington2016deep,nguyen2017personalized,zhang2017hashtag,rawat2016contagnet:}, as well as in some DL-based sequential recommender systems \cite{hidasi2016parallel,huang2018csan}. For example, p-RNN \cite{hidasi2016parallel} uses a parallel RNNs framework to process the item IDs, images and texts. Specifically, the first parallel architecture trains a GRU network (i.e., subnet) for item representation on the basis of each kind of information, respectively. The model concatenates the hidden layers of the subnets and generates the output. The second architecture has a shared hidden state to output weight matrix. The weighted sum of the hidden states is used to produce the output instead of being computed by separate subnets. In the third structure called parallel interaction, the hidden state of the item feature subnet is multiplied by the hidden state of the ID subnet in an element-wise manner before generating the final outcome.
CSAN \cite{huang2018csan} utilizes word2vec and CNN to learn the representation of texts and images respectively. Previous models have demonstrated that side information like item images and texts can alleviate the data sparsity \cite{wang2016collaborative,li2015deep,hidasi2016parallel} and cold-start \cite{wang2015collaborative,wei2016collaborative,zhou2018atrank,huang2018csan} problems.

On the other hand, side information like \emph{dwell time} partially imply a user's degrees of interest on different items. For example, when a user browses a web page for an item, the longer he/she stays, we can infer that the more he/she is interested in. Bogina et al. \cite{bogina2017incorporating} applied item boosting according to the dwell time for generating mini-batch in training. In particular, assuming that a predefined threshold of dwell time is $t_d$ seconds, if the dwell time on an object $i$ in a session is within the range of $[2t_d,3t_d)$, then the parallel mini-batch of this session will contain $2$ repeated behaviors regarding to $i$, i.e., the presence of $i$ in the session increases. This strategy (referred as \emph{item boosting}) can be considered as to re-measure the importance of behavior objects in terms of the corresponding dwell time. Zhang et al. \cite{zhang2014sequential} treated dwell time as a sequence feature, and concatenated it with other features (e.g., query text). Similarly, Dallmann et al. \cite{dallmann2017improving} proposed an extension to existing RNN approaches by adding user dwell time. Experiments in \cite{bogina2017incorporating} show that incorporating dwell time with GRU4Rec \cite{hidasi2015session} makes a great improvement (up to $153.1\%$ on MRR@20). 

\subsubsection{Behavior Type}
In the sequential recommendation, behaviors in user behavior sequences are usually heterogeneous and polysemous \cite{zhou2018atrank,huang2018csan}, and different behavior types imply users' different intents. For instance, a purchase action is a better indicator of a user's preference on an item than a click behavior. Therefore, it is critical to treat different behavior types differently \cite{twardowski2016modelling, le2018modeling,zhou2018atrank,gao2019learning}. For example, CBS \cite{le2018modeling} divides a sequence into target sequence and supporting one in terms of behavior types, where the target sequence is related to the behavior type (e.g., purchase) that has the most efficient information for prediction. Similarly, BINN \cite{li2018learning} utilizes all behavior types (e.g., click, purchase and collect) to capture a user's present interest whereas models the user's long-term preference using only purchase related information (e.g, purchase, add-to-cart and collect). \cite{zhou2018atrank} learns the representation of each behavior type and then concatenate them with the corresponding item embedding vectors.
Experiments generally support that purchase behavior can more accurately capture a user's long-term preferences, whilst other behavior types can facilitate the learning of short-term interests \cite{twardowski2016modelling,le2018modeling,zhou2018atrank,gao2019learning}.

\subsubsection{Repeat Consumption}
Repeat consumption refers to that an item is repeatedly appeared in a user's historical sequences, which is mostly ignored in the sequential recommendation. Anderson et al. \cite{anderson2014dynamics} investigated the dynamics of repeated consumption on seven real datasets and found that recency is the strongest predictor of repeated consumption. Bhagat et al. \cite{bhagat2018buy} presented four models (i.e., repeated customer probability model, aggregate time distribution model, Possion-Gamma model, and modified Possion-Gamma model) for repeat purchase recommendations. \cite{wang2019modeling} further combined collaborative filtering and Hawkes Process to build a holistic model for recommendation, and the item-specific temporal dynamics of repeat consumption are captured.
For sequential recommendation, Wan et al. \cite{wan2018representating} used \emph{loyalty} factor to model repeated consumption which can further boost the performance of next-basket recommendation. Ren et al. \cite{ren2018repeatnet}\footnote{\url{github.com/PengjieRen/RepeatNet}.} also considered this issue, and their results confirm that the consideration of repeat consumption patterns in DL network design can improve the recommendation performance. The recent study of \cite{hu2020modeling} just pointed out that RNN-based models might not well capture repeated behaviors for next-basket recommendation, and thus proposed a KNN-based model for capturing repeated consumption in next-basket recommendation.

It should be noted that, although side information and behavior types could greatly improve model performance, their collections might be either infeasible or cost-consuming.

\subsection{Data Processing}
An appropriate design of feature extraction methods (i.e., \emph{embedding design}) and \emph{data augmentation} for generating more training data have been validated to be effective in existing DL-based models.

\subsubsection{Embedding Design}
\label{subsubsec:embedding_design}
In the sequential recommendation, embedding methods are used to represent information about an item, a user, or a session.
For example, Greenstein et al. \cite{greenstein2017session} adopted the word embedding methods GloVe \cite{pennington2014glove} and Word2Vec \cite{mikolov2013efficient} (CBOW) for \emph{item embedding} in e-commerce applications.
Li et al. \cite{li2018learning} further proposed w-item2vec (inspired by item2vec \cite{barkan2016item2vec}) on the basis of the Skip-gram model and thus formed unified representations of items.
Wu et al. \cite{wu2017session} designed a session embedding for pre-training by considering different user search behaviors such as clicks and views, the target item embedding and the user embedding together to have a comprehensive session understanding (i.e., session representation).

\subsubsection{Data Augmentation}
\begin{figure}[htbp]
    \centering
    \includegraphics[width=0.5\linewidth]{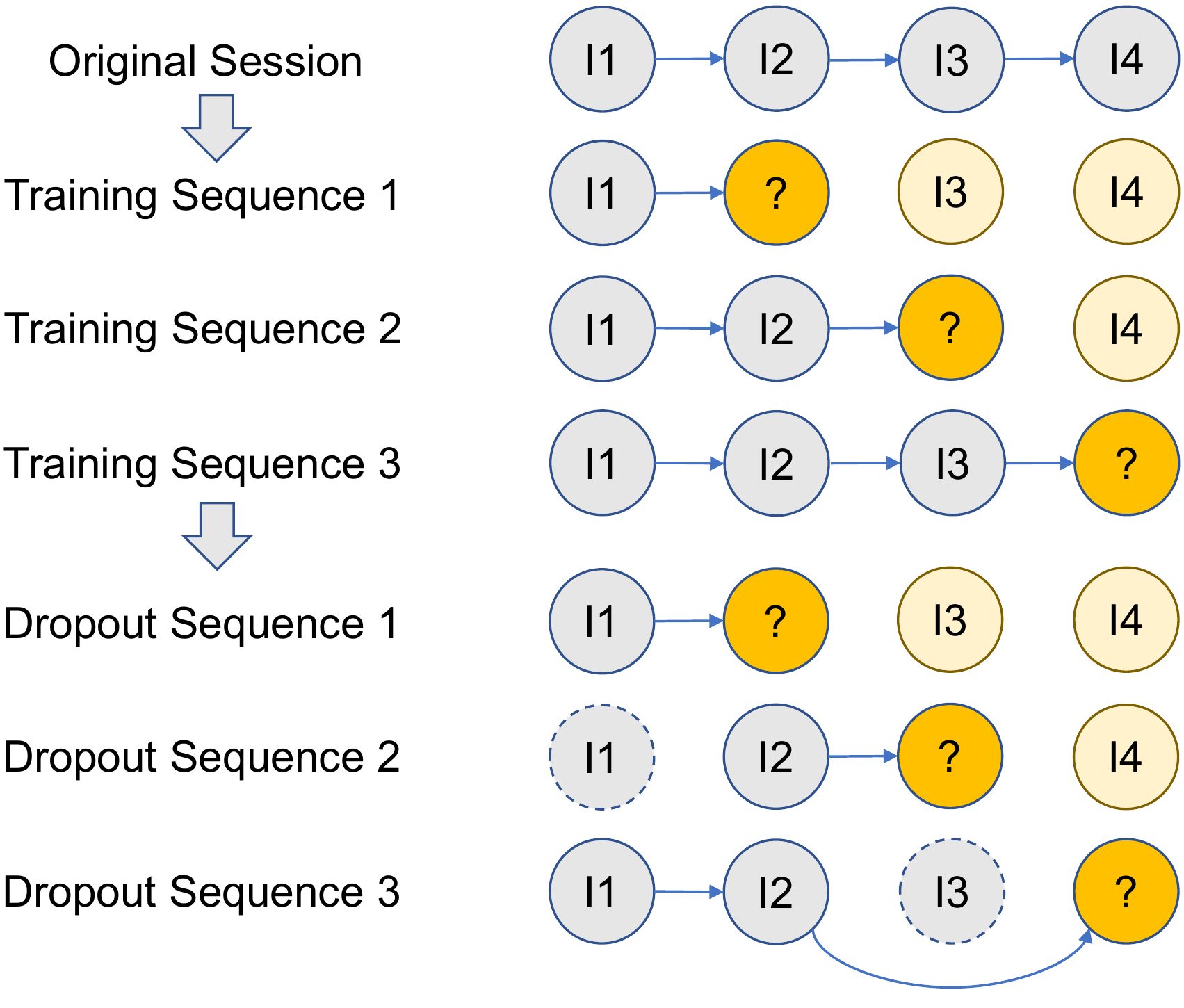}
    \caption{Data augmentation. The orange circles represent the predicted items; the dotted circles represent the item that is deleted in the dropout method, and light orange circles make up privileged information.}
    \label{fig:prefix data augmentation}
\end{figure}

In the sequential recommendation, in some scenarios there might be no user profiles or historical information for a new user, or a user who does not log in, i.e., cold-start problems. Therefore, data augmentation becomes an important technique.
For example, Tan et al. \cite{tan2016improved} proposed an augmentation method, where prefixes of the original input sessions are treated as new training sequences as shown in Figure \ref{fig:prefix data augmentation}. That is, given the original session $(l_1,l_2,l_3,l_4)$, we can generate $3$ training sequences: $(l_1,l_2),(l_1,l_2,l_3),(l_1,l_2,l_3,l_4)$, while a recommendation algorithm predicts the last item for each training sequence. With this method, a session is repeatedly utilized during training, which is demonstrated to improve $14.7\%$ over GRU4Rec on MRR@20 \cite{tan2016improved}. 
Besides, the \emph{dropout} method \cite{srivastava2014dropout} is further adopted to prevent the over-fitting problem (see Figure \ref{fig:prefix data augmentation}, a circle with the dotted line is the dropout behavior in each sequence).
Besides, considering that items after a target item may also contain valuable information,  these items are thus viewed as privileged information as \cite{vapnik2009new} to facilitate the learning process.

Similarly, with regard to two behavior types (i.e., add-to-cart and click), 3D-CNN \cite{tuan20173d} also uses \emph{data augmentation} which treats all prefixes up to the last add-to-cart item as training sequences for each session containing at least one add-to-cart item. Besides, it uses \emph{right padding} or \emph{simple dropping} methods to keep all the sequences of the same length.

\subsection{Model Structure}
We summarize the major methods to improve model structures in the previous DL-based models as: \emph{incorporating attention mechanisms}, \emph{combining with conventional models}, and \emph{adding explicit user representation}.

\subsubsection{Incorporating Attention Mechanisms}
In Section \ref{sec:Deep Learning Techniques}, we discuss that there are mainly \emph{vanilla attention} and \emph{self-attention}. Overall, we can incorporate attention mechanism with other DL models, or just build attention models to address the sequential recommendation problems.
For the first scenario, NARM \cite{li2017neural}, ATEM \cite{wang2018attention} and STAMP \cite{liu2018stamp} incorporate the \emph{vanilla attention mechanism} with RNN or MLP, aiming to capture user's main purpose in a given session. Experiments verify that their performance surpassed GRU4Rec by $25\%$, $92\%$ and $30\%$ respectively.
SASRec \cite{kang2018self-attentive} combines self-attention with feedforward network to model correlations between different behaviors, and can improve recommendation accuracy by $47.7\%$ and $4.5\%$ on HR@10 compared with GRU4Rec and Caser \cite{tang2018personalized}, respectively. 
For the second scenario, for example,
AttRec \cite{zhang2019next} simply deploys a self-attention mechanism to capture users' short-term interest, where its performance exceeds Caser by $8.5\%$ on HR@50.
ATRank \cite{zhou2018atrank} and CSAN \cite{huang2018csan} combine self-attention with vanilla attention for the sequential recommendation.
Attention mechanisms can be further employed to capture attribute-level importance level of items for modeling users' interest. For example, ANAM \cite{bai2018an} applies attention mechanism to track a user's appetite for items and their attributes.

To conclude, previous studies demonstrate that incorporating attention mechanisms can improve recommendation of the DL-based models, while mostly only using self-attention mechanisms can have better performance than some DL-based models without attention mechanisms.

\subsubsection{Combining with Conventional Methods}
DL-based models can also be combined with traditional methods to boost their performance on the sequential recommendation tasks. For example, \cite{jannach2017recurrent} combined a session-based KNN with GRU4Rec \cite{hidasi2015session} in three different ways (i.e., switching, cascading, and weighted hybrid), showing that the best combination can exceed original GRU4Rec by $9.8\%$ in some applications.
AttRec \cite{zhang2019next} combines self-attention (for short-term interest learning) and metric learning (for long-term preference modeling), and the performance exceeds Caser\cite{tang2018personalized} by $8.5\%$ on
HR@50.

\subsubsection{Adding Explicit User Representation}
Given the application scenarios where users' IDs can be recognized, we can design methods for explicitly learning user representation, i.e., users' long-term preferences can be well modeled by \emph{user embedded models} or \emph{user recurrent models}.

\textbf{User embedded models.} This fold of models explicitly learns user representations \cite{tang2018personalized,wang2015learning} via embedding methods, but not in a recurrent process as item representation. They can facilitate the performance of the sequential recommendation models \cite{tang2018personalized}. However, such models might suffer from the cold-start user problem since the long-term interest of a user with little historical information cannot be well learned.
Another issue is that, user representation via user embedded models is learned in a relatively static way, which cannot capture users evolved and dynamic preferences. In this view, user recurrent models are expected to be more effective, which learn user representations in a recurrent way as item representation learning.

\textbf{User recurrent models.} They treat both user and item representations as recurrent components in the DL-based models, which can better capture users' evolving preferences, including memory-augmented neural network \cite{chen2018sequential,ren2019lifelong}, RNN-based models \cite{quadrana2017personalizing,donkers2017sequential,li2018learning} and recurrent neural networks \cite{wu2017recurrent,bharadhwaj2018explanations}. For example, \cite{quadrana2017personalizing, donkers2017sequential,li2018learning} used RNN framework to learn users' long-term interest from their historical behavior sequences. Experiments verify that considering a user's long-term interest is critically valuable for personalized recommendation, e.g., HRNN \cite{quadrana2017personalizing} exceeds GRU4Rec by $3.5\%$ with explicit user representation in some scenarios.

In summary, model structures play an important role in the sequential recommendation, where better designs can help more effectively capture the sequential dependencies among items and behaviors, and thus better understand both users' short-term and long-term preferences.

\subsection{Model Training}
Well-designed training strategies can also facilitate the learning of DL-based sequential recommendation models. With a comprehensive investigation, we have summarized three major strategies: \emph{negative sampling}, \emph{mini-batch creation} and \emph{loss function}.

\subsubsection{Negative Sampling}
\emph{Popularity-based sampling} and \emph{uniform sampling} have been widely used in recommendation.
Popularity-based sampling assumes that the more popular an item is, the more possibly that a user knows about it. That is to say, if a user does not interact with it previously, it is more likely that the user dislikes it.
\cite{hidasi2018recurrent} further proposed a novel sampling strategy (called \emph{additional sampling}) by combining these two sampling strategies, which takes the advantages but overcomes the shortcomings of both strategies in negative sampling.
In the additional sampling strategy, negative samples are selected with a probability proportional to supp$_i^\alpha$, where supp$_i$ is the support of item $i$ and $\alpha$ is a parameter ($0\leq\alpha\leq1$). The cases of $\alpha=0$ and $\alpha=1$ are equivalent to uniform and popularity-based sampling respectively.
Experiment results show that additional sampling can surpass both the popularity-based sampling and uniform sampling methods under certain scenarios (e.g., loss functions). Besides, \emph{the size of negative samples} can also affect the performance of sequential recommendation models.

\subsubsection{Mini-batch Creation}

\emph{Session parallel mini-batch training} \cite{hidasi2015session} was proposed to accommodate sessions of varied lengths and strive to capture the dynamics of sessions over time. In particular, sessions are firstly arranged in time order. Then, the first event (behavior) of the first $X$ sessions ($X$ is the number of sessions) is used to form the input of the first mini-batch (whose desired output is the second event of the active sessions). The second mini-batch is formed from the second event of the $X$ sessions, and so on and so forth. If any of the $X$ sessions reaches its ending, the next available session out of the $X$ sessions is placed in the corresponding place to continually form the mini-batch.
Session parallel mini-batch has two variants: \emph{item boosting} and \emph{user-parallel mini-batch}. In item boosting, some items can be repeatedly used in mini-batch in terms of identified factors like the dwell time \cite{bogina2017incorporating}, while regarding the latter variant, for example, HRNN \cite{quadrana2017personalizing}
designs user-parallel mini-batch (i.e., parallel sessions belong to different users) to model the evolution of users' preferences across sessions.

\subsubsection{Loss Function Design}
Loss functions can also greatly impact the model performance. In the sequential recommendation, quite a few loss functions have been employed, including \emph{TOP1-max} (ranking-max version of \emph{TOP1}), \emph{BPR-max} (ranking-max version of \emph{BPR}), \emph{CCE} (Categorical Cross-Entropy) and \emph{Hinge}.

\textbf{TOP1} is a regularized approximation of relative rankings of positive and negative samples. As shown in Equation \ref{equ:top1 function}, it consists of two parts: the first part inclines to penalize the incorrect ranking between positive sample $i$ and any negative sample $j$ ($N_S$ is the size of negative samples), and the second part is used as the regularization.

\begin{equation}
L_{\text{TOP1}} = \frac{1}{N_S} \sum_{j=1}^{N_S}{ \sigma(r_j-r_i) + \sigma(r_j^2)}
\label{equ:top1 function}
\end{equation}
where $\sigma(.)$ is a sigmoid function, $r_i$ and $r_j$ are the ranking scores for sample $i$ and $j$ respectively. Following the same notations, \textbf{BPR} (Bayesian Personalized Ranking) \cite{rendle2009bpr} is defined as:
\begin{equation}
L_{\text{BPR}} = - \frac{1}{N_S} \sum_{j=1}^{N_S}{\log\sigma(r_i-r_j)}
\end{equation}

TOP1 and BPR loss functions might suffer from the gradients vanishing problems for DL-based models (e.g., in GRU4Rec \cite{hidasi2018recurrent}). In this view, ranking-max loss function family is proposed \cite{hidasi2018recurrent} to address this issue, where the ranking score is only compared to the negative sample which is most relevant to the target sample, i.e., the one has the highest ranking score. Accordingly, we have \textbf{TOP1-max} and \textbf{BPR-max}, which are formulated as Equations \ref{eq:top1max} and \ref{eq:bprmax} respectively. They can be considered as the weighted version of TOP1 and BPR, respectively. Previous research validates that the two loss functions largely improve the performance of RNN-based sequential recommendation models \cite{hidasi2018recurrent}.
%
\begin{equation}
\label{eq:top1max}
  L_{\text{TOP1-max}} = \sum_{j=1}^{N_S}s_j\left({ \sigma(r_j-r_i) + \sigma(r_j^2)}\right)
\end{equation}
where $s_j$ is the normalized score of $r_j$ using softmax function.
\begin{equation}
\label{eq:bprmax}
    L_{\text{BPR-max}} = - \log \sum_{j=1}^{N_S}{s_j \sigma(r_i-r_j)}
\end{equation}

In addition to the ranking-based loss functions, \emph{CCE} (categorial cross-entropy) and \emph{Hinge} loss functions have also been applied in the sequential recommendation \cite{devooght2017long}.
\textbf{CCE} is defined as:
\begin{equation}
\text{CCE}(\textbf{o},\textit{i})=-\text{log}(\text{softmax}(\textbf{o})_i)
\end{equation}
where $\textbf{o}$ is a model output and $\textit{i}$ is a target item. CCE suffers from the computation complexity issue due to the softmax function. On the contrary, \textbf{Hinge} compares the predicted results with a pre-defined threshold (e.g., $0$):
\begin{equation}
 \text{Hinge}(\textbf{o},\textit{i}) = \sum_{j \in C}\text{max}(0,1-o_j) + \gamma \sum_{j\in F}\text{max}(0,o_j)
\end{equation}
where $C$ is the set of recommendations containing item $i$, while $F$ is the set of recommendations not containing $i$  (i.e., bad recommendations). $\gamma$ is a parameter to balance the impacts of the two parts of errors (correctly recommended vs. incorrectly recommended). With Hinge loss, the recommendation task is transformed to a binary classification problem where a recommender system determines whether an item should be recommended or not.

\section{Empirical Studies on Influential Factors}
\label{sec:evaluation of deep learning based sequential recommender system}
Here, we conduct experiments\footnote{The source codes and datasets of the experiments are shared on Github: \url{https://github.com/sttich/dl-recommendation}.} on real datasets to showcase the impact of influential factors on DL-based models in terms of recommendation accuracy, where mostly the ways of incorporating influential factors are widely adopted by representative sequential recommender systems.

\subsection{Experimental Settings}
\label{sec:Evaluation Measures}

\subsubsection{Datasets} We use three real-world datasets: \emph{RSC15}, \emph{RSC19} and \emph{LastFM}. \textit{RSC15} is published by RecSys Challenge 2015\footnote{\url{www.kaggle.com/chadgostopp/recsys-challenge-2015}.}, which contains click and buy behaviors from an online shop. Only the click data is used in our evaluations. \textit{RSC19} is published by RecSys Challenge 2019\footnote{\url{www.recsyschallenge.com/2019/}.}, which contains hotel search sessions from a global hotel platform. \textit{RSC19 (user)} is a subset of \textit{RSC19}. \textit{LastFM} is collected via the LastFM API, and each sample is a 4-tuple (user, artist, song, timestamp).

\begin{figure*}[htb]
    \centering
    \flushleft
\begin{subfigure}[b]{0.3\linewidth}
\begin{tikzpicture}
\begin{axis}[
    width=5cm,
    height=4.3cm,
    legend style={at={(0.5,-0.15)},
      anchor=north,legend columns=-1},
    symbolic x coords={2,3,4,5,10,15,20},
    xtick=data,
    ]
  \addplot [mark=diamond*,thick,green] coordinates {(2, 0.44647379) (3,0.44524866 )(4,0.44422894) (5,0.44193021) (10,0.42734766) (15,0.41103468)(20,0.39742093)};
  \addplot [mark=*,thick,red] coordinates {(2, 0.5756335) (3,0.57639325)(4,0.57182623) (5,0.57344252) (10,0.55031159) (15,0.52940965)(20,0.51141364)};
  \addplot [mark=square*,thick,blue] coordinates {(2, 0.67860927) (3,0.67713141)(4,0.67255715) (5,0.67304204) (10,0.65107843) (15,0.62686111)(20,0.60457264 )};
\legend{@5, @10, @20}
\end{axis}
\end{tikzpicture}
\subcaption{Recall}
\end{subfigure}
\quad
\begin{subfigure}[b]{0.3\linewidth}
\begin{tikzpicture}
\begin{axis}[
    width=5cm,
    height=4.3cm,
    legend style={at={(0.5,-0.15)},
      anchor=north,legend columns=-1},
    symbolic x coords={2,3,4,5,10,15,20},
    xtick=data,
    ]
  \addplot [mark=diamond*,thick,green] coordinates {(2, 0.26776126)(3,0.26444051) (4,0.26323354) (5,0.2622862) (10,0.25093476) (15,0.24309674)(20,0.23026156)};
  \addplot [mark=*,thick,red] coordinates {(2, 0.28518623)(3,0.28211407) (4,0.28037168) (5,0.27976261) (10,0.26752772) (15,0.25910093)(20,0.24558889)};
  \addplot [mark=square*,thick,blue] coordinates {(2, 0.29244265)(3,0.2891858) (4,0.28744605) (5,0.28686386) (10,0.27464987) (15,0.2659783)(20,0.2521022)};
\legend{@5, @10, @20}
\end{axis}
\end{tikzpicture}
\subcaption{MRR}
\end{subfigure}
\quad
\begin{subfigure}[b]{0.3\linewidth}
\begin{tikzpicture}
\begin{axis}[
    width=5cm,
    height=4.3cm,
    legend style={at={(0.5,-0.15)},
      anchor=north,legend columns=-1},
    symbolic x coords={2,3,4,5,10,15,20},
    xtick=data,
    ]
   \addplot [mark=diamond*,thick,green] coordinates {(2, 0.31209926) (3,0.30930115)(4,0.30812437) (5,0.30683738) (10,0.29466168) (15,0.28482215)(20,0.27181734)};
  \addplot [mark=*,thick,red] coordinates {(2, 0.35405483) (3,0.35188156)(4,0.3494955) (5,0.34908491) (10,0.33460885) (15,0.32330805)(20,0.30879489)};
  \addplot [mark=square*,thick,blue] coordinates {(2, 0.38022681) (3,0.37745175)(4,0.37506784) (5,0.37463153) (10,0.36024804) (15,0.34808947)(20,0.33240854)};
\legend{@5, @10, @20}
\end{axis}
\end{tikzpicture}
\subcaption{NDCG}
\end{subfigure}

    \caption{The effect of the length of session on RSC15.}
    \label{fig:session_len15}
\end{figure*}
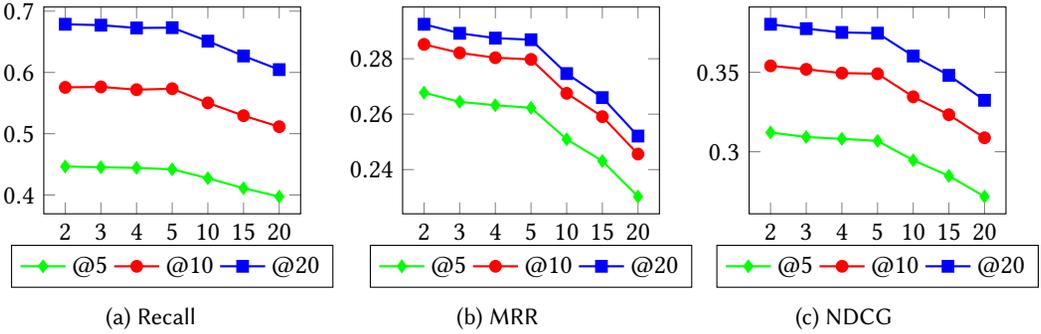

Following the common way in data pre-processing \cite{hidasi2015session,quadrana2017personalizing}, for \textit{RSC15} and \textit{RSC19}, we firstly filter out sessions which have less than $2$ behaviors, and items that appear less than $5$ times. Then we consider the sessions that end in the last day as the test set, while the others are for model training/validation. For \textit{RSC19 (user)}, we further select users with more than $10$ sessions, and consider the last session of each user as the test set. For \textit{LastFM}, due to the lack of session identities in LastFM, we manually divide the behavior sequence of each user into sessions every $30$ minutes. Then we filter out sessions which have less than $3$ behaviors, items that appear less than $5$ times, and users that appear less than $3$ times. We also consider the last session of each user as the test set.
Here, we want to emphasize that different ways of data filtering, which lead to different data scenarios, will result in varied performance. For example, we check the performance of GRU4Rec on different data scenarios by filtering out sessions less than \{2, 3, 4, 5, 10, 15, 20\} on \textit{RSC15} respectively. For fair comparisons, under all data scenarios, we use the same test set as data scenario of $20$ following the aforementioned data pre-process procedure. We have tuned the hyperparameters under each scenario, and the results are presented in Figure \ref{fig:session_len15}. As shown in Figure \ref{fig:session_len15}, the performance of GRU4Rec drops as the length of sequence gets longer. This might be caused by the decreasing number of sessions for training, i.e., the available training sessions on the RSC dataset are $7966888$, \emph{$4419603$}, $2810308$, $1876772$, $448561$, $167318$, and $78486$ under the seven data scenarios, respectively.

Besides, it should be noted that few studies have explicitly discussed their data splitting methods for model training/validation/testing. By reading the source codes publicized by the corresponding authors, we summarize the following two major forms (which are differed mainly owing to that whether user information is considered or not in model design): (1) to divide the sessions that have occurred in the latest $n$ days as the test set; (2) to treat each user's latest session as the test set. The former one is much more commonly adopted in the sequential recommendation.

The statistic information of these datasets are summarized in Table \ref{tab:datasets_descriptions}.

\begin{table}[htb]
    \centering
    \small
    \caption{The statistic information of the four datasets.}
\label{tab:datasets_descriptions}\vspace{-2mm}
    \begin{tabular}{ccccc}
        \toprule
        Feature & RSC15 & RSC19 & RSC19 (user) & LastFM \\
        \midrule
        Sessions & 7,981,581 & 356,318 & 1,885 & 23,230\\
        Items & 37,483 & 151,039 & 3,992 &122,816\\
        Behaviors & 31,708,461 & 3,452,695 & 49,747 &683,907\\
        Users & --& 279,915 & 144 & 277\\
        ABS& 3.97& 9.69& 26.39& 29.44\\
        ASU & -- & 1.27 & 13.09 & 83.86\\
        \midrule
        \multicolumn{5}{l}{ABS: Average Behaviors per Session}\\
        \multicolumn{5}{l}{ASU: Average Sessions per User}\\
        \bottomrule

    \end{tabular}
\end{table}

\begin{table}[htb]
    \centering
    \small
    \caption{Other parameters settings for different scenarios.}
\label{tab:parameters_settings}
\vspace{-2mm}
    \begin{tabular}{cccccc}
        \toprule
        \multirow{2}{*}{Model} & \multicolumn{4}{c}{RSC15} \\
        \cmidrule(r){2-5}
           & Batch Size & Lr & RNN Size& dropout rate\\
        \midrule
        Default & 32 & 0.2  & 100 & 0 \\
        GRU4Rec (Category)& 50 & 0.001& 100 & 0.5\\
        C-GRU & 50  & 0.001  & 120 &0.5 \\
        P-GRU & 50 & 0.001 & 100 (item), 20 (category) &0.5 \\
        NARM & 512 &0.001 & 100 & 0.25\\
        \bottomrule
        \toprule
        \multirow{2}{*}{Model} & \multicolumn{4}{c}{RSC19} \\
        \cmidrule(r){2-5}
         & Batch Size & Lr & RNN Size& dropout rate\\
        \midrule
        Default & 32 & 0.2 & 100 &0\\
        GRU4Rec (Behavior) & 50 & 0.001 & 100 &0.5 \\
        B-GRU&50 &0.001 & 100 & 0.5\\
        NARM & 512 & 0.001 &100 &0.25\\
        User Implicit & 50 &0.001 & 50 & 0.5 \\
        User Embedded & 50 &0.001 & 50 & 0.5 \\
        User Recurrent & 50 &0.01 & 100 (item), 100 (user) & 0\\
        \bottomrule
        \toprule
        \multirow{2}{*}{Model} & \multicolumn{4}{c}{LastFM} \\
        \cmidrule(r){2-5}
        & Batch Size & Lr & RNN Size & dropout rate\\
        \midrule
        User Implicit & 50 & 0.001  & 50 & 0.5 \\
        User Embedded & 50 & 0.001  & 50 & 0.5 \\
        User Recurrent & 200 &0.02 & 50 (item), 50 (user) & 0 \\
        \bottomrule

    \end{tabular}

\end{table}

\subsubsection{Model Settings}\label{subsec:modelsettings}
We choose GRU4Rec \cite{hidasi2015session} (Figure \ref{fig:GRU4Rec_architecture}) as our \emph{basic} model, and then consider the influential factors in Figure \ref{fig:training_testing} to check their effects on the basic model. The main reason of using GRU4Rec is that lots of algorithms in the literature make improvement on it, or recognize it as a representative and competitive baseline for the sequential recommendation tasks. This makes GRU4Rec as a perfect fit to showcase the effects of influential factors on DL-based model. \emph{Specifically, in our experiments, we focus on the widely explored transaction-based sequential recommendation task, which aims to predict the next item a user will like/purchase on the basis of transaction-based sequences.} In the future, we can consider other representative models using different DL structures, e.g., \emph{NextItNet} \cite{yuan2019simple} (CNN-based model) and \emph{NARM} \cite{li2017neural} (attention-based model).

The \emph{default} parameters for the \textbf{basic} model is no data augmentation, no user representation (i.e., implicit user representation), BPR-max loss function, uniform negative sampling with a sample size of $128$ for \emph{RSC15} and \emph{RSC19}. In the next experiments, if not being particularly figured out, other models also use these default settings.

For the input module, we choose two kinds of \textbf{side information}: \emph{item category} and \emph{dwell time}. For the item category, following the previous studies, we implement two improved versions of the basic model: \textbf{C-GRU} (concatenating item embedding with category embedding \cite{covington2016deep}) and \textbf{P-GRU} (parellelly training two basic models for item and category respectively, and then concatenating the output of the two subnets \cite{hidasi2016parallel}) with mini-batch parallel negative sampling (batch size $=50$). The corresponding control model \textbf{GRU4Rec (category)} is the basic \textbf{GRU4Rec} model with the same setup as \textbf{C-GRU} and \textbf{P-GRU} except for the RNN size.
For the dwell time, we implement the model in \cite{bogina2017incorporating}, and according to the distribution of the dwell time, we choose $75$ and $100$ seconds as thresholds for \textit{RSC15}, and $45$ and $60$ seconds for \textit{RSC19}.

\begin{figure}[htbp]
    \centering
    \includegraphics[width=1\linewidth]{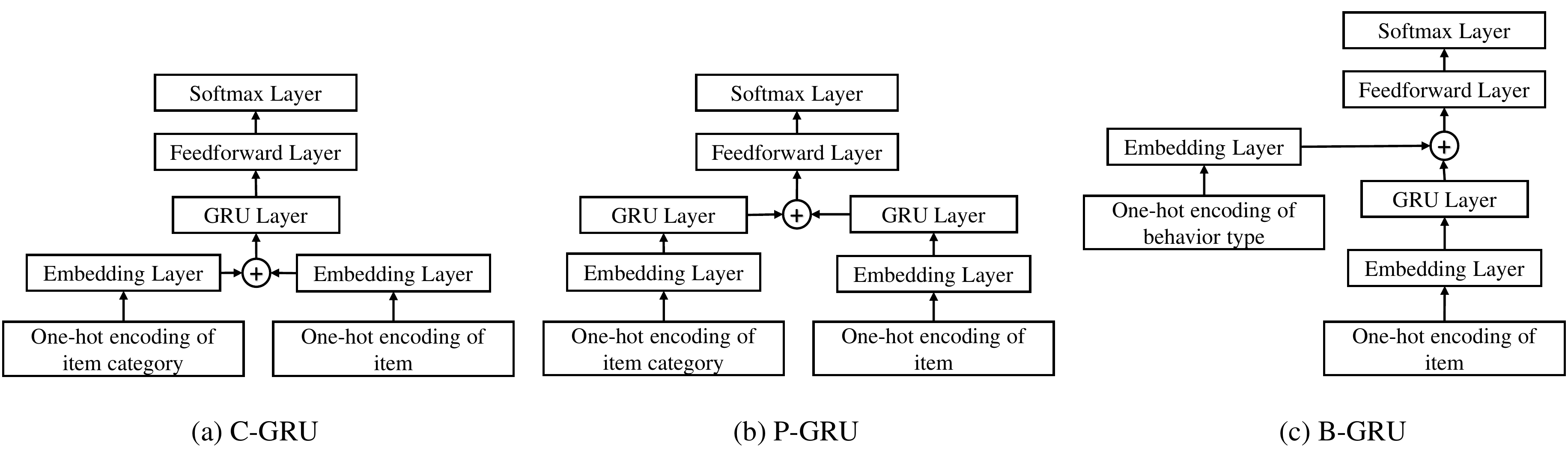}\vspace{-3mm}
    \caption{The network structures of C-GRU, P-GRU and B-GRU.}
    \label{fig:model}
\end{figure}

To verify the impact of \textbf{behavior types}, we design a new network (\textbf{B-GRU}) by adding a behavior type embedding module to the basic model. Specifically, \textbf{B-GRU} takes both the item one-hot vectors and behavior type one-hot vectors as input and converts them into embedding vectors, where item embedding vectors are fed into a GRU model, whose output is concatenated with behavior type embedding vectors for MLP layers. The current design aims to capture the intuition that a user's next behavior in the sequence is not only related to item sequence that the user has previously interacted with, but also might be impacted by the user¡¯s previous behavior type. Besides, \textbf{B-GRU} uses mini-batch parallel negative sampling method with a sample size of $50$. The respective control model \textbf{GRU4Rec (Behavior)} is the basic \textbf{GRU4Rec} model with the same setup as \textbf{B-GRU}. The structures of \textbf{C-GRU}, \textbf{P-GRU} and \textbf{B-GRU} are shown in Figure \ref{fig:model}.

For the data processing module, we implement the \textbf{data augmentation} method in \cite{tan2016improved} (see Figure \ref{fig:prefix data augmentation}) on the basic model. Specifically, we randomly select $50\%$ sessions in the training set to conduct data augmentation, and randomly treat a part of each session as new sessions.

For the model structure module, we consider three structures: \textbf{NARM} \cite{li2017neural} (\emph{incorporating the basic model with the attention mechanism}), \textbf{weighted model} in \cite{jannach2017recurrent} (\emph{combining the DL model with KNN}) and \textbf{adding an explicit user representation} in two ways, i.e., the recurrent way in \cite{quadrana2017personalizing} (referred as \emph{User Recurrent}), and the embedded way by adding a user embedding layer based on user IDs (referred as \emph{User Embedded}), which is concatenated with the output of GRU in the basic GRU4Rec model, and uses the same training method as in \cite{quadrana2017personalizing}).
Noted that we use user-parallel mini-batches \cite{quadrana2017personalizing} in training for both user recurrent and embedded models.

For the model training module, we consider three factors: \textbf{loss function} (i.e., cross-entropy, BPR-max, BPR, TOP1-max, and TOP1), \textbf{sampling method} (i.e., additional sampling in \cite{hidasi2018recurrent}), and $\alpha\in\{0, 0.25, 0.5, 0.75, 1\}$, and the size of negative samples $\{0,1,32,128,512,2,048\}$.
Other parameters in terms of different datasets are summarized in Table \ref{tab:parameters_settings}, where \textbf{Lr} refers to learning rate of these DL-based models.
It should be noted that we use the default hyper-parameters (e.g., batch size, learning rate, and RNN size) recommended in the source codes for the basic models (control models).
Besides, for demonstrating the effectiveness of the corresponding factors, in the experimental models (with factor effects), we fix the major default hyper-parameters to be the same as the corresponding control models. With this kind of setting, we aim to effectively demonstrate the impact of the influential factors (as well as the designed components) in the sequential recommendation, while maximally eliminate the impact of other factors.

\subsubsection{Evaluation Metrics}
To compare the performance of different models, we use three widely used accuracy metrics: \textbf{Recall@k}, \textbf{MRR@k} (Mean Reciprocal Rank) and \textbf{NDCG@k} (Normalized Discounted Cumulative Gain) as previous sequential recommendation models,
where $k$ is set to $5$, $10$ and $20$ respectively. For these three metrics, a larger value implies better performance. We refer interesting readers to \cite{zhang2019next} for detailed definitions of Recall and MRR evaluation metrics. Noted that GRU4Rec is to predict a user's next behavior (\emph{next-item prediction}), i.e., only one item in the recommendation list will be actually selected by the user. In this case, MRR is equivalent to Mean Average Precision (MAP), and Recall is identical to Hit Ratio (HR) \cite{BERT4Rec2019}.

\begin{itemize}

    \item \textbf{Recall}@k: it measures the coverage of the corrected recommended items in terms of ground-truth items.
    \item \textbf{MRR}@k: it refers to how well a model ranks the ground-truth items.
    \item \textbf{NDCG}@k: it rewards each ground-truth item based on its position in the recommendation list, indicating how strongly an item is recommended.

    $$\text{NDCG}@k=\frac{1}{\log_2(rank+1)}$$
    where $rank$ denotes the ranking position of the ground-truth item.

\end{itemize}
It should be noted that for each method, we run each experiment $5$ times and report the performance in the form of ``$\text{mean} \pm \text{std deviation}$" in terms of the three metrics as in Tables \ref{tab:result_of_incorporating_item_category_or_behavior_type} and \ref{Tab:recall_5}, and show the mean values in other Figures.

\subsection{Experiment Results}
\label{sec:Experiment_results}

\begin{table*}[htb]
    \centering
    \footnotesize
    \caption{Results of incorporating item category or behavior type. Statistical significance of pairwise differences of each improved model vs. basic model (GRU4Rec) is determined by a paired $t$-test (* for p-value $\leq$ 0.1, $\Diamond$ for p-value $\leq$ 0.05, $\Delta$ for p-value $\leq$ 0.01 ). }
\label{tab:result_of_incorporating_item_category_or_behavior_type}
    \begin{tabular}{ccccccc}
        \toprule
        \multirow{2}{*}{Model} & \multicolumn{6}{c}{RSC15} \\
        \cmidrule(r){2-7}
        & Recall@5 & MRR@5  & NDCG@5 & Recall@20 & MRR@20 & NDCG@20 \\
        \midrule
        GRU4Rec & 0.313$\pm$0.0047 & 0.168$\pm$0.0036& 0.203$\pm$0.0039 & 0.554$\pm$0.0043 & 0.192$\pm$0.0034 & 0.273$\pm$0.0035 \\
        C-GRU & 0.328$^{\Delta} \pm$ 0.0023 &0.178$^{\Delta}\pm$ 0.0016 & 0.215$^{\Delta}\pm$ 0.0015 & 0.564$^{\Delta} \pm$ 0.0032 &0.202$^{\Delta}\pm$ 0.0015 & 0.283$^{\Delta}\pm$ 0.0012 \\
        P-GRU & \textbf{0.335}$^{\Delta}\pm$0.0014 & \textbf{0.180}$^{\Delta}\pm$ 0.0017& \textbf{0.218}$^{\Delta}\pm$0.0014  & \textbf{0.570}$^{\Delta}\pm$0.0018 & \textbf{0.204}$^{\Delta}\pm$ 0.0017& \textbf{0.286}$^{\Delta}\pm$0.0015  \\

    \end{tabular}

    \begin{tabular}{cccccccc}
        \toprule
         \multirow{2}{*}{Model} & \multicolumn{6}{c}{RSC19} \\
        \cmidrule(r){2-7}  & Recall@5 & MRR@5  & NDCG@5 & Recall@20 & MRR@20  & NDCG@20 \\
        \midrule
        GRU4Rec & 0.568$\pm$ 0.0065& 0.489$\pm$0.001 & 0.509$\pm$ 0.0017 & 0.696$\pm$ 0.0023& 0.502$\pm$0.0019 & 0.546$\pm$ 0.0015\\
        B-GRU & \textbf{0.586}$^{\Delta}\pm$ 0.0025& \textbf{0.494}$^{\Delta}\pm$ 0.0032& \textbf{0.517}$^{\Delta}\pm$0.0030 & \textbf{0.708}$^{\Delta}\pm$ 0.0008& \textbf{0.507}$^{\Delta}\pm$ 0.0029& \textbf{0.552}$^{\Delta}\pm$0.0023  \\
        \bottomrule
    \end{tabular}

\end{table*}

\begin{table*}[htb]

\scriptsize
\centering

\caption{Results on considering different factors. Statistical significance of pairwise differences of each improved model vs. the basic \emph{GRU4Rec} is determined by a paired $t$-test (* for p-value $\leq$ 0.1, $\Diamond$ for p-value $\leq$ 0.05, and $\Delta$ for p-value $\leq$ 0.01 ). }\label{Tab:recall_5}

\begin{threeparttable}
\begin{tabular}{cccccccc}
\toprule
\multirow{2}{*}{Factor}& \multirow{2}{*}{Variable} & \multicolumn{6}{c}{RSC15} \\

\cmidrule(r){3-8}

&        &  Recall@5  &   MRR@5 &  NDCG@5 &  Recall@20  &   MRR@20 &  NDCG@20 \\
\midrule
\multirow{3}{0.3cm}{Dwell time}& 0\tnote{4} & 0.446$\pm$0.0011 & 0.268$\pm$0.0007 &  0.312$\pm$0.0008 & 0.676$\pm$0.0009 & 0.293$\pm$0.0006 &  0.380$\pm$0.0006 \\
& 75 & \textbf{0.772}$^{\Delta}\pm$0.0004& \textbf{0.692}$^{\Delta}\pm$ 0.0005& \textbf{0.712}$^{\Delta}\pm$0.0005&  \textbf{0.865}$^{\Delta}\pm$0.0005& \textbf{0.702}$^{\Delta}\pm$ 0.0005& \textbf{0.739}$^{\Delta}\pm$0.0004 \\
& 100 &0.730$^{\Delta}\pm$0.0010 &0.635$^{\Delta}\pm$0.0007 & 0.659$^{\Delta}\pm$00.0007 &0.841$^{\Delta}\pm$0.0004 &0.647$^{\Delta}\pm$0.0006 & 0.691$^{\Delta}\pm$0.0005\\

\midrule
\multirow{2}{0.3cm}{Data Aug\tnote{1} }& Off\tnote{4} & 0.446$\pm$0.0011 & \textbf{0.268}$\pm$0.0007 &  \textbf{0.312}$\pm$0.0008 & 0.676$\pm$0.0009 & \textbf{0.293}$\pm$0.0006 &  \textbf{0.380}$\pm$0.0006\\
 & On & \textbf{0.446}$\pm$0.0018 & 0.267$\pm$0.0005 & 0.312$\pm$0.0007  & \textbf{0.678}$^{\Diamond}\pm$0.0011 & 0.292$\pm$0.0005 & 0.379$\pm$0.0005 \\

\midrule
\multirow{2}{0.3cm}{Att\tnote{2}} & Off\tnote{5} &0.480$\pm$0.0005& 0.285$\pm$0.0005 & 0.334$\pm$  0.0005 &0.703$\pm$0.0001& 0.309$\pm$0.0005 & 0.400$\pm$  0.0004 \\
 & On &\textbf{0.486}$^{\Delta}\pm$0.0003& \textbf{0.290}$^{\Delta}\pm$0.0002 &\textbf{0.339}$^{\Delta}\pm$  0.0002   &\textbf{0.708}$^{\Delta}\pm$0.0002& \textbf{0.314}$^{\Delta}\pm$0.0003 &\textbf{0.404}$^{\Delta}\pm$  0.0002  \\
\midrule

\multirow{3}{0.3cm}{KNN weight }& 0\tnote{4} &  0.446$\pm$0.0011 & 0.268$\pm$0.0007 &  0.312$\pm$0.0008 & 0.676$\pm$0.0009 & 0.293$\pm$0.0006 &  0.380$\pm$0.0006 \\
& 0.1 & 0.452$^{\Delta}\pm$0.0008& 0.270$^{\Delta}\pm$0.0003& 0.315$^{\Delta}\pm$0.0002 & 0.693$^{\Delta}\pm$0.0006& 0.296$^{\Delta}\pm$0.0005& 0.386$^{\Delta}\pm$0.0004 \\
& 0.3 &\textbf{0.460}$^{\Delta}\pm$0.0007 &\textbf{0.278}$^{\Delta}\pm$0.0004 &\textbf{0.323}$^{\Delta} \pm$0.0004  &\textbf{0.698}$^{\Delta}\pm$0.0009 &\textbf{0.303}$^{\Delta}\pm$0.0005 &\textbf{0.393}$^{\Delta} \pm$0.0003\\

\toprule

\multirow{2}{*}{Factor}& \multirow{2}{*}{Variable} &  \multicolumn{6}{c}{RSC19}\\

\cmidrule(r){3-8}

&        &  Recall@5  &   MRR@5 &  NDCG@5 &  Recall@20  &   MRR@20 &  NDCG@20   \\

\midrule
\multirow{3}{0.3cm}{Dwell time}& 0\tnote{4} &0.640$\pm$0.0018&0.547$\pm$0.0028&0.571$\pm$0.0025 &0.751$\pm$0.0017&0.559$\pm$0.0028&0.602$\pm$0.0025\\
& 45 &  \textbf{0.845}$^{\Delta}\pm$0.0074& \textbf{0.783}$^{\Delta}\pm$0.0063& \textbf{0.799}$^{\Delta}\pm$0.0065 &  \textbf{0.893}$^{\Delta}\pm$0.0071& \textbf{0.788}$^{\Delta}\pm$0.0062& \textbf{0.813}$^{\Delta}\pm$0.0063 \\
& 60 &0.830$\pm$0.0007 &0.763$^{\Delta}\pm$0.0013 &0.780$^{\Delta}\pm$0.0011 &0.885$\pm$0.0010 &0.768$^{\Delta}\pm$0.0015 &0.795$^{\Delta}\pm$0.0013 \\

\midrule
\multirow{2}{0.3cm}{Data Aug\tnote{1} }& Off\tnote{4} &0.640$\pm$0.0018 &0.547$\pm$0.0028&0.571$\pm$0.0025 &0.751$\pm$0.0017& 0.559$\pm$0.0028 & 0.602$\pm$0.0025\\
 & On & \textbf{0.641} $\pm$0.0015&\textbf{0.551}$^{\Diamond}\pm$0.0019 &\textbf{0.574}$^{\Diamond}\pm$0.0013 & \textbf{0.754}$^{\Diamond} \pm$0.0004&\textbf{0.562}$^{\Diamond}\pm$0.0020 &\textbf{0.606}$^{\Diamond}\pm$0.0014 \\

\midrule
\multirow{2}{0.3cm}{Att\tnote{2}} & Off\tnote{5} &0.736$\pm$0.0015&0.569$\pm$0.0010& 0.611$\pm$0.0011 &0.905$\pm$0.0009&0.587$\pm$0.0001& 0.661$\pm$0.0001\\
 & On &\textbf{0.742}$^{\Delta}\pm$0.0023&\textbf{0.572}$^{\Delta}\pm$ 0.0024 &\textbf{0.615}$^{\Delta} \pm$0.0023 &\textbf{0.912}$^{\Delta}\pm$0.0012&\textbf{0.591}$^{\Delta}\pm$ 0.0022 &\textbf{0.665}$^{\Delta} \pm$0.0020\\
\midrule

\multirow{3}{0.3cm}{KNN weight }& 0\tnote{4}  &0.640$\pm$0.0018&0.547$\pm$0.0028&0.571$\pm$0.0025&0.751$\pm$0.0017&0.559$\pm$0.0028&0.602$\pm$0.0025 \\
& 0.1 &0.643$\pm$0.0039 & 0.549$\pm$0.0064& 0.572$\pm$0.0057&0.753$\pm$0.0034 & 0.560$\pm$0.0063& 0.604$\pm$0.0055 \\
& 0.3 &\textbf{0.657}$^{\Delta} \pm$0.0023&\textbf{0.562}$^{\Delta}\pm$0.0067 & \textbf{0.586}$^{\Delta}\pm$0.0057 &\textbf{0.765}$^{\Delta} \pm$0.0033&\textbf{0.573}$\pm$0.0067 & \textbf{0.617}$\pm$0.0057\\

\toprule

\multirow{2}{*}{Factor}& \multirow{2}{*}{Variable} & \multicolumn{6}{c}{LastFM} \\

\cmidrule(r){3-8}

&        &  Recall@5  &   MRR@5 &  NDCG@5   &  Recall@20  &   MRR@20 &  NDCG@20  \\

\midrule
\multirow{3}{0.3cm}{User Rep\tnote{3}}& Implicit\tnote{6} & \textbf{0.173}$^{\Delta}\pm$0.0021 & \textbf{0.147}$^{\Delta}\pm$0.0018 & \textbf{0.154}$^{\Delta}\pm$0.0018 & \textbf{0.193}$^{\Delta}\pm$0.0024 & \textbf{0.149}$^{\Delta}\pm$0.0018 & \textbf{0.160}$^{\Delta}\pm$0.0018\\
& Embedded & 0.006$\pm$0.0011&0.004$\pm$0.0016 & 0.004$\pm$0.0014 & 0.016$\pm$0.0026&0.005$\pm$0.0015 & 0.007$\pm$0.0012 \\
& Recurrent &0.002$\pm$0.0002 &0.001$\pm$0.0002& 0.002$\pm$0.0007 &0.003$\pm$0.0009 & 0.002$\pm$0.0005& 0.002$\pm$0.0002 \\

\toprule

\multirow{2}{*}{Factor}& \multirow{2}{*}{Variable} & \multicolumn{6}{c}{RSC19 (user)}\\

\cmidrule(r){3-8}
&        &  Recall@5  &   MRR@5    &  NDCG@5  &  Recall@20  &   MRR@20    &  NDCG@20   \\

\midrule
\multirow{3}{0.3cm}{User Rep\tnote{3}}& Implicit\tnote{6} & \textbf{0.713}$^{\Delta}\pm$0.0165  &\textbf{0.654}$^{\Delta}\pm$0.0128 &\textbf{0.668}$^{\Delta}\pm$0.0130 & \textbf{0.781}$^{\Delta}\pm$0.0134 &\textbf{0.661}$^{\Delta}\pm$0.0128 &\textbf{0.689}$^{\Delta}\pm$0.0125 \\
& Embedded &0.032$\pm$0.0098 & 0.023$\pm$0.0092& 0.025$\pm$0.0093 &0.060$\pm$0.0113 & 0.025$\pm$0.0090& 0.032$\pm$0.0094 \\
& Recurrent &0.030$\pm$0.0199&0.016$\pm$0.0113&0.019$\pm$0.0131 &0.079$\pm$0.0198&0.020$\pm$0.0111&0.033$\pm$0.0126\\

\bottomrule
    \end{tabular}

    \begin{tablenotes}
    \footnotesize
    \item[1] "Data Aug" refers to Data Augmentation.
    \item[2] "Att" refers to Attention Mechanism.
    \item[3] "User Rep" refers to User Representation.
    \item[4] ``0" and ``Off" scenarios refer to that the corresponding influential factor is not considered, i.e., being equivalent to the basic \emph{GRU4Rec}.
    \item[5] ``Off" here denotes NARM \cite{li2017neural} without attention mechanism.
    \item[6] ``Implicit" denotes no user representation.

    \end{tablenotes}

\end{threeparttable}

\end{table*}

Here, we systematically present the experimental results of different influential factors in terms of the four modules.

\subsubsection{Input Module} First, we present the experimental results regarding the factors related to the input module: side information and behavior types.

\par{\textit{Side information effects.}
Tables \ref{tab:result_of_incorporating_item_category_or_behavior_type} and \ref{Tab:recall_5} show the results of the two types of the side information on DL-based model respectively\footnote{We have consistent results when $k=10$. Due to the space limitation, we do not report it here, but on Github: \url{https://github.com/sttich/dl-recommendation}.}. As shown in Table \ref{tab:result_of_incorporating_item_category_or_behavior_type}, incorporating \textbf{item category} information into GRU4Rec can improve the model performance in terms of all the three metrics. Specifically, C-GRU and P-GRU perform better than the basic model.
As we can see in Table \ref{Tab:recall_5}, \textbf{dwell time} can greatly improve the performance, e.g., Recall@20 increases by about $28\%$ and $19\%$ on \textit{RSC15} and \textit{RSC19} datasets respectively.
To conclude, utilizing the side information can significantly improve the model performance, and the way in which the side information is incorporated also matters. Thus, it is necessary to have a calibrated design by considering the impact of side information on the final prediction.
}
\par{\textit{Behavior type effects.} The results regarding impact of behavior types (\emph{B-GRU}) are present in Table \ref{tab:result_of_incorporating_item_category_or_behavior_type}. We can see that B-GRU outperforms the basic model in terms of all metrics. If a dataset provides the \textbf{behavior type} information, it is better to integrate it into the final model with appropriately designed modules, e.g., the simple module as ours.
}

\subsubsection{Data Processing} Here, we check the impact in regard to the data augmentation.

\par{\textit{Data augmentation effects.} As shown in Table \ref{Tab:recall_5}, the model with data augmentation performs slightly better ($0.2\%$) than the basic model on Recall@20 for \textit{RSC15}, but worse in terms of MRR@20 and NDCG@20. On \textit{RSC19}, data augmentation improves the model performance in terms of all metrics but with a lower significance level (i.e., 5\%). Similar mixed results can be observed when $k=5$ (see Table \ref{Tab:recall_5}).
To conclude, simple data augmentation cannot significantly enhance the model performance for \textbf{GRU4Rec} model. We might consider to design more complex ways according to the characteristics of \textbf{GRU4Rec} model.
}

\subsubsection{Model Structure}
In this subsection, we present the experimental results with respect to varying the DL structures.

\par{\textit{Incorporating attention mechanism effects.}
As shown in Table \ref{Tab:recall_5}, incorporating \textbf{attention mechanism} enhances the performance of the model almost for all the scenarios. }

\par{\textit{Combining with conventional method effects.}
Combing the basic model with \textbf{KNN} improves model performance in terms of all metrics on both \textit{RSC15} and \textit{RSC19}, and KNN weight of $0.3$ provides better performance than that of $0.1$, manifesting that the way of combining traditional models with DL-models can have a significant effect on the sequential recommendation.
}

\par{\textit{User representation effects.}
\textit{Implicit} represents the basic GRU4Rec model with session-parallel mini-batch method. \textit{Recurrent} and \textit{Embedded} refer to adding explicit user representation in two different ways as discussed in Section \ref{subsec:modelsettings}. For user representation, we find that adding an explicit user representation module, whether embedded or recurrent one, leads to a sharp decrease on all metrics. For a complementary investigation, we tuned the major hyperparameters (e.g., batch size, learning rate, and RNN size, etc) for user recurrent and user embedded models, and found that the large gap between these two models and the basic model cannot be significantly decreased.
The main reasons might be three-folds: 1) session-parallel mini-batch (a session as a sample) is used for the implicit model while user-parallel mini-batch (a user's all historical sessions as a sample) is deployed for user embedded and recurrent models. In this case, training samples for user embedded and recurrent models are much fewer than the implicit model (since we did not find a specific model which adds user embedding in \textit{GRU4Rec} directly, we also refer to the training method in \cite{quadrana2017personalizing} for training the user embedded model); 2) as shown in Table \ref{tab:datasets_descriptions}, the number of sessions is much greater than that of users on these two datasets; 3) according to the number of items and behaviors shown in Table \ref{tab:datasets_descriptions}, the average support of items is much smaller on these two datasets. It is worth mentioning that as reported in  \cite{quadrana2017personalizing}, GRU4REC with recurrent user representation performs better than the original model on their two datasets.
Furthermore, we can see that the user embedded model outperforms user recurrent model in most scenarios, but performs worse than recurrent model in terms of Recall@20 and NDCG@20 on \textit{RSC19 (user)}. In this case, in the personalized sequential recommendation, we can infer that whether selects the user embedded model or recurrent model largely depends on the characteristics of datasets and application scenarios. However, whether considering explicit user representation model is dependent on not only the application scenarios, but also a well-designed user representation component.
}

\subsubsection{Model Training} Here, we present the experiments results from the perspectives of three factors: sampling methods, sample size, and loss functions.
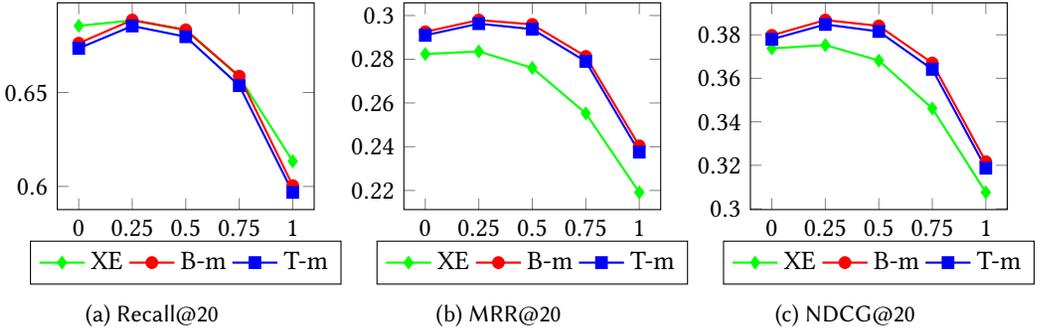
\begin{figure*}[htb]
    \centering
\begin{subfigure}[b]{0.3\linewidth}
\begin{tikzpicture}
\begin{axis}[
    width=5cm,
    height=4.3cm,
    legend style={at={(0.5,-0.15)},
      anchor=north,legend columns=-1},
    symbolic x coords={0,0.25,0.5,0.75,1},
    xtick=data,
    ]
  \addplot [mark=diamond*,thick,green] coordinates {(0, 0.6854986) (0.25,0.6883574) (0.5,0.6831622) (0.75,0.6583278) (1,0.6134028 )};
  \addplot [mark=*,thick,red] coordinates {(0, 0.676235286) (0.25,0.68857199) (0.5,0.683362744) (0.75,0.658703344) (1,0.600429354 )};
  \addplot [mark=square*,thick,blue] coordinates {(0, 0.6735232) (0.25,0.6854198) (0.5,0.6796916) (0.75,0.6536654 ) (1,0.5968334 )};

\legend{XE, B-m, T-m}
\end{axis}
\end{tikzpicture}
\subcaption{Recall@20}
\end{subfigure}
\quad
\begin{subfigure}[b]{0.3\linewidth}
\begin{tikzpicture}
\begin{axis}[
    width=5cm,
    height=4.3cm,
    legend style={at={(0.5,-0.15)},
      anchor=north,legend columns=-1},
    symbolic x coords={0,0.25,0.5,0.75,1},
    xtick=data,
    ]
  \addplot [mark=diamond*,thick,green] coordinates {(0, 0.2824) (0.25,0.2835626) (0.5,0.2760376) (0.75,0.2552634) (1,0.2191246 )};
  \addplot [mark=*,thick,red] coordinates {(0, 0.292532376) (0.25,0.297920356) (0.5,0.295934216) (0.75,0.281303866 ) (1,0.240395606 )};
  \addplot [mark=square*,thick,blue] coordinates {(0, 0.2909604) (0.25,0.296278) (0.5,0.293756) (0.75,0.279024 ) (1,0.2375098 )};

\legend{XE, B-m, T-m}
\end{axis}
\end{tikzpicture}
\subcaption{MRR@20}
\end{subfigure}
\quad
\begin{subfigure}[b]{0.3\linewidth}
\begin{tikzpicture}
\begin{axis}[
    width=5cm,
    height=4.3cm,
    legend style={at={(0.5,-0.15)},
      anchor=north,legend columns=-1},
    symbolic x coords={0,0.25,0.5,0.75,1},
    xtick=data,
    ]
  \addplot [mark=diamond*,thick,green] coordinates {(0,0.3737384) (0.25,0.3752706) (0.5,0.3681852) (0.75,0.3461604) (1,0.3075374 )};
  \addplot [mark=*,thick,red] coordinates {(0, 0.37974869) (0.25,0.38679021) (0.5,0.384095602) (0.75,0.367030728 ) (1,0.321643276 )};
  \addplot [mark=square*,thick,blue] coordinates {(0, 0.3779478) (0.25,0.3847816) (0.5,0.3815454) (0.75,0.3641304 ) (1,0.318669)};

\legend{XE, B-m, T-m}
\end{axis}
\end{tikzpicture}
\subcaption{NDCG@20}
\end{subfigure}

    \caption{The impact of $\alpha$ on additional sampling strategy on RSC15 (XE: cross-entropy; B-m: BPR-max; T-m: TOP1-max).}
    \label{fig:sample_alpha15}
\end{figure*}

\begin{figure*}[htb]
    \centering
    \flushleft
\begin{subfigure}[b]{0.3\linewidth}
\begin{tikzpicture}
\begin{axis}[
    width=5cm,
    height=4.3cm,
    legend style={at={(0.5,-0.15)},
      anchor=north,legend columns=-1},
    symbolic x coords={0,0.25,0.5,0.75,1},
    xtick=data,
    ]
  \addplot [mark=diamond*,thick,green] coordinates {(0, 0.727483872) (0.25,0.727374558) (0.5,0.72491738) (0.75,0.719866796 ) (1,0.713711242 )};
  \addplot [mark=*,thick,red] coordinates {(0, 0.751144076) (0.25,0.75925225) (0.5,0.76020922) (0.75,0.753596206 ) (1,0.74218152)};
  \addplot [mark=square*,thick,blue] coordinates {(0, 0.757770546) (0.25,0.763858828) (0.5,0.765094982) (0.75,0.759454072 ) (1,0.746598048)};

\legend{XE, B-m, T-m}
\end{axis}
\end{tikzpicture}
\subcaption{Recall@20}
\end{subfigure}
\quad
\begin{subfigure}[b]{0.3\linewidth}
\begin{tikzpicture}
\begin{axis}[
    width=5cm,
    height=4.3cm,
    legend style={at={(0.5,-0.15)},
      anchor=north,legend columns=-1},
    symbolic x coords={0,0.25,0.5,0.75,1},
    xtick=data,
    ]
  \addplot [mark=diamond*,thick,green] coordinates {(0, 0.5746851) (0.25,0.579626586) (0.5,0.580057748) (0.75,0.571986884) (1,0.554374672 )};
  \addplot [mark=*,thick,red] coordinates {(0, 0.558510686) (0.25,0.580720022) (0.5,0.596509392) (0.75,0.599929636) (1,0.57462743 )};
  \addplot [mark=square*,thick,blue] coordinates {(0, 0.547491656) (0.25,0.571766744) (0.5,0.592760482) (0.75,0.604495358 ) (1,0.592337918 )};

\legend{XE, B-m, T-m}
\end{axis}
\end{tikzpicture}
\subcaption{MRR@20}
\end{subfigure}
\quad
\begin{subfigure}[b]{0.3\linewidth}
\begin{tikzpicture}
\begin{axis}[
    width=5cm,
    height=4.3cm,
    legend style={at={(0.5,-0.15)},
      anchor=north,legend columns=-1},
    symbolic x coords={0,0.25,0.5,0.75,1},
    xtick=data,
    ]
  \addplot [mark=diamond*,thick,green] coordinates {(0,0.610195272) (0.25,0.613938856) (0.5,0.61372049) (0.75,0.60653953) (1,0.591938028 )};
  \addplot [mark=*,thick,red] coordinates {(0, 0.60248081) (0.25,0.62135775) (0.5,0.633688844) (0.75,0.634773098 ) (1,0.613040258 )};
  \addplot [mark=square*,thick,blue] coordinates {(0, 0.595313382) (0.25,0.61542794) (0.5,0.631848926) (0.75,0.639449536 ) (1,0.62729276 )};

\legend{XE, B-m, T-m}
\end{axis}
\end{tikzpicture}
\subcaption{NDCG@20}
\end{subfigure}

    \caption{The impact of $\alpha$ on additional sampling strategy on RSC19 (XE: cross-entropy; B-m: BPR-max; T-m: TOP1-max).}
    \label{fig:sample_alpha19}
\end{figure*}
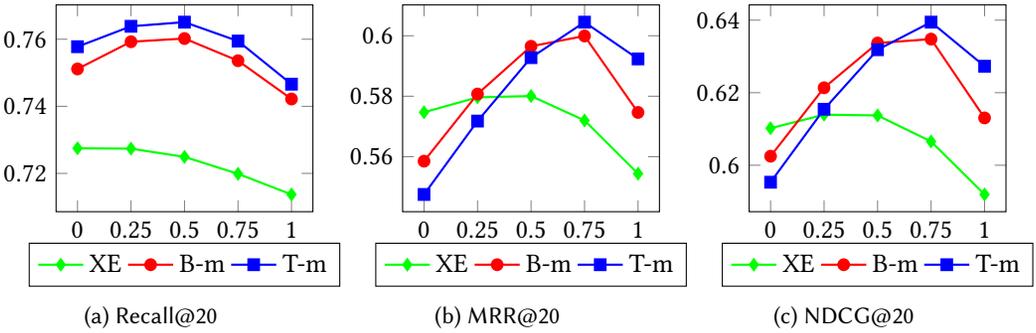

\begin{figure*}[htb]
    \centering
    \flushleft
\begin{subfigure}[b]{0.3\linewidth}
\begin{tikzpicture}
\begin{axis}[
    width=5cm,
    height=4.3cm,
    legend style={at={(0.5,-0.15)},
      anchor=north,legend columns=-1},
    symbolic x coords={0,1,32,128,512,2048},
    xtick=data,
    ]
  \addplot [mark=diamond*,thick,green] coordinates {(0, 0.10889477) (1,0.287115816)(32,0.4138572) (128,0.445897884) (512,0.4673798) (2048,0.4811694)};
  \addplot [mark=*,thick,red] coordinates {(0, 0.18998891)(1,0.393187592) (32,0.536309) (128,0.573240546) (512,0.5978424) (2048,0.6142582)};
  \addplot [mark=square*,thick,blue] coordinates {(0, 0.285491428)(1,0.488375256) (32,0.634749) (128,0.676235286) (512,0.7047586) (2048,0.7197968)};
\legend{@5, @10, @20}
\end{axis}
\end{tikzpicture}
\subcaption{Recall}
\end{subfigure}
\quad
\begin{subfigure}[b]{0.3\linewidth}
\begin{tikzpicture}
\begin{axis}[
    width=5cm,
    height=4.3cm,
    legend style={at={(0.5,-0.15)},
      anchor=north,legend columns=-1},
    symbolic x coords={0,1,32,128,512,2048},
    xtick=data,
    ]
  \addplot [mark=diamond*,thick,green] coordinates {(0, 0.050823108)(1,0.161988264) (32,0.2464506) (128,0.268122236) (512,0.2808458) (2048,0.2900052)};
  \addplot [mark=*,thick,red] coordinates {(0, 0.0614732)(1,0.176204474) (32,0.2629508) (128,0.285265824) (512,0.2984498) (2048,0.3079134)};
  \addplot [mark=square*,thick,blue] coordinates {(0, 0.068087498)(1,0.182843112) (32,0.269882) (128,0.292532376) (512,0.3059816) (2048,0.3153594)};
\legend{@5, @10, @20}
\end{axis}
\end{tikzpicture}
\subcaption{MRR}
\end{subfigure}
\quad
\begin{subfigure}[b]{0.3\linewidth}
\begin{tikzpicture}
\begin{axis}[
    width=5cm,
    height=4.3cm,
    legend style={at={(0.5,-0.15)},
      anchor=north,legend columns=-1},
    symbolic x coords={0,1,32,128,512,2048},
    xtick=data,
    ]
   \addplot [mark=diamond*,thick,green] coordinates {(0, 0.06507205) (1,0.192915576)(32,0.2879406) (128,0.312231766) (512,0.3271366) (2048,0.337497)};
  \addplot [mark=*,thick,red] coordinates {(0, 0.09112257)(1,0.227276924) (32,0.3276972) (128,0.353560236) (512,0.3695182) (2048,0.380681)};
  \addplot [mark=square*,thick,blue] coordinates {(0, 0.115252682)(1,0.251384844) (32,0.3527104) (128,0.37974869) (512,0.3966898) (2048,0.4075164)};
\legend{@5, @10, @20}
\end{axis}
\end{tikzpicture}
\subcaption{NDCG}
\end{subfigure}

    \caption{The effect of sample size on RSC15.}
    \label{fig:sample_size15}
\end{figure*}
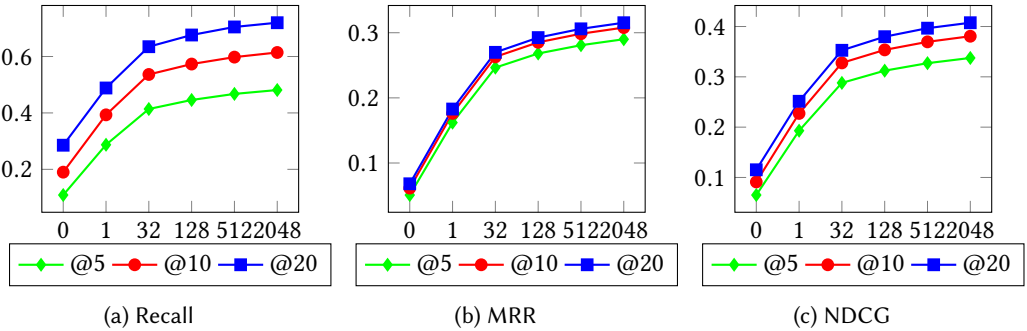

\begin{figure*}[htb]
    \centering
    \flushleft
\begin{subfigure}[b]{0.3\linewidth}
\begin{tikzpicture}
\begin{axis}[
    width=5cm,
    height=4.3cm,
    legend style={at={(0.5,-0.15)},
      anchor=north,legend columns=-1},
    symbolic x coords={0,1,32,128,512,2048},
    xtick=data,
    ]
  \addplot [mark=diamond*,thick,green] coordinates {(0, 0.5874202175) (1,0.615753005)(32,0.6406947725) (128,0.640297014) (512,0.6373605125) (2048,0.6474726075)};
  \addplot [mark=*,thick,red] coordinates {(0, 0.6347872875)(1,0.6610093625) (32,0.691082855) (128,0.695289992) (512,0.6956553725) (2048,0.7043273875)};
  \addplot [mark=square*,thick,blue] coordinates {(0, 0.68012984)(1,0.708336485) (32,0.7420869175) (128,0.751144076) (512,0.7549236025) (2048,0.7631814625)};
\legend{@5, @10, @20}
\end{axis}
\end{tikzpicture}
\subcaption{Recall}
\end{subfigure}
\quad
\begin{subfigure}[b]{0.3\linewidth}
\begin{tikzpicture}
\begin{axis}[
    width=5cm,
    height=4.3cm,
    legend style={at={(0.5,-0.15)},
      anchor=north,legend columns=-1},
    symbolic x coords={0,1,32,128,512,2048},
    xtick=data,
    ]
  \addplot [mark=diamond*,thick,green] coordinates {(0, 0.490358545)(1,0.5335505) (32,0.555903945) (128,0.547272412) (512,0.5433462325) (2048,0.559554705)};
  \addplot [mark=*,thick,red] coordinates {(0, 0.49672358)(1,0.539604975) (32,0.562653315) (128,0.554628566) (512,0.5511487325) (2048,0.567159615)};
  \addplot [mark=square*,thick,blue] coordinates {(0, 0.4998761225)(1,0.5428933325) (32,0.566189355) (128,0.558510686) (512,0.5552633375) (2048,0.57123967)};
\legend{@5, @10, @20}
\end{axis}
\end{tikzpicture}
\subcaption{MRR}
\end{subfigure}
\quad
\begin{subfigure}[b]{0.3\linewidth}
\begin{tikzpicture}
\begin{axis}[
    width=5cm,
    height=4.3cm,
    legend style={at={(0.5,-0.15)},
      anchor=north,legend columns=-1},
    symbolic x coords={0,1,32,128,512,2048},
    xtick=data,
    ]
   \addplot [mark=diamond*,thick,green] coordinates {(0, 0.5147459) (1,0.55417195)(32,0.5771273175) (128,0.570551206) (512,0.5668414375) (2048,0.5814921825)};
  \addplot [mark=*,thick,red] coordinates {(0, 0.5301069775)(1,0.5688214625) (32,0.59344616) (128,0.588351362) (512,0.5857155175) (2048,0.5998951275)};
  \addplot [mark=square*,thick,blue] coordinates {(0, 0.541578565)(1,0.58079207) (32,0.6063370375) (128,0.60248081) (512,0.6007024825) (2048,0.6147702)};
\legend{@5, @10, @20}
\end{axis}
\end{tikzpicture}
\subcaption{NDCG}
\end{subfigure}

    \caption{The effect of sample size on RSC19.}
    \label{fig:sample_size19}
\end{figure*}
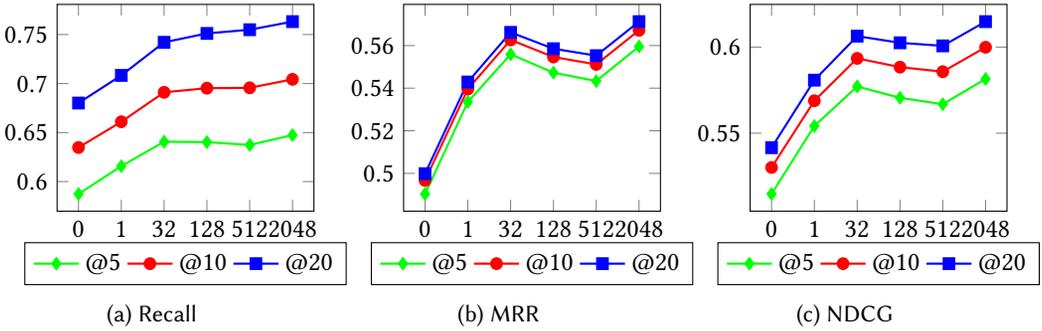

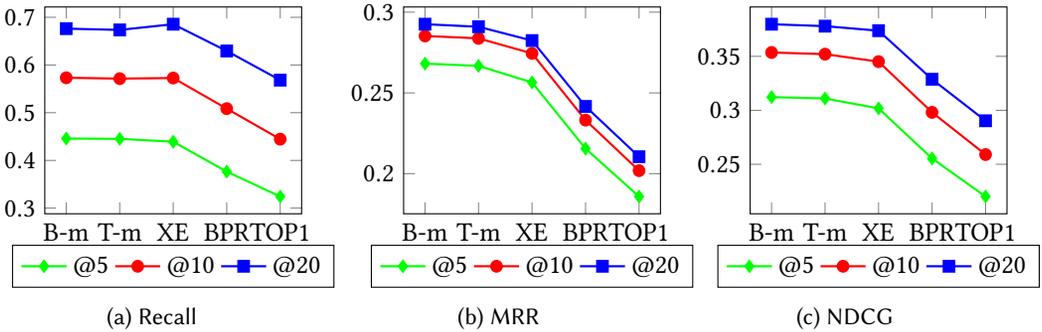
\begin{figure*}[htb]
    \centering
    \flushleft
\begin{subfigure}[b]{0.3\linewidth}
\begin{tikzpicture}
\begin{axis}[
    width=5 cm,
    height=4.3cm,
    legend style={at={(0.5,-0.15)},
      anchor=north,legend columns=-1},
    symbolic x coords={B-m,T-m,XE,BPR,TOP1},
    xtick=data,
    ]
  \addplot [mark=diamond*,thick,green] coordinates {(B-m, 0.445897884) (T-m,0.4451466) (XE,0.439193) (BPR,0.376525) (TOP1,0.3242656)};
  \addplot [mark=*,thick,red] coordinates {(B-m, 0.573240546) (T-m,0.5714408) (XE,0.5727144) (BPR,0.5085156) (TOP1,0.4445312)};
  \addplot [mark=square*,thick,blue] coordinates {(B-m, 0.676235286) (T-m,0.6735232) (XE,0.6854986) (BPR,0.6294036) (TOP1,0.5682744)};
\legend{@5, @10, @20}
\end{axis}
\end{tikzpicture}
\subcaption{Recall}
\end{subfigure}
\quad
\begin{subfigure}[b]{0.3\linewidth}
\begin{tikzpicture}
\begin{axis}[
    width=5cm,
    height=4.3cm,
    legend style={at={(0.5,-0.15)},
      anchor=north,legend columns=-1},
    symbolic x coords={B-m,T-m,XE,BPR,TOP1},
    xtick=data,
    ]
  \addplot [mark=diamond*,thick,green] coordinates {(B-m, 0.268122236) (T-m,0.2667334) (XE,0.2565068) (BPR,0.2155204) (TOP1,0.1858358)};
  \addplot [mark=*,thick,red] coordinates {(B-m, 0.285265824) (T-m,0.2837748) (XE,0.2744538) (BPR,0.2331534) (TOP1,0.2018634)};
  \addplot [mark=square*,thick,blue] coordinates {(B-m, 0.292532376) (T-m,0.2909604) (XE,0.2824) (BPR,0.2416308) (TOP1,0.2104756)};
\legend{@5, @10, @20}
\end{axis}
\end{tikzpicture}
\subcaption{MRR}
\end{subfigure}
\quad
\begin{subfigure}[b]{0.3\linewidth}
\begin{tikzpicture}
\begin{axis}[
    width=5cm,
    height=4.3cm,
    legend style={at={(0.5,-0.15)},
      anchor=north,legend columns=-1},
    symbolic x coords={B-m,T-m,XE,BPR,TOP1},
    xtick=data,
    ]
  \addplot [mark=diamond*,thick,green] coordinates {(B-m, 0.312231766) (T-m,0.3109854) (XE,0.3017688) (BPR,0.2553886) (TOP1,0.220112)};
  \addplot [mark=*,thick,red] coordinates {(B-m, 0.353560236) (T-m,0.3520124) (XE,0.3450742) (BPR,0.2980888) (TOP1,0.2589804)};
  \addplot [mark=square*,thick,blue] coordinates {(B-m, 0.37974869) (T-m,0.3779478) (XE,0.3737384) (BPR,0.3287634) (TOP1,0.2902988)};
\legend{@5, @10, @20}
\end{axis}
\end{tikzpicture}
\subcaption{NDCG}
\end{subfigure}

    \caption{Model performance for different loss functions on RSC15 (B-m: BPR-max; T-m: Top1-max; XE: cross-entropy).}
    \label{fig:loss_function15}
\end{figure*}

\begin{figure*}[htb]
    \centering
    \flushleft
\begin{subfigure}[b]{0.3\linewidth}
\begin{tikzpicture}
\begin{axis}[
    width=5cm,
    height=4.3cm,
    legend style={at={(0.5,-0.15)},
      anchor=north,legend columns=-1},
    symbolic x coords={B-m,T-m,XE,BPR,TOP1},
    xtick=data,
    ]
  \addplot [mark=diamond*,thick,green] coordinates {(B-m,0.640297014) (T-m,0.63353936) (XE,0.652184296) (BPR,0.521250956) (TOP1,0.370461748)};
  \addplot [mark=*,thick,red] coordinates {(B-m, 0.695289992) (T-m,0.695291676) (XE,0.688088328) (BPR,0.588181674) (TOP1,0.46697949)};
  \addplot [mark=square*,thick,blue] coordinates {(B-m, 0.751144076) (T-m,0.757770546) (XE,0.727483872) (BPR,0.651713378) (TOP1,0.574003716)};
\legend{@5, @10, @20}
\end{axis}
\end{tikzpicture}
\subcaption{Recall}
\end{subfigure}
\quad
\begin{subfigure}[b]{0.3\linewidth}
\begin{tikzpicture}
\begin{axis}[
    width=5cm,
    height=4.3cm,
    legend style={at={(0.5,-0.15)},
      anchor=north,legend columns=-1},
    symbolic x coords={B-m,T-m,XE,BPR,TOP1},
    xtick=data,
    ]
  \addplot [mark=diamond*,thick,green] coordinates {(B-m, 0.547272412) (T-m,0.534889964) (XE,0.567170382) (BPR,0.39950848) (TOP1,0.250755596)};
  \addplot [mark=*,thick,red] coordinates {(B-m, 0.554628566) (T-m,0.543151972) (XE,0.57196212) (BPR,0.408487154) (TOP1,0.26358219)};
  \addplot [mark=square*,thick,blue] coordinates {(B-m, 0.558510686) (T-m,0.547491656) (XE,0.5746851) (BPR,0.412911444) (TOP1,0.270978818)};
\legend{@5, @10, @20}
\end{axis}
\end{tikzpicture}
\subcaption{MRR}
\end{subfigure}
\quad
\begin{subfigure}[b]{0.3\linewidth}
\begin{tikzpicture}
\begin{axis}[
    width=5cm,
    height=4.3cm,
    legend style={at={(0.5,-0.15)},
      anchor=north,legend columns=-1},
    symbolic x coords={B-m,T-m,XE,BPR,TOP1},
    xtick=data,
    ]
  \addplot [mark=diamond*,thick,green] coordinates {(B-m, 0.570551206) (T-m,0.559522266) (XE,0.588638152) (BPR,0.430005984) (TOP1,0.280509966)};
  \addplot [mark=*,thick,red] coordinates {(B-m, 0.588351362) (T-m,0.579512126) (XE,0.600248594) (BPR,0.45169614) (TOP1,0.31166671)};
  \addplot [mark=square*,thick,blue] coordinates {(B-m, 0.60248081) (T-m,0.595313382) (XE,0.610195272) (BPR,0.467777944) (TOP1,0.338688422)};
\legend{@5, @10, @20}
\end{axis}
\end{tikzpicture}
\subcaption{NDCG}
\end{subfigure}

    \caption{Model performance for different loss functions on RSC19 (B-m: BPR-max; T-m: Top1-max; XE: cross-entropy).}
    \label{fig:loss_function19}
\end{figure*}
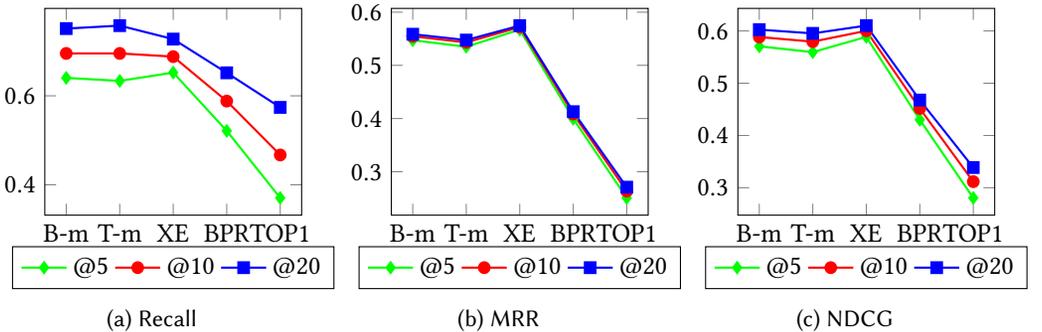

\par{\textit{Sampling method effects.} Figures \ref{fig:sample_alpha15} and \ref{fig:sample_alpha19} depict the model performance with different $\alpha$ for \textbf{additional sampling strategy} on \textit{RSC15} and \textit{RSC19} respectively, where the results on different datasets are varied. For \textit{RSC15}, the performance of cross-entropy, BPR-max and TOP1-max (on all metrics) are consistent. They firstly slowly increases as $\alpha$ increases from $0$ to $0.25$, And then decreases  as $\alpha$ is larger than $0.25$. Besides, the optimal $alpha$ on \textit{RSC15} corresponding to different loss functions and metrics are the same.
On the contrary, on \textit{RSC19}, the optimal $\alpha$ is varied for different loss functions and different evaluation metrics. For example, the optimal $\alpha$ for BPR-max loss function is $0.5$ in terms of Recall@20, but for cross-entropy that is $0$. In terms of MRR@20 and NDCG@20, the optimal $\alpha$ for BPR-max loss function is $0.75$.
Therefore, it is necessary to carry out sufficient search in validation set to figure out the optimal combination of sampling strategy and loss function with regard to the most valuable evaluation measurements in real world applications.
}
\par{\textit{The size of negative sampling effects.} As described in Figure \ref{fig:sample_size15} the larger the \textbf{size of negative sample} is, the better performance the basic model can obtain regarding all evaluation measurements. In particular, the model performance improves dramatically when the size increases from $0$ to $32$, while the increasing speed drops with the further increase of the size. Empirical results on \textit{RSC19} (Figure \ref{fig:sample_size19}) is almost similar, but there is a slight decrease when sample size varies from $32$ to $128$ in terms of MRR and NDCG.
It should be noted that additional negative sampling leads to higher computational costs. Therefore, in real world applications, we need to keep a balance between model performance and training time considering the size of negative samples.
}
\par{\textit{Loss function effects.} As depicted in Figures \ref{fig:loss_function15} and \ref{fig:loss_function19}, models with \textbf{loss functions} BPR-max, TOP1-max, and cross-entropy perform better than those with BPR and TOP1 in terms of all metrics, but the best loss function among the three is also dependent on the datasets. Overall, it is suggested to deploy these three loss functions in real-world applications.}

\subsubsection{Concluding Remarks}
The experimental results verify that the summarized influential factors in Section \ref{sec:influential factors of models} all play an important role in DL-based sequential recommendation.
Our suggestions for best in practice are summarized as follows: 1) try all possible side information (such as texts and images) when allowed, and carefully design the corresponding modules; 2)
well consider the connections between other behavior types with the target behavior, and be careful about the possible noisy information involved in final recommendation when model these connections; 3) always incorporate TOP1-max, BPR-max and cross-entropy loss functions for training, keep a balance between model performance and computational cost with regard to the size of negative samples, and
carefully choose/design data augmentation strategies to further boost the corresponding recommendation performance, especially when the available training set is relatively small; and 4) for any DL-based models, consider to further improve their performance with attention mechanism, by possibly combing with the traditional sequential learning models, and a well-designed explicit user representation module accommodating to the corresponding data scenarios.

\section{Future Directions and Conclusions}
\label{sec:future directions}

\subsection{Future Directions}
As we have discussed, DL techniques have greatly promoted the sequential recommendation studies, but also are accompanied by some challenging issues. Thus, we summarize the following open issues which can be considered as future directions for DL-based sequential recommender systems.

\subsubsection{Rigorous and Comprehensive Evaluations across Different Models}
Our empirical study can be viewed as a horizontal investigation across \emph{GRU4Rec} and its variants. In the literature, lots of \emph{GRU4Rec} variants consider \emph{GRU4Rec} \cite{hidasi2015session} for baseline, but there are few comparisons among these variants. In this case, it is difficult to judge which one is better in a specific application scenario. Besides, other competitive baselines could also be considered, e.g., \emph{NextItNet} \cite{yuan2019simple} (CNN-based model), \emph{NARM} \cite{li2017neural} (attention-based model) or further refer to the comparison framework proposed in \cite{ludewig2018evaluation}.
There is a growing consensus that only complex deep learning structures could not always guarantee better and more robust recommender systems \cite{Dacrema2019recsys,sun2020benchmarking}. Besides, one critical issue has attracted increasing attention in the field of recommender systems: there are few effective benchmarks for evaluation, especially in the sequential recommendation. Thus, benchmarking study for rigorous and comprehensive evaluations should be urgently put on the research agenda.

\subsubsection{Explainable Sequential Recommender Systems}
As has pointed out, most DL-based models are lack of explanability and considered as black-box for platform practitioners and users. As such, designing explainable sequential recommender systems is of great importance.
On the one hand, without understanding the reasons behind predictions, users will be reluctant to trust an individual prediction sufficiently and thus take actions based on it. On the other hand, model practitioners strive to fully understand the model (i.e., how the different factors like data, features and model hyper-parameters impact the model outputs?) and thus enhance their control towards the whole recommender system. Noted that our survey sheds light on the second issue, and more ideas can be borrowed from those explanability studies towards deep learning networks \cite{bach2015pixel,koh2017understanding,montavon2018methods,schnake2020xai}

For the first issue, there are mainly two categories of methods for explainable recommendation which tries to make users understand why such items or lists are recommended:
model-based and post-hoc ones \cite{zhang2018explainable}. Model-based methods aim to design interpretable algorithms to simultaneously provide accurate and explainable recommendation items for users (mostly using a multi-task learning framework). Besides user-item interactions, they usually incorporate side information (e.g., textual and visual item descriptions, textual reviews, social information) to facilitate the explainable task. For example, Huang et al. \cite{huang2018improving} leveraged knowledge graph for better explanability in the sequential recommendation.
In contrast, without directly linking explanations with side information
for recommendations, the post-hoc methods \cite{peake2018explanation} try to devise separate methods to explain the recommendation results produced by those black-box based recommendation algorithms.

\subsubsection{Better Designs on Different Components to Facilitate the Recommendation}
On the basis of the empirical results, we consider that more efforts can be devoted to the following components, which can be incorporated into every sequential recommendation model to improve the recommendation performance correspondingly.

\textbf{More designs on embedding methods.}
Most previous studies adopt the embedding methods from NLP. However, in the sequential recommendation, it is rather challenging to pre-train an embedding model (e.g., word2vec) as the item
 information and
 the dependencies relationship among items is constantly changing while the words and their connections in NLP are relatively fixed and static. On the other hand, behaviors in the sequential recommendation are also very complex than words as they involve both behavior objects and types. Furthermore, the incorporation of embedding vectors in existing sequential recommendation models are also in a relatively simple way.
In this case, more advanced and particular designs of embedding methods are needed for the sequential recommendation \cite{wan2018representating}. For example, \cite{liu2018sequential} designed a sequential embedding approach which also takes the dependencies relationships among items and their attributes into consideration for the sequential recommendation. A possible solution is to take advantage of metric learning \cite{tay2018latent} for obtaining better item, user or sequence representations by understanding sequential data. Metric learning focuses on distance metrics that capture the important relationships among data \cite{lee2018collaborative}, and has been verified to be effective in traditional static recommendation tasks \cite{yang2017yum,hsieh2017collaborative}.

\textbf{Better modeling user long-term preference.}
On the basis of our study and empirical investigation, the module in DL-based models for user representation (especially the long-term preference) is still far from satisfactory, compared to the designed modules for item representation.
In this case, further research can consider to design more favorable modules for user representation, as well as think about how to better combine a user's long-term preference with short-term preference.

\textbf{Advanced sampling strategies.}
In the sequential recommendation, most existing studies use the sampling strategies of uniform, popularity-based, or their straightforward combination (i.e., additional sampling), which are comparatively simple contrasting with the ones used in NLP.
In this view, future research could consider to borrow or extend more advanced sampling strategies from other areas (e.g., NLP or graph representation \cite{yang2020understanding,wang2020reinforced}).

\subsubsection{Personalized Recommendation Based on Polymorphic Behavior Trajectory}
We summarize behavior sequences into three types, and to the best of our knowledge, there is relatively few studies that well distinguish the behavior types and model their connections in the sequential recommendation for interaction-based sequential recommendation tasks. Our empirical evaluation also indicates that well considering another behavior type for a target type is very challenging.
In this case, more DL-models can be designed by considering the connections between polymorphic behavior types and thus for better recommendation performance in the sequential recommendation.
For example,
Qiu et al. \cite{qiu2018bprh} proposed a Bayesian personalized ranking model for heterogeneous behavior types (BPRH) that incorporates the target behavior, auxiliary behavior, and negative behavior into a unified model, and the idea might also be applicable for the  sequential recommendation.
Besides, more advanced deep learning models, such as GNNs for heterogeneous networks \cite{li2020multi} can be considered to fulfill the goal. For example, Song et al. \cite{song2019session} proposed a dynamic-graph-attention neural network to capture dynamic user preferences in sequences and social influence (social network) for better recommendation.

\subsubsection{Learning Behavior Sequences in Real Time.}
Every behavior of the user might reflect a possible interest transfer, in this case, recommender systems are expected to ideally capture this kind of information and timely justify the recommendation strategies. Reinforcement learning is a promising choice for addressing this issue. For example,
Zhao et al. \cite{zhao2019leveraging} combined MF, RNN, and GAN in film recommendations to dynamically provide movie recommendations.
Shih et al. \cite{shih2018automatic} treated the generation of music playlists as a language modeling problem and used an attention-based language model with the policy gradient in reinforcement learning.

Besides, another important issue for DL-based models is their scalability, where models should be capable of dealing with increasing amount of data. For example, in order to speed up retraining procedure, \cite{zhang2020retrain} designed a new training method (a.k.a. sequential meta-learning method) which aims to abandon the historical data through learning to transfer the past training experience. Furthermore, being validated to be effective and efficient in image recognition \cite{hinton2015distilling,yang2020distilling}, \emph{knowledge distillation} techniques have also started to be introduced in recommender systems \cite{wang2019binarized}. For example, Tang and Wang \cite{wang2019binarized} proposed ranking distillation (RD) method by incorporating knowledge distillation technique for learning to rank problem. Experimental results demonstrated that the student model using less than half of the model parameters could achieve similar or better ranking performance than the teacher model. Chen et al. \cite{chen2018adversarial} formulated the problem of recommendation with external knowledge (e.g., online reviews) into a \emph{generalized distillation framework} which simultaneously balances both the effectiveness and efficiency of recommendation model .

\subsubsection{Sequential Recommendation for Specific Domains and Cross-Domains.}
There are little research to specifically identify the suitable recommendation algorithms for different application areas, whereas most research assumes that their models are applicable to the sequential recommendation tasks in all areas. Future research can be conducted to design specific models for particular areas by capturing the characteristics of these areas, which is more valuable for real-world applications.

On the other side, more and more large companies incline to provide services/products in different domains. For example, ByteDance (\url{bytedance.com}) simultaneously offers news service and video service to users. Compared to single-domain recommendation, cross-domain recommendation addresses the problem of leveraging data in different domains to generate desirable recommendation \cite{cantador2015cross}, where the main underlying idea is to transfer knowledge from source domains to a target domain, thus further boost the recommendation performance in the target domain.
Following this idea, as deep learning is a good fit to transfer knowledge across different domains \cite{zhang2019deep}, future research can consider to well investigate the characteristics of sequential data in different domains, and design more advanced cross-domain sequential recommendation models. For example, Zhuang et al. \cite{zhuang2017sequential} exploited the sequential behavioral data of one user from different domains to mine her/his novelty-seeking traits for improving recommendation performance. Ma et al. \cite{ma2019pi} further formulated the cross-domain sequential recommendation problem as a parallel sequential recommendation problem. By considering that user behaviors on two domains are synchronously shared at each timestamp, they proposed $\pi$-Net (i.e., consisting of a shared account filter unit and a cross-domain transfer unit) to simultaneously generate recommendation for two domains.

\subsubsection{Towards Robust Sequential Recommendation Models}
It is a common sense that a single model cannot guarantee to perform consistently well in different application scenarios, as the data distributions vary. In other words, a model might be relatively fragile and vulnerable to adversarial perturbations on data samples (e.g., noisy data or purposeful user profile attacks \cite{mobasher2007toward}). To tackle the challenge, some models introduce the denoising techniques \cite{amatriain2009rate,wang2020denoising}, e.g., denoising AE-based models \cite{wu2016collaborative} by firstly corrupting the data using man-made noises.
Besides, adversarial training \cite{he2018adversarial,tang2019adversarial,yuan2019adversarial} has also been adopted to improve the robustness of DL-based recommendation algorithms. For example, Yuan et al. \cite{yuan2019adversarial} proposed a general adversarial training framework for neural network-based recommendation model and further designed a minmax game for nonlinear model optimization. They tested the effectiveness of the framework on the modified collaborative denoising AE (CDAE) model \cite{wu2016collaborative}.
Furthermore, considering that previous adversarial training techniques might ignore to consider the characteristic of sequential data, Jia et al. \cite{jia2019towards} designed a new adversarial training approach for sequential data by specifically answering when and how to perturb a sequence.

\subsubsection{Tackling Cold-start and Data Sparsity Challenges for Sequential Recommendation}
Cold-start and data sparsity are long-standing challenges for recommender systems, including sequential recommendation \cite{du2019sequential,sun2019research}.
Most of the existing sequential recommendation algorithms we have previously discussed
ignore to address these two issues, as their effective implementations rely on relatively strict requirements on sequential data (e.g., dropping sequences shorter than a minimal threshold). Therefore, future research can consider these two issues when building practical sequential recommender systems for real-world applications.

To tackle these two issues, first, machine learning techniques which are capable of learning from a limited number of data samples (e.g., few shot learning \cite{wang2019few}), can be adopted for sequential recommendation. Few-shot learning strives to bridge
the gap between artificial intelligence and human-like learning, and can learn a task with limited information by incorporating prior knowledge and deploying different
ML techniques (e.g., the meta-learning, embedding learning and generative modeling methods) \cite{wang2019few}.
For instance, Du et al. \cite{du2019sequential} proposed $s^2Meta$ (Scenario-specific Sequential Meta learner) which combines the scenario-specific learning with a model-agnostic sequential meta learning for more effective recommendation.
Secondly, sequential recommendation algorithms can be built on different types of side information (e.g., social networks, user profiles, item descriptions, and knowledge graph) to partially relieve the two issues \cite{sun2019research,meng2020incorporating}, while deep neural networks are quite suitable to process multi-modal information. For example, \emph{SDM} considers multiple types of side information, such as item ID, first level category, leaf category, brand and shop, to better model users' long-term preferences.

\subsection{Conclusions}
The study systematically investigated the DL-based sequential recommendation.
Specifically, we designed a novel taxonomy for investigating the sequential recommendation tasks in terms of the three types of behavior sequences: experienced-based, transaction-based, and interaction-based.
Based on it, we surveyed and explored a considerable amount of representative DL-based algorithms in the sequential recommendation, with the aim of a better understanding on whether sequential recommendation tasks have been sufficiently or insufficiently studied.
Thirdly, for better guiding the development of DL-based sequential recommender systems, we thoroughly identified the possible influential factors that impact the performance of DL-based models in regards to recommendation accuracy from the four perspectives with respect to learning a better model: model input, data processing, model structure and model training. We further comprehensively showcased their impacts via well-designed evaluations, which can be viewed a testbed.
Finally, we discussed the challenges and provided new potential directions for the research on DL-based sequential recommendation.

\bibliographystyle{ACM-Reference-Format}
\bibliography{sequentialSurvey}


\begin{thebibliography}{155}


\ifx \showCODEN    \undefined \def \showCODEN     #1{\unskip}     \fi
\ifx \showDOI      \undefined \def \showDOI       #1{#1}\fi
\ifx \showISBNx    \undefined \def \showISBNx     #1{\unskip}     \fi
\ifx \showISBNxiii \undefined \def \showISBNxiii  #1{\unskip}     \fi
\ifx \showISSN     \undefined \def \showISSN      #1{\unskip}     \fi
\ifx \showLCCN     \undefined \def \showLCCN      #1{\unskip}     \fi
\ifx \shownote     \undefined \def \shownote      #1{#1}          \fi
\ifx \showarticletitle \undefined \def \showarticletitle #1{#1}   \fi
\ifx \showURL      \undefined \def \showURL       {\relax}        \fi
\providecommand\bibfield[2]{#2}
\providecommand\bibinfo[2]{#2}
\providecommand\natexlab[1]{#1}
\providecommand\showeprint[2][]{arXiv:#2}

\bibitem[\protect\citeauthoryear{Amatriain, Pujol, Tintarev, and
  Oliver}{Amatriain et~al\mbox{.}}{2009}]%
        {amatriain2009rate}
\bibfield{author}{\bibinfo{person}{Xavier Amatriain}, \bibinfo{person}{Josep~M
  Pujol}, \bibinfo{person}{Nava Tintarev}, {and} \bibinfo{person}{Nuria
  Oliver}.} \bibinfo{year}{2009}\natexlab{}.
\newblock \showarticletitle{Rate it again: increasing recommendation accuracy
  by user re-rating}. In \bibinfo{booktitle}{\emph{RecSys}}.
  \bibinfo{pages}{173--180}.
\newblock


\bibitem[\protect\citeauthoryear{Anderson, Kumar, Tomkins, and
  Vassilvitskii}{Anderson et~al\mbox{.}}{2014}]%
        {anderson2014dynamics}
\bibfield{author}{\bibinfo{person}{Ashton Anderson}, \bibinfo{person}{Ravi
  Kumar}, \bibinfo{person}{Andrew Tomkins}, {and} \bibinfo{person}{Sergei
  Vassilvitskii}.} \bibinfo{year}{2014}\natexlab{}.
\newblock \showarticletitle{The dynamics of repeat consumption}. In
  \bibinfo{booktitle}{\emph{WWW}}. \bibinfo{pages}{419--430}.
\newblock


\bibitem[\protect\citeauthoryear{Bach, Binder, Montavon, Klauschen, M{\"u}ller,
  and Samek}{Bach et~al\mbox{.}}{2015}]%
        {bach2015pixel}
\bibfield{author}{\bibinfo{person}{Sebastian Bach}, \bibinfo{person}{Alexander
  Binder}, \bibinfo{person}{Gr{\'e}goire Montavon}, \bibinfo{person}{Frederick
  Klauschen}, \bibinfo{person}{Klaus-Robert M{\"u}ller}, {and}
  \bibinfo{person}{Wojciech Samek}.} \bibinfo{year}{2015}\natexlab{}.
\newblock \showarticletitle{On pixel-wise explanations for non-linear
  classifier decisions by layer-wise relevance propagation}.
\newblock \bibinfo{journal}{\emph{PloS One}} \bibinfo{volume}{10},
  \bibinfo{number}{7} (\bibinfo{year}{2015}).
\newblock


\bibitem[\protect\citeauthoryear{Bahdanau, Cho, and Bengio}{Bahdanau
  et~al\mbox{.}}{2015}]%
        {bahdanau2014neural}
\bibfield{author}{\bibinfo{person}{Dzmitry Bahdanau},
  \bibinfo{person}{Kyunghyun Cho}, {and} \bibinfo{person}{Yoshua Bengio}.}
  \bibinfo{year}{2015}\natexlab{}.
\newblock \showarticletitle{Neural machine translation by Jointly Learning to
  Align and Translate}. In \bibinfo{booktitle}{\emph{ICLR}}.
\newblock


\bibitem[\protect\citeauthoryear{Bai, Nie, Zhao, Zhu, Du, and Wen}{Bai
  et~al\mbox{.}}{2018}]%
        {bai2018an}
\bibfield{author}{\bibinfo{person}{Ting Bai}, \bibinfo{person}{Jian-Yun Nie},
  \bibinfo{person}{Wayne~Xin Zhao}, \bibinfo{person}{Yutao Zhu},
  \bibinfo{person}{Pan Du}, {and} \bibinfo{person}{Ji-Rong Wen}.}
  \bibinfo{year}{2018}\natexlab{}.
\newblock \showarticletitle{An attribute-aware neural attentive model for next
  basket recommendation}. In \bibinfo{booktitle}{\emph{SIGIR}}.
  \bibinfo{pages}{1201--1204}.
\newblock


\bibitem[\protect\citeauthoryear{Bansal, Belanger, and McCallum}{Bansal
  et~al\mbox{.}}{2016}]%
        {bansal2016ask}
\bibfield{author}{\bibinfo{person}{Trapit Bansal}, \bibinfo{person}{David
  Belanger}, {and} \bibinfo{person}{Andrew McCallum}.}
  \bibinfo{year}{2016}\natexlab{}.
\newblock \showarticletitle{Ask the {GRU}: Multi-task learning for deep text
  recommendations}. In \bibinfo{booktitle}{\emph{RecSys}}.
  \bibinfo{pages}{107--114}.
\newblock


\bibitem[\protect\citeauthoryear{Barkan and Koenigstein}{Barkan and
  Koenigstein}{2016}]%
        {barkan2016item2vec}
\bibfield{author}{\bibinfo{person}{Oren Barkan} {and} \bibinfo{person}{Noam
  Koenigstein}.} \bibinfo{year}{2016}\natexlab{}.
\newblock \showarticletitle{Item2vec: neural item embedding for collaborative
  filtering}. In \bibinfo{booktitle}{\emph{26th International Workshop on
  Machine Learning for Signal Processing (MLSP)}}. \bibinfo{pages}{1--6}.
\newblock


\bibitem[\protect\citeauthoryear{Batmaz, Yurekli, Bilge, and Kaleli}{Batmaz
  et~al\mbox{.}}{2018}]%
        {Batmaz2018}
\bibfield{author}{\bibinfo{person}{Zeynep Batmaz}, \bibinfo{person}{Ali~Ihsan
  Yurekli}, \bibinfo{person}{Alper Bilge}, {and} \bibinfo{person}{Cihan
  Kaleli}.} \bibinfo{year}{2018}\natexlab{}.
\newblock \showarticletitle{A review on deep learning for recommender systems:
  challenges and remedies}.
\newblock \bibinfo{journal}{\emph{Artificial Intelligence Review}}
  (\bibinfo{year}{2018}), \bibinfo{pages}{1--37}.
\newblock


\bibitem[\protect\citeauthoryear{Bengio, Ducharme, Vincent, and Janvin}{Bengio
  et~al\mbox{.}}{2003}]%
        {bengio2003a}
\bibfield{author}{\bibinfo{person}{Yoshua Bengio}, \bibinfo{person}{Rejean
  Ducharme}, \bibinfo{person}{Pascal Vincent}, {and} \bibinfo{person}{Christian
  Janvin}.} \bibinfo{year}{2003}\natexlab{}.
\newblock \showarticletitle{A neural probabilistic language model}.
\newblock \bibinfo{journal}{\emph{Journal of Machine Learning Research}}
  \bibinfo{volume}{3}, \bibinfo{number}{6} (\bibinfo{year}{2003}),
  \bibinfo{pages}{1137--1155}.
\newblock


\bibitem[\protect\citeauthoryear{Bhagat, Muralidharan, Lobzhanidze, and
  Vishwanath}{Bhagat et~al\mbox{.}}{2018}]%
        {bhagat2018buy}
\bibfield{author}{\bibinfo{person}{Rahul Bhagat}, \bibinfo{person}{Srevatsan
  Muralidharan}, \bibinfo{person}{Alex Lobzhanidze}, {and}
  \bibinfo{person}{Shankar Vishwanath}.} \bibinfo{year}{2018}\natexlab{}.
\newblock \showarticletitle{Buy It Again: Modeling Repeat Purchase
  Recommendations}. In \bibinfo{booktitle}{\emph{KDD}}.
  \bibinfo{pages}{62--70}.
\newblock


\bibitem[\protect\citeauthoryear{Bharadhwaj and Joshi}{Bharadhwaj and
  Joshi}{2018}]%
        {bharadhwaj2018explanations}
\bibfield{author}{\bibinfo{person}{Homanga Bharadhwaj} {and}
  \bibinfo{person}{Shruti Joshi}.} \bibinfo{year}{2018}\natexlab{}.
\newblock \showarticletitle{Explanations for temporal recommendations}.
\newblock \bibinfo{journal}{\emph{Künstliche Intelligenz}}
  \bibinfo{volume}{32}, \bibinfo{number}{4} (\bibinfo{year}{2018}),
  \bibinfo{pages}{267--272}.
\newblock


\bibitem[\protect\citeauthoryear{Bogina and Kuflik}{Bogina and Kuflik}{2017}]%
        {bogina2017incorporating}
\bibfield{author}{\bibinfo{person}{Veronika Bogina} {and} \bibinfo{person}{Tsvi
  Kuflik}.} \bibinfo{year}{2017}\natexlab{}.
\newblock \showarticletitle{Incorporating dwell time in session-based
  recommendations with recurrent Neural networks}. In
  \bibinfo{booktitle}{\emph{RecTemp@ RecSys}}. \bibinfo{pages}{57--59}.
\newblock


\bibitem[\protect\citeauthoryear{Cantador, Fern{\'a}ndez-Tob{\'\i}as,
  Berkovsky, and Cremonesi}{Cantador et~al\mbox{.}}{2015}]%
        {cantador2015cross}
\bibfield{author}{\bibinfo{person}{Iv{\'a}n Cantador}, \bibinfo{person}{Ignacio
  Fern{\'a}ndez-Tob{\'\i}as}, \bibinfo{person}{Shlomo Berkovsky}, {and}
  \bibinfo{person}{Paolo Cremonesi}.} \bibinfo{year}{2015}\natexlab{}.
\newblock \showarticletitle{Cross-domain recommender systems}.
\newblock In \bibinfo{booktitle}{\emph{Recommender systems handbook}}.
  \bibinfo{publisher}{Springer}, \bibinfo{pages}{919--959}.
\newblock


\bibitem[\protect\citeauthoryear{Chen, Ren, Cai, and de~Rijke}{Chen
  et~al\mbox{.}}{2019}]%
        {DBLP:journals/corr/abs-1908-10171}
\bibfield{author}{\bibinfo{person}{Wanyu Chen}, \bibinfo{person}{Pengjie Ren},
  \bibinfo{person}{Fei Cai}, {and} \bibinfo{person}{Maarten de Rijke}.}
  \bibinfo{year}{2019}\natexlab{}.
\newblock \showarticletitle{Improving End-to-End Sequential Recommendations
  with Intent-aware Diversification}.
\newblock \bibinfo{journal}{\emph{arXiv preprint arXiv:1908.10171}}
  (\bibinfo{year}{2019}).
\newblock


\bibitem[\protect\citeauthoryear{Chen, Xu, Zhang, Tang, Cao, Qin, and Zha}{Chen
  et~al\mbox{.}}{2018a}]%
        {chen2018sequential}
\bibfield{author}{\bibinfo{person}{Xu Chen}, \bibinfo{person}{Hongteng Xu},
  \bibinfo{person}{Yongfeng Zhang}, \bibinfo{person}{Jiaxi Tang},
  \bibinfo{person}{Yixin Cao}, \bibinfo{person}{Zheng Qin}, {and}
  \bibinfo{person}{Hongyuan Zha}.} \bibinfo{year}{2018}\natexlab{a}.
\newblock \showarticletitle{Sequential recommendation with user memory
  networks}. In \bibinfo{booktitle}{\emph{WSDM}}. \bibinfo{pages}{108--116}.
\newblock


\bibitem[\protect\citeauthoryear{Chen, Zhang, Xu, Qin, and Zha}{Chen
  et~al\mbox{.}}{2018b}]%
        {chen2018adversarial}
\bibfield{author}{\bibinfo{person}{Xu Chen}, \bibinfo{person}{Yongfeng Zhang},
  \bibinfo{person}{Hongteng Xu}, \bibinfo{person}{Zheng Qin}, {and}
  \bibinfo{person}{Hongyuan Zha}.} \bibinfo{year}{2018}\natexlab{b}.
\newblock \showarticletitle{Adversarial distillation for efficient
  recommendation with external knowledge}.
\newblock \bibinfo{journal}{\emph{TOIS}} \bibinfo{volume}{37},
  \bibinfo{number}{1} (\bibinfo{year}{2018}), \bibinfo{pages}{1--28}.
\newblock


\bibitem[\protect\citeauthoryear{Cho, Van~Merri{\"e}nboer, Bahdanau, and
  Bengio}{Cho et~al\mbox{.}}{2014}]%
        {cho2014properties}
\bibfield{author}{\bibinfo{person}{Kyunghyun Cho}, \bibinfo{person}{Bart
  Van~Merri{\"e}nboer}, \bibinfo{person}{Dzmitry Bahdanau}, {and}
  \bibinfo{person}{Yoshua Bengio}.} \bibinfo{year}{2014}\natexlab{}.
\newblock \showarticletitle{On the properties of neural machine translation:
  Encoder-decoder approaches}.
\newblock \bibinfo{journal}{\emph{arXiv preprint arXiv:1409.1259}}
  (\bibinfo{year}{2014}).
\newblock


\bibitem[\protect\citeauthoryear{Covington, Adams, and Sargin}{Covington
  et~al\mbox{.}}{2016}]%
        {covington2016deep}
\bibfield{author}{\bibinfo{person}{Paul Covington}, \bibinfo{person}{Jay
  Adams}, {and} \bibinfo{person}{Emre Sargin}.}
  \bibinfo{year}{2016}\natexlab{}.
\newblock \showarticletitle{Deep neural networks for youtube recommendations}.
  In \bibinfo{booktitle}{\emph{RecSys}}. \bibinfo{pages}{191--198}.
\newblock


\bibitem[\protect\citeauthoryear{Dacrema, Cremonesi, and Jannach}{Dacrema
  et~al\mbox{.}}{2019}]%
        {Dacrema2019recsys}
\bibfield{author}{\bibinfo{person}{Maurizio~Ferrari Dacrema},
  \bibinfo{person}{Paolo Cremonesi}, {and} \bibinfo{person}{Dietmar Jannach}.}
  \bibinfo{year}{2019}\natexlab{}.
\newblock \showarticletitle{Are We Really Making Much Progress? A Worrying
  Analysis of Recent Neural Recommendation Approaches}. In
  \bibinfo{booktitle}{\emph{RecSys}} \emph{(\bibinfo{series}{RecSys '19})}.
  \bibinfo{publisher}{ACM}, \bibinfo{address}{New York, NY, USA},
  \bibinfo{pages}{101--109}.
\newblock
\showISBNx{978-1-4503-6243-6}
\urldef\tempurl%
\url{https://doi.org/10.1145/3298689.3347058}
\showDOI{\tempurl}


\bibitem[\protect\citeauthoryear{Dallmann, Grimm, P{\"o}litz, Zoller, and
  Hotho}{Dallmann et~al\mbox{.}}{2017}]%
        {dallmann2017improving}
\bibfield{author}{\bibinfo{person}{Alexander Dallmann},
  \bibinfo{person}{Alexander Grimm}, \bibinfo{person}{Christian P{\"o}litz},
  \bibinfo{person}{Daniel Zoller}, {and} \bibinfo{person}{Andreas Hotho}.}
  \bibinfo{year}{2017}\natexlab{}.
\newblock \showarticletitle{Improving session recommendation with recurrent
  neural networks by exploiting dwell time}.
\newblock \bibinfo{journal}{\emph{arXiv preprint arXiv:1706.10231}}
  (\bibinfo{year}{2017}).
\newblock


\bibitem[\protect\citeauthoryear{Davidson, Liebald, Liu, Nandy, Vleet, Gargi,
  Gupta, He, Lambert, and Livingston}{Davidson et~al\mbox{.}}{2010}]%
        {Davidson2010The}
\bibfield{author}{\bibinfo{person}{James Davidson}, \bibinfo{person}{Benjamin
  Liebald}, \bibinfo{person}{Junning Liu}, \bibinfo{person}{Palash Nandy},
  \bibinfo{person}{Taylor~Van Vleet}, \bibinfo{person}{Ullas Gargi},
  \bibinfo{person}{Sujoy Gupta}, \bibinfo{person}{Yu He}, \bibinfo{person}{Mike
  Lambert}, {and} \bibinfo{person}{Blake Livingston}.}
  \bibinfo{year}{2010}\natexlab{}.
\newblock \showarticletitle{The YouTube video recommendation system}. In
  \bibinfo{booktitle}{\emph{RecSys}}. \bibinfo{pages}{293--296}.
\newblock


\bibitem[\protect\citeauthoryear{Devooght and Bersini}{Devooght and
  Bersini}{2017}]%
        {devooght2017long}
\bibfield{author}{\bibinfo{person}{Robin Devooght} {and}
  \bibinfo{person}{Hugues Bersini}.} \bibinfo{year}{2017}\natexlab{}.
\newblock \showarticletitle{Long and short-term recommendations with recurrent
  neural networks}. In \bibinfo{booktitle}{\emph{UMAP}}.
  \bibinfo{pages}{13--21}.
\newblock


\bibitem[\protect\citeauthoryear{Ding, Yu, He, Quan, Li, Chua, Jin, and
  Yu}{Ding et~al\mbox{.}}{2018}]%
        {ding2018improving}
\bibfield{author}{\bibinfo{person}{Jingtao Ding}, \bibinfo{person}{Guanghui
  Yu}, \bibinfo{person}{Xiangnan He}, \bibinfo{person}{Yuhan Quan},
  \bibinfo{person}{Yong Li}, \bibinfo{person}{Tat-Seng Chua},
  \bibinfo{person}{Depeng Jin}, {and} \bibinfo{person}{Jiajie Yu}.}
  \bibinfo{year}{2018}\natexlab{}.
\newblock \showarticletitle{Improving implicit recommender systems with view
  data.}. In \bibinfo{booktitle}{\emph{IJCAI}}. \bibinfo{pages}{3343--3349}.
\newblock


\bibitem[\protect\citeauthoryear{Donkers, Loepp, and Ziegler}{Donkers
  et~al\mbox{.}}{2017}]%
        {donkers2017sequential}
\bibfield{author}{\bibinfo{person}{Tim Donkers}, \bibinfo{person}{Benedikt
  Loepp}, {and} \bibinfo{person}{J{\"u}rgen Ziegler}.}
  \bibinfo{year}{2017}\natexlab{}.
\newblock \showarticletitle{Sequential user-based recurrent neural network
  recommendations}. In \bibinfo{booktitle}{\emph{RecSys}}.
  \bibinfo{pages}{152--160}.
\newblock


\bibitem[\protect\citeauthoryear{Du, Wang, Yang, Zhou, and Tang}{Du
  et~al\mbox{.}}{2019}]%
        {du2019sequential}
\bibfield{author}{\bibinfo{person}{Zhengxiao Du}, \bibinfo{person}{Xiaowei
  Wang}, \bibinfo{person}{Hongxia Yang}, \bibinfo{person}{Jingren Zhou}, {and}
  \bibinfo{person}{Jie Tang}.} \bibinfo{year}{2019}\natexlab{}.
\newblock \showarticletitle{Sequential Scenario-Specific Meta Learner for
  Online Recommendation}. In \bibinfo{booktitle}{\emph{KDD}}.
  \bibinfo{pages}{2895--2904}.
\newblock


\bibitem[\protect\citeauthoryear{Gao, He, Gan, Chen, Feng, Li, Chua, Yao, Song,
  and Jin}{Gao et~al\mbox{.}}{2019}]%
        {gao2019learning}
\bibfield{author}{\bibinfo{person}{Chen Gao}, \bibinfo{person}{Xiangnan He},
  \bibinfo{person}{Danhua Gan}, \bibinfo{person}{Xiangning Chen},
  \bibinfo{person}{Fuli Feng}, \bibinfo{person}{Yong Li},
  \bibinfo{person}{Tat-Seng Chua}, \bibinfo{person}{Lina Yao},
  \bibinfo{person}{Yang Song}, {and} \bibinfo{person}{Depeng Jin}.}
  \bibinfo{year}{2019}\natexlab{}.
\newblock \showarticletitle{Learning to Recommend with Multiple Cascading
  Behaviors}.
\newblock \bibinfo{journal}{\emph{IEEE Transactions on Knowledge and Data
  Engineering}} (\bibinfo{year}{2019}).
\newblock


\bibitem[\protect\citeauthoryear{Greenstein-Messica, Rokach, and
  Friedman}{Greenstein-Messica et~al\mbox{.}}{2017}]%
        {greenstein2017session}
\bibfield{author}{\bibinfo{person}{Asnat Greenstein-Messica},
  \bibinfo{person}{Lior Rokach}, {and} \bibinfo{person}{Michael Friedman}.}
  \bibinfo{year}{2017}\natexlab{}.
\newblock \showarticletitle{Session-based recommendations using item
  embedding}. In \bibinfo{booktitle}{\emph{IUI}}. \bibinfo{pages}{629--633}.
\newblock


\bibitem[\protect\citeauthoryear{Guo, Qiu, Tan, Liu, Ma, and Wang}{Guo
  et~al\mbox{.}}{2017}]%
        {guo2017resolving}
\bibfield{author}{\bibinfo{person}{Guibing Guo}, \bibinfo{person}{Huihuai Qiu},
  \bibinfo{person}{Zhenhua Tan}, \bibinfo{person}{Yuan Liu},
  \bibinfo{person}{Jing Ma}, {and} \bibinfo{person}{Xingwei Wang}.}
  \bibinfo{year}{2017}\natexlab{}.
\newblock \showarticletitle{Resolving data sparsity by multi-type auxiliary
  implicit feedback for recommender systems}.
\newblock \bibinfo{journal}{\emph{Knowledge-Based Systems}}
  \bibinfo{volume}{138} (\bibinfo{year}{2017}), \bibinfo{pages}{202--207}.
\newblock


\bibitem[\protect\citeauthoryear{He, Fang, Wang, and McAuley}{He
  et~al\mbox{.}}{2016}]%
        {he2016vista}
\bibfield{author}{\bibinfo{person}{Ruining He}, \bibinfo{person}{Chen Fang},
  \bibinfo{person}{Zhaowen Wang}, {and} \bibinfo{person}{Julian McAuley}.}
  \bibinfo{year}{2016}\natexlab{}.
\newblock \showarticletitle{Vista: A visually, socially, and temporally-aware
  model for artistic recommendation}. In \bibinfo{booktitle}{\emph{RecSys}}.
  \bibinfo{pages}{309--316}.
\newblock


\bibitem[\protect\citeauthoryear{He, Kang, and McAuley}{He
  et~al\mbox{.}}{2017}]%
        {he2017translation}
\bibfield{author}{\bibinfo{person}{Ruining He}, \bibinfo{person}{Wang-Cheng
  Kang}, {and} \bibinfo{person}{Julian McAuley}.}
  \bibinfo{year}{2017}\natexlab{}.
\newblock \showarticletitle{Translation-based recommendation}. In
  \bibinfo{booktitle}{\emph{RecSys}}. \bibinfo{pages}{161--169}.
\newblock


\bibitem[\protect\citeauthoryear{He and McAuley}{He and McAuley}{2016}]%
        {he2016fusing}
\bibfield{author}{\bibinfo{person}{Ruining He} {and} \bibinfo{person}{Julian
  McAuley}.} \bibinfo{year}{2016}\natexlab{}.
\newblock \showarticletitle{Fusing similarity models with markov chains for
  sparse sequential recommendation}. In \bibinfo{booktitle}{\emph{ICDM}}.
  \bibinfo{pages}{191--200}.
\newblock


\bibitem[\protect\citeauthoryear{He, He, Du, and Chua}{He
  et~al\mbox{.}}{2018}]%
        {he2018adversarial}
\bibfield{author}{\bibinfo{person}{Xiangnan He}, \bibinfo{person}{Zhankui He},
  \bibinfo{person}{Xiaoyu Du}, {and} \bibinfo{person}{Tat-Seng Chua}.}
  \bibinfo{year}{2018}\natexlab{}.
\newblock \showarticletitle{Adversarial personalized ranking for
  recommendation}. In \bibinfo{booktitle}{\emph{SIGIR}}.
  \bibinfo{pages}{355--364}.
\newblock


\bibitem[\protect\citeauthoryear{Hidasi and Karatzoglou}{Hidasi and
  Karatzoglou}{2018}]%
        {hidasi2018recurrent}
\bibfield{author}{\bibinfo{person}{Bal{\'a}zs Hidasi} {and}
  \bibinfo{person}{Alexandros Karatzoglou}.} \bibinfo{year}{2018}\natexlab{}.
\newblock \showarticletitle{Recurrent neural networks with top-k gains for
  session-based recommendations}. In \bibinfo{booktitle}{\emph{CIKM}}.
  \bibinfo{pages}{843--852}.
\newblock


\bibitem[\protect\citeauthoryear{Hidasi, Karatzoglou, Baltrunas, and
  Tikk}{Hidasi et~al\mbox{.}}{2016a}]%
        {hidasi2015session}
\bibfield{author}{\bibinfo{person}{Bal{\'a}zs Hidasi},
  \bibinfo{person}{Alexandros Karatzoglou}, \bibinfo{person}{Linas Baltrunas},
  {and} \bibinfo{person}{Domonkos Tikk}.} \bibinfo{year}{2016}\natexlab{a}.
\newblock \showarticletitle{Session-based recommendations with recurrent neural
  networks}. In \bibinfo{booktitle}{\emph{ICLR}}.
\newblock


\bibitem[\protect\citeauthoryear{Hidasi, Quadrana, Karatzoglou, and
  Tikk}{Hidasi et~al\mbox{.}}{2016b}]%
        {hidasi2016parallel}
\bibfield{author}{\bibinfo{person}{Bal{\'a}zs Hidasi}, \bibinfo{person}{Massimo
  Quadrana}, \bibinfo{person}{Alexandros Karatzoglou}, {and}
  \bibinfo{person}{Domonkos Tikk}.} \bibinfo{year}{2016}\natexlab{b}.
\newblock \showarticletitle{Parallel recurrent neural network architectures for
  feature-rich session-based recommendations}. In
  \bibinfo{booktitle}{\emph{RecSys}}. \bibinfo{pages}{241--248}.
\newblock


\bibitem[\protect\citeauthoryear{Hinton, Vinyals, and Dean}{Hinton
  et~al\mbox{.}}{2015}]%
        {hinton2015distilling}
\bibfield{author}{\bibinfo{person}{Geoffrey Hinton}, \bibinfo{person}{Oriol
  Vinyals}, {and} \bibinfo{person}{Jeff Dean}.}
  \bibinfo{year}{2015}\natexlab{}.
\newblock \showarticletitle{Distilling the knowledge in a neural network}.
\newblock \bibinfo{journal}{\emph{arXiv preprint arXiv:1503.02531}}
  (\bibinfo{year}{2015}).
\newblock


\bibitem[\protect\citeauthoryear{Hsieh, Yang, Cui, Lin, Belongie, and
  Estrin}{Hsieh et~al\mbox{.}}{2017}]%
        {hsieh2017collaborative}
\bibfield{author}{\bibinfo{person}{Cheng-Kang Hsieh}, \bibinfo{person}{Longqi
  Yang}, \bibinfo{person}{Yin Cui}, \bibinfo{person}{Tsung-Yi Lin},
  \bibinfo{person}{Serge Belongie}, {and} \bibinfo{person}{Deborah Estrin}.}
  \bibinfo{year}{2017}\natexlab{}.
\newblock \showarticletitle{Collaborative metric learning}. In
  \bibinfo{booktitle}{\emph{WWW}}. \bibinfo{pages}{193--201}.
\newblock


\bibitem[\protect\citeauthoryear{Hsu, Chou, Yang, and Chi}{Hsu
  et~al\mbox{.}}{2016}]%
        {hsu2016neural}
\bibfield{author}{\bibinfo{person}{Kai-Chun Hsu}, \bibinfo{person}{Szu-Yu
  Chou}, \bibinfo{person}{Yi-Hsuan Yang}, {and} \bibinfo{person}{Tai-Shih
  Chi}.} \bibinfo{year}{2016}\natexlab{}.
\newblock \showarticletitle{Neural network based next-song recommendation}.
\newblock \bibinfo{journal}{\emph{arXiv preprint arXiv:1606.07722}}
  (\bibinfo{year}{2016}).
\newblock


\bibitem[\protect\citeauthoryear{Hu, He, Gao, and Zhang}{Hu
  et~al\mbox{.}}{2020}]%
        {hu2020modeling}
\bibfield{author}{\bibinfo{person}{Haoji Hu}, \bibinfo{person}{Xiangnan He},
  \bibinfo{person}{Jinyang Gao}, {and} \bibinfo{person}{Zhi-Li Zhang}.}
  \bibinfo{year}{2020}\natexlab{}.
\newblock \showarticletitle{Modeling Personalized Item Frequency Information
  for Next-basket Recommendation}. In \bibinfo{booktitle}{\emph{SIGIR}}.
\newblock


\bibitem[\protect\citeauthoryear{Huang, Zhao, Dou, Wen, and Chang}{Huang
  et~al\mbox{.}}{2018b}]%
        {huang2018improving}
\bibfield{author}{\bibinfo{person}{Jin Huang}, \bibinfo{person}{Wayne~Xin
  Zhao}, \bibinfo{person}{Hongjian Dou}, \bibinfo{person}{Ji-Rong Wen}, {and}
  \bibinfo{person}{Edward~Y Chang}.} \bibinfo{year}{2018}\natexlab{b}.
\newblock \showarticletitle{Improving sequential recommendation with
  knowledge-enhanced memory networks}. In \bibinfo{booktitle}{\emph{SIGIR}}.
  \bibinfo{pages}{505--514}.
\newblock


\bibitem[\protect\citeauthoryear{Huang, Qian, Fang, Sang, and Xu}{Huang
  et~al\mbox{.}}{2018a}]%
        {huang2018csan}
\bibfield{author}{\bibinfo{person}{Xiaowen Huang}, \bibinfo{person}{Shengsheng
  Qian}, \bibinfo{person}{Quan Fang}, \bibinfo{person}{Jitao Sang}, {and}
  \bibinfo{person}{Changsheng Xu}.} \bibinfo{year}{2018}\natexlab{a}.
\newblock \showarticletitle{CSAN: Contextual self-attention network for user
  sequential recommendation}. In \bibinfo{booktitle}{\emph{MM}}.
  \bibinfo{pages}{447--455}.
\newblock


\bibitem[\protect\citeauthoryear{Jannach and Ludewig}{Jannach and
  Ludewig}{2017}]%
        {jannach2017recurrent}
\bibfield{author}{\bibinfo{person}{Dietmar Jannach} {and}
  \bibinfo{person}{Malte Ludewig}.} \bibinfo{year}{2017}\natexlab{}.
\newblock \showarticletitle{When recurrent neural networks meet the
  neighborhood for session-based recommendation}. In
  \bibinfo{booktitle}{\emph{RecSys}}. \bibinfo{pages}{306--310}.
\newblock


\bibitem[\protect\citeauthoryear{Jia, Li, Zhao, Kim, and Kumar}{Jia
  et~al\mbox{.}}{2019}]%
        {jia2019towards}
\bibfield{author}{\bibinfo{person}{Xiaowei Jia}, \bibinfo{person}{Sheng Li},
  \bibinfo{person}{Handong Zhao}, \bibinfo{person}{Sungchul Kim}, {and}
  \bibinfo{person}{Vipin Kumar}.} \bibinfo{year}{2019}\natexlab{}.
\newblock \showarticletitle{Towards robust and discriminative sequential data
  learning: When and how to perform adversarial training?}. In
  \bibinfo{booktitle}{\emph{KDD}}. \bibinfo{pages}{1665--1673}.
\newblock


\bibitem[\protect\citeauthoryear{Kabbur, Ning, and Karypis}{Kabbur
  et~al\mbox{.}}{2013}]%
        {kabbur2013fism}
\bibfield{author}{\bibinfo{person}{Santosh Kabbur}, \bibinfo{person}{Xia Ning},
  {and} \bibinfo{person}{George Karypis}.} \bibinfo{year}{2013}\natexlab{}.
\newblock \showarticletitle{Fism: factored item similarity models for top-n
  recommender systems}. In \bibinfo{booktitle}{\emph{KDD}}.
  \bibinfo{pages}{659--667}.
\newblock


\bibitem[\protect\citeauthoryear{Kang and McAuley}{Kang and McAuley}{2018}]%
        {kang2018self-attentive}
\bibfield{author}{\bibinfo{person}{Wang-Cheng Kang} {and}
  \bibinfo{person}{Julian McAuley}.} \bibinfo{year}{2018}\natexlab{}.
\newblock \showarticletitle{Self-attentive sequential recommendation}. In
  \bibinfo{booktitle}{\emph{ICDM}}. IEEE, \bibinfo{pages}{197--206}.
\newblock


\bibitem[\protect\citeauthoryear{Koh and Liang}{Koh and Liang}{2017}]%
        {koh2017understanding}
\bibfield{author}{\bibinfo{person}{Pang~Wei Koh} {and} \bibinfo{person}{Percy
  Liang}.} \bibinfo{year}{2017}\natexlab{}.
\newblock \showarticletitle{Understanding black-box predictions via influence
  functions}. In \bibinfo{booktitle}{\emph{ICML}}. JMLR. org,
  \bibinfo{pages}{1885--1894}.
\newblock


\bibitem[\protect\citeauthoryear{Koren}{Koren}{2009}]%
        {koren2009collaborative}
\bibfield{author}{\bibinfo{person}{Yehuda Koren}.}
  \bibinfo{year}{2009}\natexlab{}.
\newblock \showarticletitle{Collaborative filtering with temporal dynamics}. In
  \bibinfo{booktitle}{\emph{KDD}}. ACM, \bibinfo{pages}{447--456}.
\newblock


\bibitem[\protect\citeauthoryear{Krohn-Grimberghe, Drumond, Freudenthaler, and
  Schmidt-Thieme}{Krohn-Grimberghe et~al\mbox{.}}{2012}]%
        {krohn2012multi}
\bibfield{author}{\bibinfo{person}{Artus Krohn-Grimberghe},
  \bibinfo{person}{Lucas Drumond}, \bibinfo{person}{Christoph Freudenthaler},
  {and} \bibinfo{person}{Lars Schmidt-Thieme}.}
  \bibinfo{year}{2012}\natexlab{}.
\newblock \showarticletitle{Multi-relational matrix factorization using
  bayesian personalized ranking for social network data}. In
  \bibinfo{booktitle}{\emph{WSDM}}. \bibinfo{pages}{173--182}.
\newblock


\bibitem[\protect\citeauthoryear{Le, Lauw, and Fang}{Le et~al\mbox{.}}{2018}]%
        {le2018modeling}
\bibfield{author}{\bibinfo{person}{Duc-Trong Le}, \bibinfo{person}{Hady~W
  Lauw}, {and} \bibinfo{person}{Yuan Fang}.} \bibinfo{year}{2018}\natexlab{}.
\newblock \showarticletitle{Modeling contemporaneous basket sequences with twin
  networks for next-item recommendation}. In \bibinfo{booktitle}{\emph{IJCAI}}.
  \bibinfo{pages}{3414--3420}.
\newblock


\bibitem[\protect\citeauthoryear{Lee, Abu-El-Haija, Varadarajan, and
  Natsev}{Lee et~al\mbox{.}}{2018}]%
        {lee2018collaborative}
\bibfield{author}{\bibinfo{person}{Joonseok Lee}, \bibinfo{person}{Sami
  Abu-El-Haija}, \bibinfo{person}{Balakrishnan Varadarajan}, {and}
  \bibinfo{person}{Apostol Natsev}.} \bibinfo{year}{2018}\natexlab{}.
\newblock \showarticletitle{Collaborative deep metric learning for video
  understanding}. In \bibinfo{booktitle}{\emph{KDD}}.
  \bibinfo{pages}{481--490}.
\newblock


\bibitem[\protect\citeauthoryear{Lerche, Jannach, and Ludewig}{Lerche
  et~al\mbox{.}}{2016}]%
        {lerche2016value}
\bibfield{author}{\bibinfo{person}{Lukas Lerche}, \bibinfo{person}{Dietmar
  Jannach}, {and} \bibinfo{person}{Malte Ludewig}.}
  \bibinfo{year}{2016}\natexlab{}.
\newblock \showarticletitle{On the value of reminders within e-commerce
  recommendations}. In \bibinfo{booktitle}{\emph{UMAP}}.
  \bibinfo{pages}{27--35}.
\newblock


\bibitem[\protect\citeauthoryear{Li, Wang, Lyu, and Shi}{Li
  et~al\mbox{.}}{2020b}]%
        {li2020multi}
\bibfield{author}{\bibinfo{person}{Hui Li}, \bibinfo{person}{Yanlin Wang},
  \bibinfo{person}{Ziyu Lyu}, {and} \bibinfo{person}{Jieming Shi}.}
  \bibinfo{year}{2020}\natexlab{b}.
\newblock \showarticletitle{Multi-task Learning for Recommendation over
  Heterogeneous Information Network}.
\newblock \bibinfo{journal}{\emph{TKDE}} (\bibinfo{year}{2020}).
\newblock


\bibitem[\protect\citeauthoryear{Li, Ren, Chen, Ren, Lian, and Ma}{Li
  et~al\mbox{.}}{2017}]%
        {li2017neural}
\bibfield{author}{\bibinfo{person}{Jing Li}, \bibinfo{person}{Pengjie Ren},
  \bibinfo{person}{Zhumin Chen}, \bibinfo{person}{Zhaochun Ren},
  \bibinfo{person}{Tao Lian}, {and} \bibinfo{person}{Jun Ma}.}
  \bibinfo{year}{2017}\natexlab{}.
\newblock \showarticletitle{Neural attentive session-based recommendation}. In
  \bibinfo{booktitle}{\emph{CIKM}}. \bibinfo{pages}{1419--1428}.
\newblock


\bibitem[\protect\citeauthoryear{Li, Wang, and McAuley}{Li
  et~al\mbox{.}}{2020a}]%
        {li2020time}
\bibfield{author}{\bibinfo{person}{Jiacheng Li}, \bibinfo{person}{Yujie Wang},
  {and} \bibinfo{person}{Julian McAuley}.} \bibinfo{year}{2020}\natexlab{a}.
\newblock \showarticletitle{Time Interval Aware Self-Attention for Sequential
  Recommendation}. In \bibinfo{booktitle}{\emph{WSDM}}.
  \bibinfo{pages}{322--330}.
\newblock


\bibitem[\protect\citeauthoryear{Li, Kawale, and Fu}{Li et~al\mbox{.}}{2015}]%
        {li2015deep}
\bibfield{author}{\bibinfo{person}{Sheng Li}, \bibinfo{person}{Jaya Kawale},
  {and} \bibinfo{person}{Yun Fu}.} \bibinfo{year}{2015}\natexlab{}.
\newblock \showarticletitle{Deep collaborative filtering via marginalized
  denoising auto-encoder}. In \bibinfo{booktitle}{\emph{CIKM}}.
  \bibinfo{pages}{811--820}.
\newblock


\bibitem[\protect\citeauthoryear{Li, Zhao, Liu, Huang, Mei, and Chen}{Li
  et~al\mbox{.}}{2018}]%
        {li2018learning}
\bibfield{author}{\bibinfo{person}{Zhi Li}, \bibinfo{person}{Hongke Zhao},
  \bibinfo{person}{Qi Liu}, \bibinfo{person}{Zhenya Huang},
  \bibinfo{person}{Tao Mei}, {and} \bibinfo{person}{Enhong Chen}.}
  \bibinfo{year}{2018}\natexlab{}.
\newblock \showarticletitle{Learning from history and present: next-item
  recommendation via discriminatively exploiting user behaviors}. In
  \bibinfo{booktitle}{\emph{KDD}}. \bibinfo{pages}{1734--1743}.
\newblock


\bibitem[\protect\citeauthoryear{Linden, Smith, and York}{Linden
  et~al\mbox{.}}{2003}]%
        {Linden2003Amazon}
\bibfield{author}{\bibinfo{person}{Greg Linden}, \bibinfo{person}{Brent Smith},
  {and} \bibinfo{person}{Jeremy York}.} \bibinfo{year}{2003}\natexlab{}.
\newblock \showarticletitle{Amazon.com recommendations: item-to-item
  collaborative filtering}.
\newblock \bibinfo{journal}{\emph{IEEE Internet Computing}}
  \bibinfo{volume}{7}, \bibinfo{number}{1} (\bibinfo{year}{2003}),
  \bibinfo{pages}{76--80}.
\newblock


\bibitem[\protect\citeauthoryear{Liu, Shi, and Natarajan}{Liu
  et~al\mbox{.}}{2018a}]%
        {liu2018sequential}
\bibfield{author}{\bibinfo{person}{Kuan Liu}, \bibinfo{person}{Xing Shi}, {and}
  \bibinfo{person}{Prem Natarajan}.} \bibinfo{year}{2018}\natexlab{a}.
\newblock \bibinfo{title}{A Sequential Embedding Approach for Item
  Recommendation with Heterogeneous Attributes}.
\newblock
\newblock
\showeprint[arxiv]{cs.IR/1805.11008}


\bibitem[\protect\citeauthoryear{Liu, Wu, Wang, Li, and Wang}{Liu
  et~al\mbox{.}}{2016}]%
        {Liu2016ContextAwareSR}
\bibfield{author}{\bibinfo{person}{Qiang Liu}, \bibinfo{person}{Shu Wu},
  \bibinfo{person}{Diyi Wang}, \bibinfo{person}{Zhaokang Li}, {and}
  \bibinfo{person}{Liang Wang}.} \bibinfo{year}{2016}\natexlab{}.
\newblock \showarticletitle{Context-aware sequential recommendation}. In
  \bibinfo{booktitle}{\emph{ICDM}}. \bibinfo{pages}{1053--1058}.
\newblock


\bibitem[\protect\citeauthoryear{Liu, Wu, and Wang}{Liu et~al\mbox{.}}{2017}]%
        {liu2017multi}
\bibfield{author}{\bibinfo{person}{Qiang Liu}, \bibinfo{person}{Shu Wu}, {and}
  \bibinfo{person}{Liang Wang}.} \bibinfo{year}{2017}\natexlab{}.
\newblock \showarticletitle{Multi-behavioral sequential prediction with
  recurrent log-bilinear model}.
\newblock \bibinfo{journal}{\emph{TKDE}} \bibinfo{volume}{29},
  \bibinfo{number}{6} (\bibinfo{year}{2017}), \bibinfo{pages}{1254--1267}.
\newblock


\bibitem[\protect\citeauthoryear{Liu, Zeng, Mokhosi, and Zhang}{Liu
  et~al\mbox{.}}{2018b}]%
        {liu2018stamp}
\bibfield{author}{\bibinfo{person}{Qiao Liu}, \bibinfo{person}{Yifu Zeng},
  \bibinfo{person}{Refuoe Mokhosi}, {and} \bibinfo{person}{Haibin Zhang}.}
  \bibinfo{year}{2018}\natexlab{b}.
\newblock \showarticletitle{STAMP: short-term attention/memory priority model
  for session-based recommendation}. In \bibinfo{booktitle}{\emph{KDD}}.
  \bibinfo{pages}{1831--1839}.
\newblock


\bibitem[\protect\citeauthoryear{Loni, Pagano, Larson, and Hanjalic}{Loni
  et~al\mbox{.}}{2016}]%
        {loni2016bayesian}
\bibfield{author}{\bibinfo{person}{Babak Loni}, \bibinfo{person}{Roberto
  Pagano}, \bibinfo{person}{Martha Larson}, {and} \bibinfo{person}{Alan
  Hanjalic}.} \bibinfo{year}{2016}\natexlab{}.
\newblock \showarticletitle{Bayesian personalized ranking with multi-channel
  user feedback}. In \bibinfo{booktitle}{\emph{RecSys}}.
  \bibinfo{pages}{361--364}.
\newblock


\bibitem[\protect\citeauthoryear{Loyola, Liu, and Hirate}{Loyola
  et~al\mbox{.}}{2017}]%
        {loyola2017modeling}
\bibfield{author}{\bibinfo{person}{Pablo Loyola}, \bibinfo{person}{Chen Liu},
  {and} \bibinfo{person}{Yu Hirate}.} \bibinfo{year}{2017}\natexlab{}.
\newblock \showarticletitle{Modeling user session and intent with an
  attention-based encoder-decoder architecture}. In
  \bibinfo{booktitle}{\emph{RecSys}}. \bibinfo{pages}{147--151}.
\newblock


\bibitem[\protect\citeauthoryear{Ludewig and Jannach}{Ludewig and
  Jannach}{2018}]%
        {ludewig2018evaluation}
\bibfield{author}{\bibinfo{person}{Malte Ludewig} {and}
  \bibinfo{person}{Dietmar Jannach}.} \bibinfo{year}{2018}\natexlab{}.
\newblock \showarticletitle{Evaluation of session-based recommendation
  algorithms}.
\newblock \bibinfo{journal}{\emph{User Modeling and User-Adapted Interaction}}
  \bibinfo{volume}{28}, \bibinfo{number}{4-5} (\bibinfo{year}{2018}),
  \bibinfo{pages}{331--390}.
\newblock


\bibitem[\protect\citeauthoryear{Lv, Jin, Yu, Sun, Lin, Yang, and Ng}{Lv
  et~al\mbox{.}}{2019}]%
        {lv2019sdm}
\bibfield{author}{\bibinfo{person}{Fuyu Lv}, \bibinfo{person}{Taiwei Jin},
  \bibinfo{person}{Changlong Yu}, \bibinfo{person}{Fei Sun},
  \bibinfo{person}{Quan Lin}, \bibinfo{person}{Keping Yang}, {and}
  \bibinfo{person}{Wilfred Ng}.} \bibinfo{year}{2019}\natexlab{}.
\newblock \showarticletitle{SDM: Sequential deep matching model for online
  large-scale recommender system}. In \bibinfo{booktitle}{\emph{CIKM}}.
  \bibinfo{pages}{2635--2643}.
\newblock


\bibitem[\protect\citeauthoryear{Ma, Kang, and Liu}{Ma et~al\mbox{.}}{2019a}]%
        {ma2019hierarchical}
\bibfield{author}{\bibinfo{person}{Chen Ma}, \bibinfo{person}{Peng Kang}, {and}
  \bibinfo{person}{Xue Liu}.} \bibinfo{year}{2019}\natexlab{a}.
\newblock \showarticletitle{Hierarchical gating networks for sequential
  recommendation}. In \bibinfo{booktitle}{\emph{KDD}}.
  \bibinfo{pages}{825--833}.
\newblock


\bibitem[\protect\citeauthoryear{Ma, Na, Xu, and Fan}{Ma et~al\mbox{.}}{2018}]%
        {ma2018graph}
\bibfield{author}{\bibinfo{person}{Mingyuan Ma}, \bibinfo{person}{Sen Na},
  \bibinfo{person}{Cong Xu}, {and} \bibinfo{person}{Xin Fan}.}
  \bibinfo{year}{2018}\natexlab{}.
\newblock \showarticletitle{The graph-based broad behavior-aware recommendation
  system for interactive news}.
\newblock \bibinfo{journal}{\emph{arXiv preprint arXiv:1812.00002}}
  (\bibinfo{year}{2018}).
\newblock


\bibitem[\protect\citeauthoryear{Ma, Ren, Lin, Chen, Ma, and Rijke}{Ma
  et~al\mbox{.}}{2019b}]%
        {ma2019pi}
\bibfield{author}{\bibinfo{person}{Muyang Ma}, \bibinfo{person}{Pengjie Ren},
  \bibinfo{person}{Yujie Lin}, \bibinfo{person}{Zhumin Chen},
  \bibinfo{person}{Jun Ma}, {and} \bibinfo{person}{Maarten~de Rijke}.}
  \bibinfo{year}{2019}\natexlab{b}.
\newblock \showarticletitle{$\pi$-Net: A Parallel Information-sharing Network
  for Shared-account Cross-domain Sequential Recommendations}. In
  \bibinfo{booktitle}{\emph{SIGIR}}. \bibinfo{pages}{685--694}.
\newblock


\bibitem[\protect\citeauthoryear{Meng, Yang, and Xiao}{Meng
  et~al\mbox{.}}{2020}]%
        {meng2020incorporating}
\bibfield{author}{\bibinfo{person}{Wenjing Meng}, \bibinfo{person}{Deqing
  Yang}, {and} \bibinfo{person}{Yanghua Xiao}.}
  \bibinfo{year}{2020}\natexlab{}.
\newblock \showarticletitle{Incorporating User Micro-behaviors and Item
  Knowledge into Multi-task Learning for Session-based Recommendation}. In
  \bibinfo{booktitle}{\emph{SIGIR}}.
\newblock


\bibitem[\protect\citeauthoryear{Mikolov, Chen, Corrado, and Dean}{Mikolov
  et~al\mbox{.}}{2013}]%
        {mikolov2013efficient}
\bibfield{author}{\bibinfo{person}{Tomas Mikolov}, \bibinfo{person}{Kai Chen},
  \bibinfo{person}{Greg Corrado}, {and} \bibinfo{person}{Jeffrey Dean}.}
  \bibinfo{year}{2013}\natexlab{}.
\newblock \showarticletitle{Efficient estimation of word representations in
  vector space}.
\newblock \bibinfo{journal}{\emph{arXiv preprint arXiv:1301.3781}}
  (\bibinfo{year}{2013}).
\newblock


\bibitem[\protect\citeauthoryear{Mobasher, Burke, Bhaumik, and
  Williams}{Mobasher et~al\mbox{.}}{2007}]%
        {mobasher2007toward}
\bibfield{author}{\bibinfo{person}{Bamshad Mobasher}, \bibinfo{person}{Robin
  Burke}, \bibinfo{person}{Runa Bhaumik}, {and} \bibinfo{person}{Chad
  Williams}.} \bibinfo{year}{2007}\natexlab{}.
\newblock \showarticletitle{Toward trustworthy recommender systems: An analysis
  of attack models and algorithm robustness}.
\newblock \bibinfo{journal}{\emph{TOIT}} \bibinfo{volume}{7},
  \bibinfo{number}{4} (\bibinfo{year}{2007}), \bibinfo{pages}{23--es}.
\newblock


\bibitem[\protect\citeauthoryear{Montavon, Samek, and M{\"u}ller}{Montavon
  et~al\mbox{.}}{2018}]%
        {montavon2018methods}
\bibfield{author}{\bibinfo{person}{Gr{\'e}goire Montavon},
  \bibinfo{person}{Wojciech Samek}, {and} \bibinfo{person}{Klaus-Robert
  M{\"u}ller}.} \bibinfo{year}{2018}\natexlab{}.
\newblock \showarticletitle{Methods for interpreting and understanding deep
  neural networks}.
\newblock \bibinfo{journal}{\emph{Digital Signal Processing}}
  \bibinfo{volume}{73} (\bibinfo{year}{2018}), \bibinfo{pages}{1--15}.
\newblock


\bibitem[\protect\citeauthoryear{Nguyen, Wistuba, Grabocka, Drumond, and
  Schmidt-Thieme}{Nguyen et~al\mbox{.}}{2017}]%
        {nguyen2017personalized}
\bibfield{author}{\bibinfo{person}{Hanh~TH Nguyen}, \bibinfo{person}{Martin
  Wistuba}, \bibinfo{person}{Josif Grabocka}, \bibinfo{person}{Lucas~Rego
  Drumond}, {and} \bibinfo{person}{Lars Schmidt-Thieme}.}
  \bibinfo{year}{2017}\natexlab{}.
\newblock \showarticletitle{Personalized deep learning for tag recommendation}.
  In \bibinfo{booktitle}{\emph{PAKDD}}. \bibinfo{pages}{186--197}.
\newblock


\bibitem[\protect\citeauthoryear{Peake and Wang}{Peake and Wang}{2018}]%
        {peake2018explanation}
\bibfield{author}{\bibinfo{person}{Georgina Peake} {and} \bibinfo{person}{Jun
  Wang}.} \bibinfo{year}{2018}\natexlab{}.
\newblock \showarticletitle{Explanation mining: Post hoc interpretability of
  latent factor models for recommendation systems}. In
  \bibinfo{booktitle}{\emph{KDD}}. \bibinfo{pages}{2060--2069}.
\newblock


\bibitem[\protect\citeauthoryear{Pennington, Socher, and Manning}{Pennington
  et~al\mbox{.}}{2014}]%
        {pennington2014glove}
\bibfield{author}{\bibinfo{person}{Jeffrey Pennington},
  \bibinfo{person}{Richard Socher}, {and} \bibinfo{person}{Christopher
  Manning}.} \bibinfo{year}{2014}\natexlab{}.
\newblock \showarticletitle{Glove: Global vectors for word representation}. In
  \bibinfo{booktitle}{\emph{EMNLP}}. \bibinfo{pages}{1532--1543}.
\newblock


\bibitem[\protect\citeauthoryear{Qiu, Liu, Guo, Sun, Zhang, and Nguyen}{Qiu
  et~al\mbox{.}}{2018}]%
        {qiu2018bprh}
\bibfield{author}{\bibinfo{person}{Huihuai Qiu}, \bibinfo{person}{Yun Liu},
  \bibinfo{person}{Guibing Guo}, \bibinfo{person}{Zhu Sun},
  \bibinfo{person}{Jie Zhang}, {and} \bibinfo{person}{Hai~Thanh Nguyen}.}
  \bibinfo{year}{2018}\natexlab{}.
\newblock \showarticletitle{BPRH: Bayesian personalized ranking for
  heterogeneous implicit feedback}.
\newblock \bibinfo{journal}{\emph{Information Sciences}}  \bibinfo{volume}{453}
  (\bibinfo{year}{2018}), \bibinfo{pages}{80--98}.
\newblock


\bibitem[\protect\citeauthoryear{Quadrana, Cremonesi, and Jannach}{Quadrana
  et~al\mbox{.}}{2018}]%
        {quadrana2018sequence}
\bibfield{author}{\bibinfo{person}{Massimo Quadrana}, \bibinfo{person}{Paolo
  Cremonesi}, {and} \bibinfo{person}{Dietmar Jannach}.}
  \bibinfo{year}{2018}\natexlab{}.
\newblock \showarticletitle{Sequence-aware recommender systems}.
\newblock \bibinfo{journal}{\emph{ACM Computing Surveys (CSUR)}}
  \bibinfo{volume}{51}, \bibinfo{number}{4} (\bibinfo{year}{2018}),
  \bibinfo{pages}{1--36}.
\newblock


\bibitem[\protect\citeauthoryear{Quadrana, Karatzoglou, Hidasi, and
  Cremonesi}{Quadrana et~al\mbox{.}}{2017}]%
        {quadrana2017personalizing}
\bibfield{author}{\bibinfo{person}{Massimo Quadrana},
  \bibinfo{person}{Alexandros Karatzoglou}, \bibinfo{person}{Bal{\'a}zs
  Hidasi}, {and} \bibinfo{person}{Paolo Cremonesi}.}
  \bibinfo{year}{2017}\natexlab{}.
\newblock \showarticletitle{Personalizing session-based recommendations with
  hierarchical recurrent neural networks}. In
  \bibinfo{booktitle}{\emph{RecSys}}. \bibinfo{pages}{130--137}.
\newblock


\bibitem[\protect\citeauthoryear{Rawat and Kankanhalli}{Rawat and
  Kankanhalli}{2016}]%
        {rawat2016contagnet:}
\bibfield{author}{\bibinfo{person}{Yogesh~Singh Rawat} {and}
  \bibinfo{person}{Mohan~S Kankanhalli}.} \bibinfo{year}{2016}\natexlab{}.
\newblock \showarticletitle{ConTagNet: Exploiting user context for image tag
  recommendation}. In \bibinfo{booktitle}{\emph{MM}}.
  \bibinfo{pages}{1102--1106}.
\newblock


\bibitem[\protect\citeauthoryear{Ren, Qin, Fang, Zhang, Zheng, Bian, Zhou, Xu,
  Yu, Zhu, et~al\mbox{.}}{Ren et~al\mbox{.}}{2019b}]%
        {ren2019lifelong}
\bibfield{author}{\bibinfo{person}{Kan Ren}, \bibinfo{person}{Jiarui Qin},
  \bibinfo{person}{Yuchen Fang}, \bibinfo{person}{Weinan Zhang},
  \bibinfo{person}{Lei Zheng}, \bibinfo{person}{Weijie Bian},
  \bibinfo{person}{Guorui Zhou}, \bibinfo{person}{Jian Xu},
  \bibinfo{person}{Yong Yu}, \bibinfo{person}{Xiaoqiang Zhu}, {et~al\mbox{.}}}
  \bibinfo{year}{2019}\natexlab{b}.
\newblock \showarticletitle{Lifelong Sequential Modeling with Personalized
  Memorization for User Response Prediction}. In
  \bibinfo{booktitle}{\emph{SIGIR}}. \bibinfo{pages}{565--574}.
\newblock


\bibitem[\protect\citeauthoryear{Ren, Chen, Li, Ren, Ma, and de~Rijke}{Ren
  et~al\mbox{.}}{2019a}]%
        {ren2018repeatnet}
\bibfield{author}{\bibinfo{person}{Pengjie Ren}, \bibinfo{person}{Zhumin Chen},
  \bibinfo{person}{Jing Li}, \bibinfo{person}{Zhaochun Ren},
  \bibinfo{person}{Jun Ma}, {and} \bibinfo{person}{Maarten de Rijke}.}
  \bibinfo{year}{2019}\natexlab{a}.
\newblock \showarticletitle{RepeatNet: A repeat aware neural recommendation
  machine for session-based recommendation}. In
  \bibinfo{booktitle}{\emph{AAAI}}, Vol.~\bibinfo{volume}{33}.
  \bibinfo{pages}{4806--4813}.
\newblock


\bibitem[\protect\citeauthoryear{Rendle, Freudenthaler, Gantner, and
  Schmidt-Thieme}{Rendle et~al\mbox{.}}{2009}]%
        {rendle2009bpr}
\bibfield{author}{\bibinfo{person}{Steffen Rendle}, \bibinfo{person}{Christoph
  Freudenthaler}, \bibinfo{person}{Zeno Gantner}, {and} \bibinfo{person}{Lars
  Schmidt-Thieme}.} \bibinfo{year}{2009}\natexlab{}.
\newblock \showarticletitle{BPR: Bayesian personalized ranking from implicit
  feedback}. In \bibinfo{booktitle}{\emph{UAI}}. \bibinfo{pages}{452--461}.
\newblock


\bibitem[\protect\citeauthoryear{Rendle, Freudenthaler, and
  Schmidt-Thieme}{Rendle et~al\mbox{.}}{2010}]%
        {rendle2010factorizing}
\bibfield{author}{\bibinfo{person}{Steffen Rendle}, \bibinfo{person}{Christoph
  Freudenthaler}, {and} \bibinfo{person}{Lars Schmidt-Thieme}.}
  \bibinfo{year}{2010}\natexlab{}.
\newblock \showarticletitle{Factorizing personalized markov chains for
  next-basket recommendation}. In \bibinfo{booktitle}{\emph{WWW}}.
  \bibinfo{pages}{811--820}.
\newblock


\bibitem[\protect\citeauthoryear{Rendle, Gantner, Freudenthaler, and
  Schmidt-Thieme}{Rendle et~al\mbox{.}}{2011}]%
        {rendle2011fast}
\bibfield{author}{\bibinfo{person}{Steffen Rendle}, \bibinfo{person}{Zeno
  Gantner}, \bibinfo{person}{Christoph Freudenthaler}, {and}
  \bibinfo{person}{Lars Schmidt-Thieme}.} \bibinfo{year}{2011}\natexlab{}.
\newblock \showarticletitle{Fast context-aware recommendations with
  factorization machines}. In \bibinfo{booktitle}{\emph{SIGIR}}.
  \bibinfo{pages}{635--644}.
\newblock


\bibitem[\protect\citeauthoryear{Ruocco, Skrede, and Langseth}{Ruocco
  et~al\mbox{.}}{2017}]%
        {ruocco2017inter-session}
\bibfield{author}{\bibinfo{person}{Massimiliano Ruocco}, \bibinfo{person}{Ole
  Steinar~Lillest{\o}l Skrede}, {and} \bibinfo{person}{Helge Langseth}.}
  \bibinfo{year}{2017}\natexlab{}.
\newblock \showarticletitle{Inter-session modeling for session-based
  recommendation}. In \bibinfo{booktitle}{\emph{the 2nd Workshop on Deep
  Learning for Recommender Systems}}. \bibinfo{pages}{24--31}.
\newblock


\bibitem[\protect\citeauthoryear{Sachdeva, Gupta, and Pudi}{Sachdeva
  et~al\mbox{.}}{2018}]%
        {sachdeva2018attentive}
\bibfield{author}{\bibinfo{person}{Noveen Sachdeva}, \bibinfo{person}{Kartik
  Gupta}, {and} \bibinfo{person}{Vikram Pudi}.}
  \bibinfo{year}{2018}\natexlab{}.
\newblock \showarticletitle{Attentive neural architecture incorporating song
  features for music recommendation}. In \bibinfo{booktitle}{\emph{RecSys}}.
  \bibinfo{pages}{417--421}.
\newblock


\bibitem[\protect\citeauthoryear{Sachdeva, Manco, Ritacco, and Pudi}{Sachdeva
  et~al\mbox{.}}{2019}]%
        {sachdeva2019sequential}
\bibfield{author}{\bibinfo{person}{Noveen Sachdeva}, \bibinfo{person}{Giuseppe
  Manco}, \bibinfo{person}{Ettore Ritacco}, {and} \bibinfo{person}{Vikram
  Pudi}.} \bibinfo{year}{2019}\natexlab{}.
\newblock \showarticletitle{Sequential Variational Autoencoders for
  Collaborative Filtering}. In \bibinfo{booktitle}{\emph{WSDM}}.
\newblock


\bibitem[\protect\citeauthoryear{Schnake, Eberle, Lederer, Nakajima,
  Sch{\"u}tt, M{\"u}ller, and Montavon}{Schnake et~al\mbox{.}}{2020}]%
        {schnake2020xai}
\bibfield{author}{\bibinfo{person}{Thomas Schnake}, \bibinfo{person}{Oliver
  Eberle}, \bibinfo{person}{Jonas Lederer}, \bibinfo{person}{Shinichi
  Nakajima}, \bibinfo{person}{Kristof~T Sch{\"u}tt},
  \bibinfo{person}{Klaus-Robert M{\"u}ller}, {and}
  \bibinfo{person}{Gr{\'e}goire Montavon}.} \bibinfo{year}{2020}\natexlab{}.
\newblock \showarticletitle{XAI for Graphs: Explaining Graph Neural Network
  Predictions by Identifying Relevant Walks}.
\newblock \bibinfo{journal}{\emph{arXiv preprint arXiv:2006.03589}}
  (\bibinfo{year}{2020}).
\newblock


\bibitem[\protect\citeauthoryear{Shi, Keneshloo, Ramakrishnan, and Reddy}{Shi
  et~al\mbox{.}}{2018}]%
        {shi2018neural}
\bibfield{author}{\bibinfo{person}{Tian Shi}, \bibinfo{person}{Yaser
  Keneshloo}, \bibinfo{person}{Naren Ramakrishnan}, {and}
  \bibinfo{person}{Chandan~K Reddy}.} \bibinfo{year}{2018}\natexlab{}.
\newblock \showarticletitle{Neural abstractive text summarization with
  sequence-to-sequence models}.
\newblock \bibinfo{journal}{\emph{arXiv preprint arXiv:1812.02303}}
  (\bibinfo{year}{2018}).
\newblock


\bibitem[\protect\citeauthoryear{Shih and Chi}{Shih and Chi}{2018}]%
        {shih2018automatic}
\bibfield{author}{\bibinfo{person}{Shun-Yao Shih} {and}
  \bibinfo{person}{Heng-Yu Chi}.} \bibinfo{year}{2018}\natexlab{}.
\newblock \showarticletitle{Automatic, personalized, and flexible playlist
  generation using reinforcement learning}.
\newblock \bibinfo{journal}{\emph{arXiv preprint arXiv:1809.04214}}
  (\bibinfo{year}{2018}).
\newblock


\bibitem[\protect\citeauthoryear{Singh and Gordon}{Singh and Gordon}{2008}]%
        {singh2008relational}
\bibfield{author}{\bibinfo{person}{Ajit~P Singh} {and}
  \bibinfo{person}{Geoffrey~J Gordon}.} \bibinfo{year}{2008}\natexlab{}.
\newblock \showarticletitle{Relational learning via collective matrix
  factorization}. In \bibinfo{booktitle}{\emph{KDD}}.
  \bibinfo{pages}{650--658}.
\newblock


\bibitem[\protect\citeauthoryear{Singhal, Sinha, and Pant}{Singhal
  et~al\mbox{.}}{2017}]%
        {singhal2017use}
\bibfield{author}{\bibinfo{person}{Ayush Singhal}, \bibinfo{person}{Pradeep
  Sinha}, {and} \bibinfo{person}{Rakesh Pant}.}
  \bibinfo{year}{2017}\natexlab{}.
\newblock \showarticletitle{Use of deep learning in modern recommendation
  system: A summary of recent works}.
\newblock \bibinfo{journal}{\emph{IJCA}} (\bibinfo{year}{2017}).
\newblock


\bibitem[\protect\citeauthoryear{Smirnova and Vasile}{Smirnova and
  Vasile}{2017}]%
        {smirnova2017contextual}
\bibfield{author}{\bibinfo{person}{Elena Smirnova} {and}
  \bibinfo{person}{Flavian Vasile}.} \bibinfo{year}{2017}\natexlab{}.
\newblock \showarticletitle{Contextual sequence modeling for recommendation
  with recurrent neural networks}. In \bibinfo{booktitle}{\emph{the 2nd
  Workshop on Deep Learning for Recommender Systems}}. \bibinfo{pages}{2--9}.
\newblock


\bibitem[\protect\citeauthoryear{Soh, Sanner, White, and Jamieson}{Soh
  et~al\mbox{.}}{2017}]%
        {soh2017deep}
\bibfield{author}{\bibinfo{person}{Harold Soh}, \bibinfo{person}{Scott Sanner},
  \bibinfo{person}{Madeleine White}, {and} \bibinfo{person}{Greg Jamieson}.}
  \bibinfo{year}{2017}\natexlab{}.
\newblock \showarticletitle{Deep sequential recommendation for personalized
  adaptive user interfaces}. In \bibinfo{booktitle}{\emph{IUI}}.
  \bibinfo{pages}{589--593}.
\newblock


\bibitem[\protect\citeauthoryear{Song, Xiao, Wang, Charlin, Zhang, and
  Tang}{Song et~al\mbox{.}}{2019}]%
        {song2019session}
\bibfield{author}{\bibinfo{person}{Weiping Song}, \bibinfo{person}{Zhiping
  Xiao}, \bibinfo{person}{Yifan Wang}, \bibinfo{person}{Laurent Charlin},
  \bibinfo{person}{Ming Zhang}, {and} \bibinfo{person}{Jian Tang}.}
  \bibinfo{year}{2019}\natexlab{}.
\newblock \showarticletitle{Session-based social recommendation via dynamic
  graph attention networks}. In \bibinfo{booktitle}{\emph{WSDM}}.
  \bibinfo{pages}{555--563}.
\newblock


\bibitem[\protect\citeauthoryear{Song and Lee}{Song and Lee}{2018}]%
        {song2018augmenting}
\bibfield{author}{\bibinfo{person}{Younghun Song} {and}
  \bibinfo{person}{Jae-Gil Lee}.} \bibinfo{year}{2018}\natexlab{}.
\newblock \showarticletitle{Augmenting recurrent neural networks with
  high-order user-contextual preference for session-based recommendation}.
\newblock \bibinfo{journal}{\emph{arXiv preprint arXiv:1805.02983}}
  (\bibinfo{year}{2018}).
\newblock


\bibitem[\protect\citeauthoryear{Sottocornola, Symeonidis, and
  Zanker}{Sottocornola et~al\mbox{.}}{2018}]%
        {Sottocornola2018session-based}
\bibfield{author}{\bibinfo{person}{Gabriele Sottocornola},
  \bibinfo{person}{Panagiotis Symeonidis}, {and} \bibinfo{person}{Markus
  Zanker}.} \bibinfo{year}{2018}\natexlab{}.
\newblock \showarticletitle{Session-based news recommendations}. In
  \bibinfo{booktitle}{\emph{WWW}}. \bibinfo{pages}{1395--1399}.
\newblock


\bibitem[\protect\citeauthoryear{Srivastava, Hinton, Krizhevsky, Sutskever, and
  Salakhutdinov}{Srivastava et~al\mbox{.}}{2014}]%
        {srivastava2014dropout}
\bibfield{author}{\bibinfo{person}{Nitish Srivastava},
  \bibinfo{person}{Geoffrey Hinton}, \bibinfo{person}{Alex Krizhevsky},
  \bibinfo{person}{Ilya Sutskever}, {and} \bibinfo{person}{Ruslan
  Salakhutdinov}.} \bibinfo{year}{2014}\natexlab{}.
\newblock \showarticletitle{Dropout: a simple way to prevent neural networks
  from overfitting}.
\newblock \bibinfo{journal}{\emph{The Journal of Machine Learning Research}}
  \bibinfo{volume}{15}, \bibinfo{number}{1} (\bibinfo{year}{2014}),
  \bibinfo{pages}{1929--1958}.
\newblock


\bibitem[\protect\citeauthoryear{Sun, Liu, Wu, Pei, Lin, Ou, and Jiang}{Sun
  et~al\mbox{.}}{2019b}]%
        {BERT4Rec2019}
\bibfield{author}{\bibinfo{person}{Fei Sun}, \bibinfo{person}{Jun Liu},
  \bibinfo{person}{Jian Wu}, \bibinfo{person}{Changhua Pei},
  \bibinfo{person}{Xiao Lin}, \bibinfo{person}{Wenwu Ou}, {and}
  \bibinfo{person}{Peng Jiang}.} \bibinfo{year}{2019}\natexlab{b}.
\newblock \showarticletitle{BERT4Rec: Sequential recommendation with
  bidirectional encoder representations from transformer}. In
  \bibinfo{booktitle}{\emph{CIKM}}. \bibinfo{pages}{1441--1450}.
\newblock


\bibitem[\protect\citeauthoryear{Sun, Tang, Dai, and Zhou}{Sun
  et~al\mbox{.}}{2019c}]%
        {DBLP:journals/access/SunTDZ19}
\bibfield{author}{\bibinfo{person}{Shiming Sun}, \bibinfo{person}{Yuanhe Tang},
  \bibinfo{person}{Zemei Dai}, {and} \bibinfo{person}{Fu Zhou}.}
  \bibinfo{year}{2019}\natexlab{c}.
\newblock \showarticletitle{Self-Attention Network for Session-Based
  Recommendation With Streaming Data Input}.
\newblock \bibinfo{journal}{\emph{IEEE Access}} (\bibinfo{year}{2019}).
\newblock


\bibitem[\protect\citeauthoryear{Sun, Guo, Yang, Fang, Guo, Zhang, and
  Burke}{Sun et~al\mbox{.}}{2019a}]%
        {sun2019research}
\bibfield{author}{\bibinfo{person}{Zhu Sun}, \bibinfo{person}{Qing Guo},
  \bibinfo{person}{Jie Yang}, \bibinfo{person}{Hui Fang},
  \bibinfo{person}{Guibing Guo}, \bibinfo{person}{Jie Zhang}, {and}
  \bibinfo{person}{Robin Burke}.} \bibinfo{year}{2019}\natexlab{a}.
\newblock \showarticletitle{Research commentary on recommendations with side
  information: A survey and research directions}.
\newblock \bibinfo{journal}{\emph{Electronic Commerce Research and
  Applications}}  \bibinfo{volume}{37} (\bibinfo{year}{2019}),
  \bibinfo{pages}{100879}.
\newblock


\bibitem[\protect\citeauthoryear{Sun, Yu, Fang, Yang, Qu, and Geng}{Sun
  et~al\mbox{.}}{2020}]%
        {sun2020benchmarking}
\bibfield{author}{\bibinfo{person}{Zhu Sun}, \bibinfo{person}{Di Yu},
  \bibinfo{person}{Hui Fang}, \bibinfo{person}{Jie Yang},
  \bibinfo{person}{Zhang~Jie Qu, Xinghua}, {and} \bibinfo{person}{Cong Geng}.}
  \bibinfo{year}{2020}\natexlab{}.
\newblock \showarticletitle{re We Evaluating Rigorously? {B}enchmarking
  Recommendation for Reproducible Evaluation and Fair Comparison}. In
  \bibinfo{booktitle}{\emph{Recsys}}.
\newblock


\bibitem[\protect\citeauthoryear{Tan, Xu, and Liu}{Tan et~al\mbox{.}}{2016}]%
        {tan2016improved}
\bibfield{author}{\bibinfo{person}{Yong~Kiam Tan}, \bibinfo{person}{Xinxing
  Xu}, {and} \bibinfo{person}{Yong Liu}.} \bibinfo{year}{2016}\natexlab{}.
\newblock \showarticletitle{Improved recurrent neural networks for
  session-based recommendations}. In \bibinfo{booktitle}{\emph{the 1st Workshop
  on Deep Learning for Recommender Systems}}. \bibinfo{pages}{17--22}.
\newblock


\bibitem[\protect\citeauthoryear{Tang, Du, He, Yuan, Tian, and Chua}{Tang
  et~al\mbox{.}}{2019}]%
        {tang2019adversarial}
\bibfield{author}{\bibinfo{person}{Jinhui Tang}, \bibinfo{person}{Xiaoyu Du},
  \bibinfo{person}{Xiangnan He}, \bibinfo{person}{Fajie Yuan},
  \bibinfo{person}{Qi Tian}, {and} \bibinfo{person}{Tat-Seng Chua}.}
  \bibinfo{year}{2019}\natexlab{}.
\newblock \showarticletitle{Adversarial training towards robust multimedia
  recommender system}.
\newblock \bibinfo{journal}{\emph{TKDE}} (\bibinfo{year}{2019}).
\newblock


\bibitem[\protect\citeauthoryear{Tang and Wang}{Tang and Wang}{2018}]%
        {tang2018personalized}
\bibfield{author}{\bibinfo{person}{Jiaxi Tang} {and} \bibinfo{person}{Ke
  Wang}.} \bibinfo{year}{2018}\natexlab{}.
\newblock \showarticletitle{Personalized top-n sequential recommendation via
  convolutional sequence embedding}. In \bibinfo{booktitle}{\emph{WSDM}}.
  \bibinfo{pages}{565--573}.
\newblock


\bibitem[\protect\citeauthoryear{Tay, Anh~Tuan, and Hui}{Tay
  et~al\mbox{.}}{2018}]%
        {tay2018latent}
\bibfield{author}{\bibinfo{person}{Yi Tay}, \bibinfo{person}{Luu Anh~Tuan},
  {and} \bibinfo{person}{Siu~Cheung Hui}.} \bibinfo{year}{2018}\natexlab{}.
\newblock \showarticletitle{Latent relational metric learning via memory-based
  attention for collaborative ranking}. In \bibinfo{booktitle}{\emph{WWW}}.
  \bibinfo{pages}{729--739}.
\newblock


\bibitem[\protect\citeauthoryear{Tuan and Phuong}{Tuan and Phuong}{2017}]%
        {tuan20173d}
\bibfield{author}{\bibinfo{person}{Trinh~Xuan Tuan} {and}
  \bibinfo{person}{Tu~Minh Phuong}.} \bibinfo{year}{2017}\natexlab{}.
\newblock \showarticletitle{3D convolutional networks for session-based
  recommendation with content features}. In \bibinfo{booktitle}{\emph{RecSys}}.
  \bibinfo{pages}{138--146}.
\newblock


\bibitem[\protect\citeauthoryear{Twardowski}{Twardowski}{2016}]%
        {twardowski2016modelling}
\bibfield{author}{\bibinfo{person}{Bart{\l}omiej Twardowski}.}
  \bibinfo{year}{2016}\natexlab{}.
\newblock \showarticletitle{Modelling contextual information in session-aware
  recommender systems with neural networks}. In
  \bibinfo{booktitle}{\emph{RecSys}}. \bibinfo{pages}{273--276}.
\newblock


\bibitem[\protect\citeauthoryear{Vapnik and Vashist}{Vapnik and
  Vashist}{2009}]%
        {vapnik2009new}
\bibfield{author}{\bibinfo{person}{Vladimir Vapnik} {and}
  \bibinfo{person}{Akshay Vashist}.} \bibinfo{year}{2009}\natexlab{}.
\newblock \showarticletitle{A new learning paradigm: Learning using privileged
  information}.
\newblock \bibinfo{journal}{\emph{Neural Networks}} \bibinfo{volume}{22},
  \bibinfo{number}{5-6} (\bibinfo{year}{2009}), \bibinfo{pages}{544--557}.
\newblock


\bibitem[\protect\citeauthoryear{Vaswani, Shazeer, Parmar, Uszkoreit, Jones,
  Gomez, Kaiser, and Polosukhin}{Vaswani et~al\mbox{.}}{2017}]%
        {vaswani2017attention}
\bibfield{author}{\bibinfo{person}{Ashish Vaswani}, \bibinfo{person}{Noam
  Shazeer}, \bibinfo{person}{Niki Parmar}, \bibinfo{person}{Jakob Uszkoreit},
  \bibinfo{person}{Llion Jones}, \bibinfo{person}{Aidan~N Gomez},
  \bibinfo{person}{{\L}ukasz Kaiser}, {and} \bibinfo{person}{Illia
  Polosukhin}.} \bibinfo{year}{2017}\natexlab{}.
\newblock \showarticletitle{Attention is all you need}. In
  \bibinfo{booktitle}{\emph{NIPS}}. \bibinfo{pages}{5998--6008}.
\newblock


\bibitem[\protect\citeauthoryear{Wan, Wang, Liu, Bennett, and McAuley}{Wan
  et~al\mbox{.}}{2018}]%
        {wan2018representating}
\bibfield{author}{\bibinfo{person}{Mengting Wan}, \bibinfo{person}{Di Wang},
  \bibinfo{person}{Jie Liu}, \bibinfo{person}{Paul Bennett}, {and}
  \bibinfo{person}{Julian McAuley}.} \bibinfo{year}{2018}\natexlab{}.
\newblock \showarticletitle{Representing and Recommending Shopping Baskets with
  Complementarity, Compatibility and Loyalty}. In
  \bibinfo{booktitle}{\emph{CIKM}}. \bibinfo{publisher}{Association for
  Computing Machinery}, \bibinfo{address}{New York, NY, USA}.
\newblock
\urldef\tempurl%
\url{https://doi.org/10.1145/3269206.3271786}
\showDOI{\tempurl}


\bibitem[\protect\citeauthoryear{Wan, Lan, Wang, Guo, Xu, and Cheng}{Wan
  et~al\mbox{.}}{2015}]%
        {wan2015next}
\bibfield{author}{\bibinfo{person}{Shengxian Wan}, \bibinfo{person}{Yanyan
  Lan}, \bibinfo{person}{Pengfei Wang}, \bibinfo{person}{Jiafeng Guo},
  \bibinfo{person}{Jun Xu}, {and} \bibinfo{person}{Xueqi Cheng}.}
  \bibinfo{year}{2015}\natexlab{}.
\newblock \showarticletitle{Next basket recommendation with neural networks}.
  In \bibinfo{booktitle}{\emph{RecSys Posters}}.
\newblock


\bibitem[\protect\citeauthoryear{Wang, Zhang, Ma, Liu, and Ma}{Wang
  et~al\mbox{.}}{2019f}]%
        {wang2019modeling}
\bibfield{author}{\bibinfo{person}{Chenyang Wang}, \bibinfo{person}{Min Zhang},
  \bibinfo{person}{Weizhi Ma}, \bibinfo{person}{Yiqun Liu}, {and}
  \bibinfo{person}{Shaoping Ma}.} \bibinfo{year}{2019}\natexlab{f}.
\newblock \showarticletitle{Modeling Item-Specific Temporal Dynamics of Repeat
  Consumption for Recommender Systems}. In \bibinfo{booktitle}{\emph{WWW}}.
  \bibinfo{pages}{1977--1987}.
\newblock


\bibitem[\protect\citeauthoryear{Wang, Lian, and Ge}{Wang
  et~al\mbox{.}}{2019e}]%
        {wang2019binarized}
\bibfield{author}{\bibinfo{person}{Haoyu Wang}, \bibinfo{person}{Defu Lian},
  {and} \bibinfo{person}{Yong Ge}.} \bibinfo{year}{2019}\natexlab{e}.
\newblock \showarticletitle{Binarized collaborative filtering with distilling
  graph convolutional networks}.
\newblock \bibinfo{journal}{\emph{arXiv preprint arXiv:1906.01829}}
  (\bibinfo{year}{2019}).
\newblock


\bibitem[\protect\citeauthoryear{Wang, Wang, and Yeung}{Wang
  et~al\mbox{.}}{2015b}]%
        {wang2015collaborative}
\bibfield{author}{\bibinfo{person}{Hao Wang}, \bibinfo{person}{Naiyan Wang},
  {and} \bibinfo{person}{Dit-Yan Yeung}.} \bibinfo{year}{2015}\natexlab{b}.
\newblock \showarticletitle{Collaborative deep learning for recommender
  systems}. In \bibinfo{booktitle}{\emph{KDD}}. \bibinfo{pages}{1235--1244}.
\newblock


\bibitem[\protect\citeauthoryear{Wang, Xingjian, and Yeung}{Wang
  et~al\mbox{.}}{2016}]%
        {wang2016collaborative}
\bibfield{author}{\bibinfo{person}{Hao Wang}, \bibinfo{person}{SHI Xingjian},
  {and} \bibinfo{person}{Dit-Yan Yeung}.} \bibinfo{year}{2016}\natexlab{}.
\newblock \showarticletitle{Collaborative recurrent autoencoder: Recommend
  while learning to fill in the blanks}. In \bibinfo{booktitle}{\emph{NIPS}}.
  \bibinfo{pages}{415--423}.
\newblock


\bibitem[\protect\citeauthoryear{Wang, Chen, Zhu, Shen, and Zhang}{Wang
  et~al\mbox{.}}{2019b}]%
        {wang2019unified}
\bibfield{author}{\bibinfo{person}{Pengfei Wang}, \bibinfo{person}{Hanxiong
  Chen}, \bibinfo{person}{Yadong Zhu}, \bibinfo{person}{Huawei Shen}, {and}
  \bibinfo{person}{Yongfeng Zhang}.} \bibinfo{year}{2019}\natexlab{b}.
\newblock \showarticletitle{Unified Collaborative Filtering over Graph
  Embeddings}. In \bibinfo{booktitle}{\emph{SIGIR}}. \bibinfo{pages}{155--164}.
\newblock


\bibitem[\protect\citeauthoryear{Wang, Guo, Lan, Xu, Wan, and Cheng}{Wang
  et~al\mbox{.}}{2015a}]%
        {wang2015learning}
\bibfield{author}{\bibinfo{person}{Pengfei Wang}, \bibinfo{person}{Jiafeng
  Guo}, \bibinfo{person}{Yanyan Lan}, \bibinfo{person}{Jun Xu},
  \bibinfo{person}{Shengxian Wan}, {and} \bibinfo{person}{Xueqi Cheng}.}
  \bibinfo{year}{2015}\natexlab{a}.
\newblock \showarticletitle{Learning hierarchical representation model for next
  basket recommendation}. In \bibinfo{booktitle}{\emph{SIGIR}}.
  \bibinfo{pages}{403--412}.
\newblock


\bibitem[\protect\citeauthoryear{Wang, Cao, and Wang}{Wang
  et~al\mbox{.}}{2019a}]%
        {wang2019asurvey}
\bibfield{author}{\bibinfo{person}{Shoujin Wang}, \bibinfo{person}{Longbing
  Cao}, {and} \bibinfo{person}{Yan Wang}.} \bibinfo{year}{2019}\natexlab{a}.
\newblock \showarticletitle{A survey on session-based recommender systems}.
\newblock \bibinfo{journal}{\emph{arXiv preprint arXiv:1902.04864}}
  (\bibinfo{year}{2019}).
\newblock


\bibitem[\protect\citeauthoryear{Wang, Hu, Cao, Huang, Lian, and Liu}{Wang
  et~al\mbox{.}}{2018}]%
        {wang2018attention}
\bibfield{author}{\bibinfo{person}{Shoujin Wang}, \bibinfo{person}{Liang Hu},
  \bibinfo{person}{Longbing Cao}, \bibinfo{person}{Xiaoshui Huang},
  \bibinfo{person}{Defu Lian}, {and} \bibinfo{person}{Wei Liu}.}
  \bibinfo{year}{2018}\natexlab{}.
\newblock \showarticletitle{Attention-based transactional context embedding for
  next-item recommendation}. In \bibinfo{booktitle}{\emph{AAAI}}.
  \bibinfo{pages}{2532--2539}.
\newblock


\bibitem[\protect\citeauthoryear{Wang, Hu, Wang, Cao, Sheng, and Orgun}{Wang
  et~al\mbox{.}}{2019c}]%
        {wang2019sequential}
\bibfield{author}{\bibinfo{person}{Shoujin Wang}, \bibinfo{person}{Liang Hu},
  \bibinfo{person}{Yan Wang}, \bibinfo{person}{Longbing Cao},
  \bibinfo{person}{Quan~Z Sheng}, {and} \bibinfo{person}{Mehmet Orgun}.}
  \bibinfo{year}{2019}\natexlab{c}.
\newblock \showarticletitle{Sequential recommender systems: challenges,
  progress and prospects}. In \bibinfo{booktitle}{\emph{IJCAI}}. AAAI Press,
  \bibinfo{pages}{6332--6338}.
\newblock


\bibitem[\protect\citeauthoryear{Wang, Hu, Wang, Sheng, Orgun, and Cao}{Wang
  et~al\mbox{.}}{2019d}]%
        {DBLP:conf/ijcai/Wang0WSOC19}
\bibfield{author}{\bibinfo{person}{Shoujin Wang}, \bibinfo{person}{Liang Hu},
  \bibinfo{person}{Yan Wang}, \bibinfo{person}{Quan~Z. Sheng},
  \bibinfo{person}{Mehmet~A. Orgun}, {and} \bibinfo{person}{Longbing Cao}.}
  \bibinfo{year}{2019}\natexlab{d}.
\newblock \showarticletitle{Modeling Multi-Purpose Sessions for Next-Item
  Recommendations via Mixture-Channel Purpose Routing Networks}. In
  \bibinfo{booktitle}{\emph{IJCAI}}. \bibinfo{pages}{3771--3777}.
\newblock


\bibitem[\protect\citeauthoryear{Wang, Feng, He, Nie, and Chua}{Wang
  et~al\mbox{.}}{2020a}]%
        {wang2020denoising}
\bibfield{author}{\bibinfo{person}{Wenjie Wang}, \bibinfo{person}{Fuli Feng},
  \bibinfo{person}{Xiangnan He}, \bibinfo{person}{Liqiang Nie}, {and}
  \bibinfo{person}{Tat-Seng Chua}.} \bibinfo{year}{2020}\natexlab{a}.
\newblock \showarticletitle{Denoising Implicit Feedback for Recommendation}.
\newblock \bibinfo{journal}{\emph{arXiv preprint arXiv:2006.04153}}
  (\bibinfo{year}{2020}).
\newblock


\bibitem[\protect\citeauthoryear{Wang, Xu, He, Cao, Wang, and Chua}{Wang
  et~al\mbox{.}}{2020c}]%
        {wang2020reinforced}
\bibfield{author}{\bibinfo{person}{Xiang Wang}, \bibinfo{person}{Yaokun Xu},
  \bibinfo{person}{Xiangnan He}, \bibinfo{person}{Yixin Cao},
  \bibinfo{person}{Meng Wang}, {and} \bibinfo{person}{Tat-Seng Chua}.}
  \bibinfo{year}{2020}\natexlab{c}.
\newblock \showarticletitle{Reinforced Negative Sampling over Knowledge Graph
  for Recommendation}. In \bibinfo{booktitle}{\emph{WWW}}.
  \bibinfo{pages}{99--109}.
\newblock


\bibitem[\protect\citeauthoryear{Wang, Yao, Kwok, and Ni}{Wang
  et~al\mbox{.}}{2020d}]%
        {wang2019few}
\bibfield{author}{\bibinfo{person}{Yaqing Wang}, \bibinfo{person}{Quanming
  Yao}, \bibinfo{person}{James~T Kwok}, {and} \bibinfo{person}{Lionel~M Ni}.}
  \bibinfo{year}{2020}\natexlab{d}.
\newblock \showarticletitle{Generalizing from a few examples: A survey on
  few-shot learning}.
\newblock \bibinfo{journal}{\emph{CSUR}} (\bibinfo{year}{2020}),
  \bibinfo{pages}{1--34}.
\newblock


\bibitem[\protect\citeauthoryear{Wang, Wei, Cong, Li, Mao, and Qiu}{Wang
  et~al\mbox{.}}{2020b}]%
        {wang2020global}
\bibfield{author}{\bibinfo{person}{Ziyang Wang}, \bibinfo{person}{Wei Wei},
  \bibinfo{person}{Gao Cong}, \bibinfo{person}{Xiao-Li Li},
  \bibinfo{person}{Xian-Ling Mao}, {and} \bibinfo{person}{Minghui Qiu}.}
  \bibinfo{year}{2020}\natexlab{b}.
\newblock \showarticletitle{Global Context Enhanced Graph Neural Networks for
  Session-based Recommendation}. In \bibinfo{booktitle}{\emph{CIKM}}.
  \bibinfo{pages}{169--178}.
\newblock


\bibitem[\protect\citeauthoryear{Wei, He, Chen, Zhou, and Tang}{Wei
  et~al\mbox{.}}{2016}]%
        {wei2016collaborative}
\bibfield{author}{\bibinfo{person}{Jian Wei}, \bibinfo{person}{Jianhua He},
  \bibinfo{person}{Kai Chen}, \bibinfo{person}{Yi Zhou}, {and}
  \bibinfo{person}{Zuoyin Tang}.} \bibinfo{year}{2016}\natexlab{}.
\newblock \showarticletitle{Collaborative filtering and deep learning based
  hybrid recommendation for cold start problem}. In
  \bibinfo{booktitle}{\emph{DASC/PiCom/DataCom/CyberSciTech}}.
  \bibinfo{pages}{874--877}.
\newblock


\bibitem[\protect\citeauthoryear{Wu and Yan}{Wu and Yan}{2017}]%
        {wu2017session}
\bibfield{author}{\bibinfo{person}{Chen Wu} {and} \bibinfo{person}{Ming Yan}.}
  \bibinfo{year}{2017}\natexlab{}.
\newblock \showarticletitle{Session-aware information embedding for e-commerce
  product recommendation}. In \bibinfo{booktitle}{\emph{CIKM}}.
  \bibinfo{pages}{2379--2382}.
\newblock


\bibitem[\protect\citeauthoryear{Wu, Ahmed, Beutel, Smola, and Jing}{Wu
  et~al\mbox{.}}{2017}]%
        {wu2017recurrent}
\bibfield{author}{\bibinfo{person}{Chao-Yuan Wu}, \bibinfo{person}{Amr Ahmed},
  \bibinfo{person}{Alex Beutel}, \bibinfo{person}{Alexander~J Smola}, {and}
  \bibinfo{person}{How Jing}.} \bibinfo{year}{2017}\natexlab{}.
\newblock \showarticletitle{Recurrent recommender networks}. In
  \bibinfo{booktitle}{\emph{WSDM}}. \bibinfo{pages}{495--503}.
\newblock


\bibitem[\protect\citeauthoryear{Wu, Gao, Gao, Weng, and Chen}{Wu
  et~al\mbox{.}}{2019a}]%
        {wu2019dual}
\bibfield{author}{\bibinfo{person}{Qitian Wu}, \bibinfo{person}{Yirui Gao},
  \bibinfo{person}{Xiaofeng Gao}, \bibinfo{person}{Paul Weng}, {and}
  \bibinfo{person}{Guihai Chen}.} \bibinfo{year}{2019}\natexlab{a}.
\newblock \showarticletitle{Dual Sequential Prediction Models Linking
  Sequential Recommendation and Information Dissemination}. In
  \bibinfo{booktitle}{\emph{KDD}}. \bibinfo{pages}{447--457}.
\newblock


\bibitem[\protect\citeauthoryear{Wu, Tang, Zhu, Wang, Xie, and Tan}{Wu
  et~al\mbox{.}}{2019b}]%
        {wu2019session}
\bibfield{author}{\bibinfo{person}{Shu Wu}, \bibinfo{person}{Yuyuan Tang},
  \bibinfo{person}{Yanqiao Zhu}, \bibinfo{person}{Liang Wang},
  \bibinfo{person}{Xing Xie}, {and} \bibinfo{person}{Tieniu Tan}.}
  \bibinfo{year}{2019}\natexlab{b}.
\newblock \showarticletitle{Session-based recommendation with graph neural
  networks}. In \bibinfo{booktitle}{\emph{AAAI}}, Vol.~\bibinfo{volume}{33}.
  \bibinfo{pages}{346--353}.
\newblock


\bibitem[\protect\citeauthoryear{Wu, DuBois, Zheng, and Ester}{Wu
  et~al\mbox{.}}{2016}]%
        {wu2016collaborative}
\bibfield{author}{\bibinfo{person}{Yao Wu}, \bibinfo{person}{Christopher
  DuBois}, \bibinfo{person}{Alice~X Zheng}, {and} \bibinfo{person}{Martin
  Ester}.} \bibinfo{year}{2016}\natexlab{}.
\newblock \showarticletitle{Collaborative denoising auto-encoders for top-n
  recommender systems}. In \bibinfo{booktitle}{\emph{WSDM}}.
  \bibinfo{pages}{153--162}.
\newblock


\bibitem[\protect\citeauthoryear{Xia, Jiang, Sun, Zhang, Wang, and Sui}{Xia
  et~al\mbox{.}}{2018}]%
        {xia2018modeling}
\bibfield{author}{\bibinfo{person}{Qiaolin Xia}, \bibinfo{person}{Peng Jiang},
  \bibinfo{person}{Fei Sun}, \bibinfo{person}{Yi Zhang},
  \bibinfo{person}{Xiaobo Wang}, {and} \bibinfo{person}{Zhifang Sui}.}
  \bibinfo{year}{2018}\natexlab{}.
\newblock \showarticletitle{Modeling consumer buying decision for
  recommendation based on multi-task deep learning}. In
  \bibinfo{booktitle}{\emph{CIKM}}. \bibinfo{pages}{1703--1706}.
\newblock


\bibitem[\protect\citeauthoryear{Yang, Hsieh, Yang, Pollak, Dell, Belongie,
  Cole, and Estrin}{Yang et~al\mbox{.}}{2017}]%
        {yang2017yum}
\bibfield{author}{\bibinfo{person}{Longqi Yang}, \bibinfo{person}{Cheng-Kang
  Hsieh}, \bibinfo{person}{Hongjian Yang}, \bibinfo{person}{John~P Pollak},
  \bibinfo{person}{Nicola Dell}, \bibinfo{person}{Serge Belongie},
  \bibinfo{person}{Curtis Cole}, {and} \bibinfo{person}{Deborah Estrin}.}
  \bibinfo{year}{2017}\natexlab{}.
\newblock \showarticletitle{Yum-me: a personalized nutrient-based meal
  recommender system}.
\newblock \bibinfo{journal}{\emph{TOIS}} \bibinfo{volume}{36},
  \bibinfo{number}{1} (\bibinfo{year}{2017}), \bibinfo{pages}{1--31}.
\newblock


\bibitem[\protect\citeauthoryear{Yang, Qiu, Song, Tao, and Wang}{Yang
  et~al\mbox{.}}{2020b}]%
        {yang2020distilling}
\bibfield{author}{\bibinfo{person}{Yiding Yang}, \bibinfo{person}{Jiayan Qiu},
  \bibinfo{person}{Mingli Song}, \bibinfo{person}{Dacheng Tao}, {and}
  \bibinfo{person}{Xinchao Wang}.} \bibinfo{year}{2020}\natexlab{b}.
\newblock \showarticletitle{Distilling Knowledge From Graph Convolutional
  Networks}. In \bibinfo{booktitle}{\emph{CVPR}}. \bibinfo{pages}{7074--7083}.
\newblock


\bibitem[\protect\citeauthoryear{Yang, Ding, Zhou, Yang, Zhou, and Tang}{Yang
  et~al\mbox{.}}{2020a}]%
        {yang2020understanding}
\bibfield{author}{\bibinfo{person}{Zhen Yang}, \bibinfo{person}{Ming Ding},
  \bibinfo{person}{Chang Zhou}, \bibinfo{person}{Hongxia Yang},
  \bibinfo{person}{Jingren Zhou}, {and} \bibinfo{person}{Jie Tang}.}
  \bibinfo{year}{2020}\natexlab{a}.
\newblock \showarticletitle{Understanding Negative Sampling in Graph
  Representation Learning}. In \bibinfo{booktitle}{\emph{KDD}}.
\newblock


\bibitem[\protect\citeauthoryear{Ying, Zhuang, Zhang, Liu, Xu, Xie, Xiong, and
  Wu}{Ying et~al\mbox{.}}{2018}]%
        {ying2018sequential}
\bibfield{author}{\bibinfo{person}{Haochao Ying}, \bibinfo{person}{Fuzhen
  Zhuang}, \bibinfo{person}{Fuzheng Zhang}, \bibinfo{person}{Yanchi Liu},
  \bibinfo{person}{Guandong Xu}, \bibinfo{person}{Xing Xie},
  \bibinfo{person}{Hui Xiong}, {and} \bibinfo{person}{Jian Wu}.}
  \bibinfo{year}{2018}\natexlab{}.
\newblock \showarticletitle{Sequential recommender system based on hierarchical
  attention network}. In \bibinfo{booktitle}{\emph{IJCAI}}.
\newblock


\bibitem[\protect\citeauthoryear{You, Wang, Pal, Eksombatchai, Rosenburg, and
  Leskovec}{You et~al\mbox{.}}{2019}]%
        {you2019hierarchical}
\bibfield{author}{\bibinfo{person}{Jiaxuan You}, \bibinfo{person}{Yichen Wang},
  \bibinfo{person}{Aditya Pal}, \bibinfo{person}{Pong Eksombatchai},
  \bibinfo{person}{Chuck Rosenburg}, {and} \bibinfo{person}{Jure Leskovec}.}
  \bibinfo{year}{2019}\natexlab{}.
\newblock \showarticletitle{Hierarchical temporal convolutional networks for
  dynamic recommender systems}. In \bibinfo{booktitle}{\emph{WWW}}.
  \bibinfo{pages}{2236--2246}.
\newblock


\bibitem[\protect\citeauthoryear{Yu, Liu, Wu, Wang, and Tan}{Yu
  et~al\mbox{.}}{2016}]%
        {yu2016dynamic}
\bibfield{author}{\bibinfo{person}{Feng Yu}, \bibinfo{person}{Qiang Liu},
  \bibinfo{person}{Shu Wu}, \bibinfo{person}{Liang Wang}, {and}
  \bibinfo{person}{Tieniu Tan}.} \bibinfo{year}{2016}\natexlab{}.
\newblock \showarticletitle{A dynamic recurrent model for next basket
  recommendation}. In \bibinfo{booktitle}{\emph{SIGIR}}.
  \bibinfo{pages}{729--732}.
\newblock


\bibitem[\protect\citeauthoryear{Yu, Zhang, Liang, and Zhang}{Yu
  et~al\mbox{.}}{2019}]%
        {yu2019multi}
\bibfield{author}{\bibinfo{person}{Lu Yu}, \bibinfo{person}{Chuxu Zhang},
  \bibinfo{person}{Shangsong Liang}, {and} \bibinfo{person}{Xiangliang Zhang}.}
  \bibinfo{year}{2019}\natexlab{}.
\newblock \showarticletitle{Multi-order attentive ranking model for sequential
  recommendation}. In \bibinfo{booktitle}{\emph{AAAI}},
  Vol.~\bibinfo{volume}{33}. \bibinfo{pages}{5709--5716}.
\newblock


\bibitem[\protect\citeauthoryear{Yuan, He, Jiang, Guo, Xiong, Xu, and
  Xiong}{Yuan et~al\mbox{.}}{2020}]%
        {yuan2020future}
\bibfield{author}{\bibinfo{person}{Fajie Yuan}, \bibinfo{person}{Xiangnan He},
  \bibinfo{person}{Haochuan Jiang}, \bibinfo{person}{Guibing Guo},
  \bibinfo{person}{Jian Xiong}, \bibinfo{person}{Zhezhao Xu}, {and}
  \bibinfo{person}{Yilin Xiong}.} \bibinfo{year}{2020}\natexlab{}.
\newblock \showarticletitle{Future Data Helps Training: Modeling Future
  Contexts for Session-based Recommendation}. In
  \bibinfo{booktitle}{\emph{WWW}}. \bibinfo{pages}{303--313}.
\newblock


\bibitem[\protect\citeauthoryear{Yuan, Karatzoglou, Arapakis, Jose, and
  He}{Yuan et~al\mbox{.}}{2019a}]%
        {yuan2019simple}
\bibfield{author}{\bibinfo{person}{Fajie Yuan}, \bibinfo{person}{Alexandros
  Karatzoglou}, \bibinfo{person}{Ioannis Arapakis}, \bibinfo{person}{Joemon~M
  Jose}, {and} \bibinfo{person}{Xiangnan He}.}
  \bibinfo{year}{2019}\natexlab{a}.
\newblock \showarticletitle{A simple convolutional generative network for next
  item recommendation}. In \bibinfo{booktitle}{\emph{WSDM}}.
  \bibinfo{pages}{582--590}.
\newblock


\bibitem[\protect\citeauthoryear{Yuan, Yao, and Benatallah}{Yuan
  et~al\mbox{.}}{2019b}]%
        {yuan2019adversarial}
\bibfield{author}{\bibinfo{person}{Feng Yuan}, \bibinfo{person}{Lina Yao},
  {and} \bibinfo{person}{Boualem Benatallah}.}
  \bibinfo{year}{2019}\natexlab{b}.
\newblock \showarticletitle{Adversarial Collaborative Neural Network for Robust
  Recommendation}. In \bibinfo{booktitle}{\emph{SIGIR}}.
\newblock


\bibitem[\protect\citeauthoryear{Zhang and Yang}{Zhang and Yang}{2019}]%
        {GACOforRec}
\bibfield{author}{\bibinfo{person}{Mingge Zhang} {and} \bibinfo{person}{Zhenyu
  Yang}.} \bibinfo{year}{2019}\natexlab{}.
\newblock \showarticletitle{GACOforRec: Session-Based Graph Convolutional
  Neural Networks Recommendation Model}.
\newblock \bibinfo{journal}{\emph{IEEE Access}} (\bibinfo{year}{2019}).
\newblock


\bibitem[\protect\citeauthoryear{Zhang, Wang, Huang, Huang, and Gong}{Zhang
  et~al\mbox{.}}{2017}]%
        {zhang2017hashtag}
\bibfield{author}{\bibinfo{person}{Qi Zhang}, \bibinfo{person}{Jiawen Wang},
  \bibinfo{person}{Haoran Huang}, \bibinfo{person}{Xuanjing Huang}, {and}
  \bibinfo{person}{Yeyun Gong}.} \bibinfo{year}{2017}\natexlab{}.
\newblock \showarticletitle{Hashtag recommendation for multimodal microblog
  using co-attention network.}. In \bibinfo{booktitle}{\emph{IJCAI}}.
  \bibinfo{pages}{3420--3426}.
\newblock


\bibitem[\protect\citeauthoryear{Zhang, Tay, Yao, Sun, and An}{Zhang
  et~al\mbox{.}}{2019a}]%
        {zhang2019next}
\bibfield{author}{\bibinfo{person}{Shuai Zhang}, \bibinfo{person}{Yi Tay},
  \bibinfo{person}{Lina Yao}, \bibinfo{person}{Aixin Sun}, {and}
  \bibinfo{person}{Jake An}.} \bibinfo{year}{2019}\natexlab{a}.
\newblock \showarticletitle{Next item recommendation with self-attentive metric
  learning}. In \bibinfo{booktitle}{\emph{AAAI}}, Vol.~\bibinfo{volume}{9}.
\newblock


\bibitem[\protect\citeauthoryear{Zhang, Yao, Sun, and Tay}{Zhang
  et~al\mbox{.}}{2019b}]%
        {zhang2019deep}
\bibfield{author}{\bibinfo{person}{Shuai Zhang}, \bibinfo{person}{Lina Yao},
  \bibinfo{person}{Aixin Sun}, {and} \bibinfo{person}{Yi Tay}.}
  \bibinfo{year}{2019}\natexlab{b}.
\newblock \showarticletitle{Deep learning based recommender system: A survey
  and new perspectives}.
\newblock \bibinfo{journal}{\emph{ACM Computing Surveys (CSUR)}}
  \bibinfo{volume}{52}, \bibinfo{number}{1} (\bibinfo{year}{2019}),
  \bibinfo{pages}{5}.
\newblock


\bibitem[\protect\citeauthoryear{Zhang and Chen}{Zhang and Chen}{2018}]%
        {zhang2018explainable}
\bibfield{author}{\bibinfo{person}{Yongfeng Zhang} {and} \bibinfo{person}{Xu
  Chen}.} \bibinfo{year}{2018}\natexlab{}.
\newblock \showarticletitle{Explainable recommendation: A survey and new
  perspectives}.
\newblock \bibinfo{journal}{\emph{arXiv preprint arXiv:1804.11192}}
  (\bibinfo{year}{2018}).
\newblock


\bibitem[\protect\citeauthoryear{Zhang, Dai, Xu, Feng, Wang, Bian, Wang, and
  Liu}{Zhang et~al\mbox{.}}{2014}]%
        {zhang2014sequential}
\bibfield{author}{\bibinfo{person}{Yuyu Zhang}, \bibinfo{person}{Hanjun Dai},
  \bibinfo{person}{Chang Xu}, \bibinfo{person}{Jun Feng},
  \bibinfo{person}{Taifeng Wang}, \bibinfo{person}{Jiang Bian},
  \bibinfo{person}{Bin Wang}, {and} \bibinfo{person}{Tie-Yan Liu}.}
  \bibinfo{year}{2014}\natexlab{}.
\newblock \showarticletitle{Sequential click prediction for sponsored search
  with recurrent neural networks}. In \bibinfo{booktitle}{\emph{AAAI}}.
\newblock


\bibitem[\protect\citeauthoryear{Zhang, Feng, Wang, He, Wang, Li, and
  Zhang}{Zhang et~al\mbox{.}}{2020}]%
        {zhang2020retrain}
\bibfield{author}{\bibinfo{person}{Yang Zhang}, \bibinfo{person}{Fuli Feng},
  \bibinfo{person}{Chenxu Wang}, \bibinfo{person}{Xiangnan He},
  \bibinfo{person}{Meng Wang}, \bibinfo{person}{Yan Li}, {and}
  \bibinfo{person}{Yongdong Zhang}.} \bibinfo{year}{2020}\natexlab{}.
\newblock \showarticletitle{How to Retrain Recommender System? A Sequential
  Meta-Learning Method}. In \bibinfo{booktitle}{\emph{SIGIR}}.
\newblock


\bibitem[\protect\citeauthoryear{Zhao, Wang, Yang, Ye, Zhao, Chen, and
  Shen}{Zhao et~al\mbox{.}}{2019}]%
        {zhao2019leveraging}
\bibfield{author}{\bibinfo{person}{Wei Zhao}, \bibinfo{person}{Benyou Wang},
  \bibinfo{person}{Min Yang}, \bibinfo{person}{Jianbo Ye},
  \bibinfo{person}{Zhou Zhao}, \bibinfo{person}{Xiaojun Chen}, {and}
  \bibinfo{person}{Ying Shen}.} \bibinfo{year}{2019}\natexlab{}.
\newblock \showarticletitle{Leveraging Long and Short-Term Information in
  Content-Aware Movie Recommendation via Adversarial Training}.
\newblock \bibinfo{journal}{\emph{IEEE Transactions on Cybernetics}}
  (\bibinfo{year}{2019}).
\newblock


\bibitem[\protect\citeauthoryear{Zhao, Cheng, Hong, and Chi}{Zhao
  et~al\mbox{.}}{2015}]%
        {zhao2015improving}
\bibfield{author}{\bibinfo{person}{Zhe Zhao}, \bibinfo{person}{Zhiyuan Cheng},
  \bibinfo{person}{Lichan Hong}, {and} \bibinfo{person}{Ed~H Chi}.}
  \bibinfo{year}{2015}\natexlab{}.
\newblock \showarticletitle{Improving user topic interest profiles by behavior
  factorization}. In \bibinfo{booktitle}{\emph{WWW}}.
  \bibinfo{pages}{1406--1416}.
\newblock


\bibitem[\protect\citeauthoryear{Zhou, Bai, Song, Liu, Zhao, Chen, and
  Gao}{Zhou et~al\mbox{.}}{2018a}]%
        {zhou2018atrank}
\bibfield{author}{\bibinfo{person}{Chang Zhou}, \bibinfo{person}{Jinze Bai},
  \bibinfo{person}{Junshuai Song}, \bibinfo{person}{Xiaofei Liu},
  \bibinfo{person}{Zhengchao Zhao}, \bibinfo{person}{Xiusi Chen}, {and}
  \bibinfo{person}{Jun Gao}.} \bibinfo{year}{2018}\natexlab{a}.
\newblock \showarticletitle{{ATR}ank: An attention-Based user behavior modeling
  framework for recommendation}. In \bibinfo{booktitle}{\emph{AAAI}}.
\newblock


\bibitem[\protect\citeauthoryear{Zhou, Cui, Zhang, Yang, Liu, and Sun}{Zhou
  et~al\mbox{.}}{2018b}]%
        {zhou2018graph}
\bibfield{author}{\bibinfo{person}{Jie Zhou}, \bibinfo{person}{Ganqu Cui},
  \bibinfo{person}{Zhengyan Zhang}, \bibinfo{person}{Cheng Yang},
  \bibinfo{person}{Zhiyuan Liu}, {and} \bibinfo{person}{Maosong Sun}.}
  \bibinfo{year}{2018}\natexlab{b}.
\newblock \showarticletitle{Graph neural networks: a review of methods and
  applications}.
\newblock \bibinfo{journal}{\emph{arXiv preprint arXiv:1812.08434}}
  (\bibinfo{year}{2018}).
\newblock


\bibitem[\protect\citeauthoryear{Zhuang, Zhou, Zhang, Ao, Xie, and He}{Zhuang
  et~al\mbox{.}}{2017}]%
        {zhuang2017sequential}
\bibfield{author}{\bibinfo{person}{Fuzhen Zhuang}, \bibinfo{person}{Yingmin
  Zhou}, \bibinfo{person}{Fuzheng Zhang}, \bibinfo{person}{Xiang Ao},
  \bibinfo{person}{Xing Xie}, {and} \bibinfo{person}{Qing He}.}
  \bibinfo{year}{2017}\natexlab{}.
\newblock \showarticletitle{Sequential transfer learning: cross-domain novelty
  seeking trait mining for recommendation}. In \bibinfo{booktitle}{\emph{WWW
  Companion}}. \bibinfo{pages}{881--882}.
\newblock


\end{thebibliography}

\end{document}